\documentclass[12pt]{article}
\usepackage{amsmath,amssymb,amsthm,amsxtra,overpic,bbm,bm,epsfig,subfigure}
\usepackage{hyperref}

\usepackage{mathrsfs}
\usepackage{enumitem}
\usepackage{graphicx}
\usepackage{color}
\usepackage{comment}
\usepackage{epstopdf}
\usepackage{float}
\usepackage{appendix}
\usepackage{cite}
\usepackage{multirow}

\usepackage{slashed,stmaryrd}

\def\thefootnote{\fnsymbol{footnote}}

\begin{document}
	
\vspace{0.2cm}
\begin{center}
{\Large \bf Effective neutrino masses in KATRIN and future tritium beta-decay experiments}
\end{center}

\vspace{0.2cm}

\begin{center}
{\bf Guo-yuan Huang$^{a,\,b}$}\,\footnote{E-mail: huanggy@ihep.ac.cn},
\quad
{\bf Werner Rodejohann$^{c}$}\,\footnote{E-mail: werner.rodejohann@mpi-hd.mpg.de},
\quad
{\bf Shun Zhou$^{a,\,b}$}\,\footnote{E-mail: zhoush@ihep.ac.cn} \\
\vspace{0.2cm}
{\small
$^a$Institute of High Energy Physics, Chinese Academy of
Sciences, Beijing 100049, China \\
$^b$School of Physical Sciences,
University of Chinese Academy of Sciences, Beijing 100049, China \\
$^c$Max-Planck-Institut f\"ur Kernphysik, Postfach
103980, D-69029 Heidelberg, Germany}
\end{center}

\vspace{1.5cm}

\begin{abstract}
\noindent
Past and current direct neutrino mass experiments set limits on the so-called effective
neutrino mass, which is an incoherent sum of neutrino masses and lepton mixing matrix elements.
The electron energy spectrum which neglects the relativistic and nuclear recoil effects is often assumed. Alternative definitions of effective masses exist, and an exact relativistic spectrum is calculable. We quantitatively compare the validity of those different approximations as function of energy resolution and exposure in view of tritium beta decays in the KATRIN, Project 8 and PTOLEMY experiments. Furthermore, adopting the Bayesian approach, we present the posterior distributions of the effective neutrino mass by including current experimental information from neutrino oscillations, beta decay, neutrinoless double-beta decay and cosmological observations. Both linear and logarithmic priors for the smallest neutrino mass are assumed.

%The absolute scale of neutrino masses has to be determined from  observations other than neutrino oscillations, since the latter are sensitive to two independent neutrino mass-squared differences. The latest measurement of the electron energy spectrum in tritium beta decays from KATRIN sets the most stringent upper bound on the effective neutrino mass  $m^{}_\beta \equiv (|U^{}_{e1}|^2 m^2_1 + |U^{}_{e2}|^2 m^2_2 + |U^{}_{e3}|^2 m^2_3)^{1/2} < 1.1~{\rm eV}$ at $90\%$ confidence level, where $U^{}_{ei}$ and $m^{}_i$ (for $i = 1, 2, 3$) stand for the first-row elements of the leptonic flavor mixing matrix $U$ and neutrino masses, respectively. In this work, we reexamine the validity of the effective neutrino mass $m^{}_\beta$ and its variants in current and future tritium beta-decay experiments in a quantitative way. In addition, adopting the Bayesian approach, we present the posterior distribution of the effective neutrino mass by including current experimental information from neutrino oscillations, beta decays, neutrinoless double-beta decays and cosmological observations.
\end{abstract}

\def\thefootnote{\arabic{footnote}}
\setcounter{footnote}{0}

\newpage

\section{Introduction}

Neutrino oscillation experiments have measured with  very good precision the three leptonic flavor mixing angles $\{\theta^{}_{12}, \theta^{}_{13}, \theta^{}_{23}\}$ and two independent neutrino mass-squared differences $\Delta m^2_{21} \equiv m^2_2 - m^2_1$ and $|\Delta m^2_{31}| \equiv |m^2_3 - m^2_1|$. The absolute scale of neutrino masses, however, has to be determined from non-oscillation approaches, using beta decay \cite{Otten:2008zz},  neutrinoless double-beta decay \cite{Dolinski:2019nrj}, or cosmological observations \cite{Lattanzi:2017ubx}.
%According to the latest global-fit analysis of neutrino oscillation data~\cite{Esteban:2018azc}, we have $\Delta m^2_{21} \approx 7.4\times 10^{-5}~{\rm eV}^2$ and $|\Delta m^2_{31}| \approx 2.5\times 10^{-3}~{\rm eV}^2$. Therefore,
Once the neutrino mass scale is established, one knows
the lightest neutrino mass, which is $m^{}_1$ in the case of normal neutrino mass ordering (NO) with $m^{}_1 < m^{}_2 < m^{}_3$, or $m^{}_3$ in the case of inverted neutrino mass ordering (IO) with $m^{}_3 < m^{}_1 < m^{}_2$.
%is one of the most important yet-unknown parameters in nature and should be pinned down in order to understand the origin of neutrino masses.

As first suggested by Enrico Fermi and Francis Perrin~\cite{Fermi:1934hr, Fermi:1934sk,Perrin:1933}, the precise measurement of the electron energy spectrum in nuclear beta decays ${^{A}_{Z}N} \to {^{A}_{Z+1}N} + e^- + \overline{\nu}^{}_e$, where $A$ and $Z$ denote the mass and atomic number of the decaying nucleus, can be utilized to probe absolute neutrino masses.   Since the energy released in beta decays is distributed to massive  neutrinos, the energy spectrum of electrons in the region close to its endpoint will be distorted in comparison to that in the limit of zero neutrino masses. This  kinematic effect is usually described by the effective neutrino mass \cite{Shrock:1980vy}
\begin{eqnarray}\label{eq:mbeta}
m^{}_\beta \equiv \sqrt{|U^{}_{e1}|^2 m^2_1 + |U^{}_{e2}|^2 m^2_2 + |U^{}_{e3}|^2 m^2_3} \; ,
%     (1)
\end{eqnarray}
where $U^{}_{ei}$ (for $i = 1, 2, 3$) stand for the first-row elements of the leptonic flavor mixing matrix $U$,
i.e., $|U^{}_{e1}| = \cos{\theta^{}_{13}} \cos{\theta^{}_{12}}$, $|U^{}_{e2}| = \cos{\theta^{}_{13}} \sin{\theta^{}_{12}}$, $|U^{}_{e3}| = \sin{\theta^{}_{13}}$ in the standard parametrization~\cite{Tanabashi:2018oca}, and $m^{}_i$ (for $i = 1, 2, 3$) for the absolute neutrino masses.
Very recently, the KATRIN collaboration has reported its first result on the effective neutrino mass using tritium beta decay ${^3{\rm H}} \to {^3{\rm He}} + e^- + \overline{\nu}^{}_e$, and reached the currently most stringent upper bound~\cite{Aker:2019uuj, Aker:2019qfn}
\begin{eqnarray}\label{eq:katrin}
m^{}_\beta < 1.1~{\rm eV} \; ,
%     (2)
\end{eqnarray}
at the $90\%$ confidence level (CL). With the full exposure in the near future, KATRIN aims for an ultimate limit of $m^{}_\beta < 0.2~{\rm eV}$ at the same CL~\cite{Angrik:2005ep}, which is an order of magnitude better than the result $m^{}_\beta \lesssim 2~{\rm eV}$ from the Mainz~\cite{Kraus:2004zw} and Troitsk~\cite{Aseev:2011dq} experiments.

Motivated by this impressive achievement of the KATRIN experiment, we revisit the validity of the effective neutrino mass $m^{}_\beta$ in Eq.~(\ref{eq:mbeta}) and clarify how it depends on the energy resolution and the sensitivity of a realistic experiment for tritium beta decays.\footnote{In this work, we focus only on tritium beta-decay experiments. Similar analyses of the effective neutrino mass can be performed for the electron-capture decays of holmium, namely, $e^- + {^{163}{\rm Ho}} \to \nu^{}_e + {^{163}{\rm Dy}}$, which are and will be investigated in the ECHo~\cite{Gastaldo:2017edk} and HOLMES~\cite{Alpert:2014lfa}, NuMECS~\cite{Engle:2013qka} experiments.}
More explicitly, we shall consider the KATRIN \cite{Angrik:2005ep}, Project 8 \cite{Esfahani:2017dmu} and PTOLEMY \cite{Betts:2013uya, Baracchini:2018wwj, Betti:2019ouf} experiments. Their main features and projected sensitivities  have been summarized in Appendix~\ref{sec:appA} and Table~\ref{table:betaDecaysChi2}.
We compare $m^{}_\beta$ in Eq.~(\ref{eq:mbeta}) with other effective neutrino masses proposed in the literature, and also consider the exact relativistic spectrum of tritium beta decays.
A measure of the validity of $m^{}_\beta$ in terms of the exposure and energy resolution for a beta-decay experiment can be set.
As a result, we find that the standard effective mass $m^{}_\beta$ and the classical spectrum form can be used for KATRIN and Project 8 essentially without losing accuracy.

Furthermore, it is of interest to estimate how likely a signal in upcoming neutrino mass experiments, including those using electron-capture, is. Towards this end, we perform a Bayesian analysis to obtain the posterior distributions of $m_\beta$. 
The probability to find the beta-decay signal depends on the experimental likelihood input one considers, in particular the neutrino mass  information from cosmology  and neutrinoless double beta decays. The cosmological constraints on neutrino masses reply on the datasets one has included in generating the likelihood, whereas the constraints from neutrinoless double beta decays are subject to the assumption whether neutrinos are Dirac or Majorana particles.
It is thus quantified what the consequences of adding more and more additional mass information are. Moreover, the prior on the smallest neutrino mass, which could be linear or logarithmic, is important for the final posteriors. 

The remaining part of our paper is organized as follows. In Sec.~\ref{sec:effmass}, we make a comparison between the exact relativistic spectrum of electrons from tritium beta decays with the ordinary one with an effective neutrino mass $m^{}_\beta$. Then, a quantitative assessment of the validity of the effective neutrino mass is carried out. The posterior distributions of the effective neutrino mass are calculated in Sec.~\ref{sec:Bayes}, where the present experimental information from neutrino oscillations, neutrinoless double beta decays and cosmology are included. Finally, we summarize our main results in Sec.~\ref{sec:conclusion}. Technical details on the considered experiments and on the likelihoods used for our Bayesian analysis are delegated to appendices.

\section{The Effective Neutrino Mass}\label{sec:effmass}

\subsection{The relativistic electron spectrum}
Before introducing the effective neutrino mass for beta decays, we present the exact relativistic energy spectrum of the outgoing electrons for tritium beta decays (or equivalently the differential decay rate), which can be calculated within standard electroweak theory~\cite{Masood:2007rc,Simkovic:2007yi,Long:2014zva,Ludl:2016ane}, the result being
\begin{eqnarray}\label{eq:betaExact}
\frac{\mathrm{d} \Gamma^{}_{\rm rel}}{\mathrm{d} K^{}_{e}} & = & N^{}_{\rm T} \frac{\overline{\sigma}(E^{}_e)}{\pi^2} \sum^3_{i=1} |U^{}_{ei}|^2 H(E^{}_{e},m^{}_{i}) \; .
%     (3)
\end{eqnarray}
Here $N^{}_{\rm T}$ is the target mass of $^3{\rm H}$ and $E^{}_{e} = K^{}_e + m^{}_e$ is the electron energy with $K^{}_e$ being its kinetic energy. In Eq.~(\ref{eq:betaExact}), the reduced cross section is given by
\begin{eqnarray}\label{eq:sigmabar}
\overline{\sigma}(E^{}_e) \equiv \frac{G^2_{\rm F}}{2\pi}|V^{}_{\rm ud}|^2 F(Z, E^{}_{e})\frac{m^{}_{^3 {\rm  He}}}{m^{}_{^3 {\rm H}}} E^{}_e \sqrt{E^2_{e} - m^2_e} \left[\langle f^{}_{\rm F}\rangle^2 + \left(\frac{C^{}_{\rm A}}{C^{}_{\rm V}}\right)^2 \langle g^{}_{\rm GT}\rangle^2\right]  ,
%     (4)
\end{eqnarray}
where $G^{}_{\rm F} = 1.166\times 10^{-5}~{\rm GeV}^{-2}$ is the Fermi constant, $|V^{}_{\rm ud}| \approx \cos \theta^{}_{\rm C}$ is determined by the Cabibbo angle $\theta^{}_{\rm C} \approx 12.8^\circ$, $F(Z, E^{}_e)$ is the ordinary Fermi function with $Z = 1$ for tritium taking account of the distortion of the electron wave function in the Coulomb potential of the decaying nucleus\footnote{The Fermi function is given by $F(Z, E^{}_e) = 2\pi \eta/(1 - e^{-2\pi \eta})$, where $\eta \equiv Z \alpha E^{}_e/p^{}_e$ with $p^{}_e = \sqrt{E^2_e - m^2_e}$ being the electron momentum and $\alpha \approx 1/137$  the fine-structure constant.}, $C^{}_{\rm V} \approx 1$ and $C^{}_{\rm A} \approx 1.2695$ stand for the vector and axial-vector coupling constants of the charged-current weak interaction of nucleons, respectively. In addition, $\langle f^{}_{\rm F}\rangle^2 \approx 0.9987$ and $\langle g^{}_{\rm GT}\rangle^2 \approx 2.788$ are the squared nuclear matrix elements of the allowed Fermi and Gamow-Teller transitions. The kinematics of the tritium beta decays is encoded in the function $H(E^{}_{e}, m^{}_{i})$ in Eq.~(\ref{eq:betaExact}), namely,
\begin{eqnarray}\label{eq:Hfunction}
H(E^{}_{e},m^{}_{i}) \equiv \frac{1-m^2_e/\left(m^{}_{^3{\rm H}}E^{}_{e}\right)}{\left(1-2E^{}_{e}/m^{}_{^3{\rm H}}+ m^2_e/m^2_{^3{\rm H}}\right)^2}\sqrt{y\left(y + \frac{2 m^{}_{i} m^{}_{^3{\rm He}}}{m^{}_{^3{\rm H}}}\right)} \left[y + \frac{m^{}_{i}}{m^{}_{^3{\rm H}}}(m^{}_{^3{\rm He}} + m^{}_{i})\right]  ,
%     (5)
\end{eqnarray}
where $y \equiv K^{}_{\rm end} - K^{}_{e}$ with $K^{}_{\rm end}= \left[(m^{}_{^3{\rm H}}-m^{}_{e})^2-(m^{}_{^3{\rm He}} + m^{}_{i})^2\right]/(2m^{}_{^3{\rm H}})$ being the endpoint energy corresponding to the neutrino mass $m^{}_{i}$. Some comments on the kinematics are in order:
\begin{itemize}
\item Given the nuclear masses $m^{}_{^3{\rm H}} \approx 2~808~920.8205~{\rm keV}$ and $m^{}_{^3{\rm He}} \approx 2~808~391.2193~{\rm keV}$~\cite{Long:2014zva}, as well as the electron mass $m^{}_e \approx 510.9989~{\rm keV}$, one can obtain the $Q$-value of tritium beta decay $Q \equiv m^{}_{{^3{\rm H}}} - m^{}_{{^3{\rm He}}} - m^{}_e \approx 18.6023~{\rm keV}$. In the limit of vanishing neutrino masses, the endpoint energy $K^{}_{\rm end}$ turns out to be
    \begin{eqnarray}\label{eq:K0end}
    K^{}_{\rm end,0} \equiv \left[(m^{}_{^3{\rm H}}-m^{}_{e})^2-m^{2}_{^3{\rm He}}\right]/(2m^{}_{^3{\rm H}}) \approx 18.5989~{\rm keV} \; ,
    %     (6)
    \end{eqnarray}
    which is lower than the $Q$-value by a small amount of $Q - K^{}_{\rm end, 0} \approx 3.4~{\rm eV}$. This difference arises from the recoil energies of the final-state particles and is naturally included when one  considers fully relativistic kinematics. Since the electron spectrum near its endpoint is sensitive to absolute neutrino masses, which are much smaller than this energy difference of $3.4~{\rm eV}$, it is not  appropriate to treat the $Q$-value as the endpoint energy. 
\item It is straightforward to verify that $K^{}_{\rm end} \approx K^{}_{\rm end, 0} - m^{}_i m^{}_{^3{\rm He}}/m^{}_{^3{\rm H}}$ and thus $y + m^{}_i m^{}_{^3{\rm He}}/m^{}_{^3{\rm H}} \approx K^{}_{\rm end, 0} - K^{}_e$, where a tiny term $m^2_i/(2m^{}_{^3{\rm H}}) < 1.78\times 10^{-10}~{\rm eV}$ for $m^{}_i < 1~{\rm eV}$ can be safely ignored. Taking this approximation on the right-hand side of Eq.~(\ref{eq:Hfunction}), we can recast the kinematical function into
    \begin{eqnarray}\label{eq:Hfunctionapp}
    H(E^{}_e, m^{}_i) \approx \frac{1 - m^2_e/\left(m^{}_{^3{\rm H}}E^{}_{e}\right)}{\left(1 - 2E^{}_{e}/m^{}_{^3{\rm H}} + m^2_e/m^2_{^3{\rm H}}\right)^2} \sqrt{(K^{}_{\rm end,0} - K^{}_e)^2 - \left(m^{}_{i}\frac{m^{}_{^3 {\rm  He}}}{m^{}_{^3 {\rm H}}}\right)^2} (K^{}_{\rm end,0} - K^{}_{e}) \; , \quad
    %     (7)
    \end{eqnarray}
    from which it is interesting to observe that the absolute neutrino mass $m^{}_i$ in the square root receives a correction factor ${m^{}_{^3 {\rm He}}}/{m^{}_{^3 {\rm H}}} \approx 0.999811$. The difference between Eqs.~(\ref{eq:Hfunctionapp}) and (\ref{eq:Hfunction}) is negligibly small, so the former will be used in the following discussions.
\end{itemize}

Furthermore, given $1 - {m^{}_{^3 {\rm He}}}/{m^{}_{^3 {\rm H}}} \approx 1.89\times 10^{-4}$ and ${m^{}_{e}}/{m^{}_{^3 {\rm H}}} \approx 1.82\times 10^{-4}$, the relativistic electron spectrum ${\mathrm{d} \Gamma^{}_{\rm rel}}/{\mathrm{d} K^{}_{e}}$ approximates to the classical one
\begin{eqnarray}\label{eq:betaclassic}
   \frac{{\rm d}\Gamma^{}_{\rm cl}}{{\rm d}K^{}_e} = N^{}_{\rm T} \frac{\overline{\sigma}^{}_{\rm cl}(E^{}_e)}{\pi^2} \sum^3_{i=1} |U^{}_{ei}|^2 \sqrt{(K^{}_{\rm end,0}-K^{}_{e})^2-m^{2}_{i}}
\left(K^{}_{\rm end,0}-K^{}_{e}\right) ,
%     (8)
\end{eqnarray}
where $\overline{\sigma}^{}_{\rm cl}(E^{}_e) = \overline{\sigma}(E^{}_e)/ (m^{}_{^3{\rm He}}/m^{}_{^3{\rm H}})$ and $\overline{\sigma}(E^{}_e)$ has been given in Eq.~(\ref{eq:sigmabar}). Comparing the classical spectrum in Eq.~(\ref{eq:betaclassic}) with the relativistic one in Eq.~(\ref{eq:betaExact}), one can observe that the endpoint energy in the former case deviates from the true one by an amount of $(1 - m^{}_{^3{\rm He}}/m^{}_{^3{\rm H}})\, m_{i} \approx 10^{-4}\,m^{}_{i}$. As the PTOLEMY experiment could achieve a relative precision of $10^{-6}$ for the determination of the lightest neutrino mass \cite{Betti:2019ouf}, it would be no longer appropriate to use the classical spectrum in PTOLEMY.
However, it is rather safe for KATRIN and Project 8 to neglect the factor $1- m^{}_{^3{\rm He}}/m^{}_{^3{\rm H}}$, as their sensitivities to the neutrino mass are weaker than for PTOLEMY. To be more specific, the $1\sigma$ sensitivities of KATRIN and Project 8 to $m^2_{\beta}$ are $\sigma(m^2_{\beta}) \approx 0.025~{\rm eV^2}$ \cite{Angrik:2005ep} and $\sigma(m^2_{\beta}) \approx 0.001~{\rm eV^2}$ \cite{Esfahani:2017dmu}, respectively, corresponding to $\sigma(m^{}_{\beta})/m^{}_{\beta} \approx 0.05~(0.5~{\rm eV}/m^{}_{\beta})^2$  and $\sigma(m^{}_{\beta})/m^{}_{\beta} \approx 0.002~(0.5~{\rm eV}/m^{}_{\beta})^2$. Both values are much larger than the correction of order $10^{-4}$ from the factor $m^{}_{^3{\rm He}}/m^{}_{^3{\rm H}}$.
To have an expression for the electron spectrum applicable to experiments beyond KATRIN and Project 8, i.e., leading
to PTOLEMY, we can slightly modify the classical energy spectrum as follows:
\begin{eqnarray}\label{eq:Gammaprime}
\frac{{\rm d}\Gamma^\prime_{\rm cl}}{{\rm d}K^{}_e} = N^{}_{\rm T} \frac{\overline{\sigma}^{}_{\rm cl}(E^{}_e)}{\pi^2} \sum^3_{i=1} |U^{}_{ei}|^2 \sqrt{(K^{}_{\rm end,0}-K^{}_{e})^2 - \left(m^{}_{i} \frac{m^{}_{^3 {\rm  He}}}{m^{}_{^3 {\rm H}}}\right)^2} \left(K^{}_{\rm end,0} - K^{}_{e}\right) .
%     (9)
\end{eqnarray}
Let us now check whether the difference between the exact relativistic spectrum ${\mathrm{d} \Gamma^{}_{\rm rel}}/{\mathrm{d} K^{}_{e}}$ and the modified classical one ${\mathrm{d} \Gamma^{\prime}_{\rm cl}}/{\mathrm{d} K^{}_{e}}$ affects the determination of absolute neutrino masses in future beta-decay experiments with a target mass of tritium ranging from $10^{-4}~{\rm g}$ in KATRIN to $100~{\rm g}$ in PTOLEMY. In other words, we examine whether these two spectra are statistically distinguishable in realistic experiments. Consider the ratio of these two energy spectra
\begin{eqnarray}\label{eq:spectrumratio}
\frac{{\mathrm{d} \Gamma^{}_{\rm rel}}/{\mathrm{d} K^{}_{e}}}{{\mathrm{d} \Gamma^{\prime}_{\rm cl}}/{\mathrm{d} K^{}_{e}}} = \frac{1 - m^2_e/\left(m^{}_{^3{\rm H}}E^{}_{e}\right)} {\left(1 - 2E^{}_{e}/m^{}_{^3{\rm H}} + m^2_e/m^2_{^3{\rm H}}\right)^2} \cdot \frac{m^{}_{^3 {\rm  He}}}{m^{}_{^3 {\rm H}}} \approx 1.0036 + 1.7 \times 10^{-9} \left(\frac{K^{}_{e}}{{\rm eV}}\right)  ,
%     (10)
\end{eqnarray}
where an expansion in terms of the electron kinetic energy $K^{}_e = E^{}_e - m^{}_e$ has been carried out. First of all, the constant on the rightmost side of Eq.~(\ref{eq:spectrumratio}) can be absorbed into the uncertainty of the overall normalization factor $A^{}_{\beta}$ in the statistical analysis, so it is irrelevant for our discussions.\footnote{For instance, in the statistical analysis of the simulated data for the PTOLEMY experiment~\cite{Betti:2019ouf}, the prior of the normalization factor $A^{}_{\beta}$ is set to be in the range of $\left(0\cdots 2\right)$, which is wide enough to take account of the difference corresponding to the constant term in the ratio in Eq.~(\ref{eq:spectrumratio}).} Considering the term proportional to the electron kinetic energy $K^{}_e$ in Eq.~(\ref{eq:spectrumratio}), it could potentially disturb the determination of the normalization factor of the spectrum. The distortion amplitude induced by the $K^{}_e$-dependent term can be characterized by the specified energy window $\Delta K^{}_e$ below the endpoint. For example, we have $\Delta K^{}_{e} = 30~{\rm eV}$ for KATRIN while $\Delta K^{}_{e} = 5~{\rm eV}$ for PTOLEMY, which is limited by the detector performance, see Appendix \ref{sec:appA}.  In this way, we can obtain the distortion amplitude of $5.1\times 10^{-8}$ for KATRIN and $8.5\times 10^{-9}$ for PTOLEMY, respectively. To examine the impact of this distortion, one can compare it with the statistical fluctuation of the events within the corresponding energy window. The integrated number of beta-decay events within the energy window below the endpoint can be calculated via
\begin{eqnarray}\label{eq:Nint}
N^{}_{\rm int} = T \int^{K^{}_{\rm end,0}}_{K^{}_{\rm end,0} - \Delta K^{}_{e}} \frac{\mathrm{d} \Gamma}{\mathrm{d} K^{}_{e}} {\mathrm{d} K^{}_{e}} \approx 3.2 \times 10^{11} \cdot \left(\frac{\Delta K^{}_{e}}{\rm eV}\right)^3 \cdot \left(\frac{\mathcal{E}}{100~{\rm g\cdot yr}}\right)  ,
%     (11)
\end{eqnarray}
where $T$ is the operation time and $\mathcal{E} \equiv N^{}_{\rm T} \cdot T$ is the total exposure. The statistical fluctuation of the beta-decay events within the energy window is estimated as $\sqrt{N^{}_{\rm int}}/N^{}_{\rm int} \approx 10^{-5}$ for KATRIN with $\Delta K^{}_e = 30~{\rm eV}$ and ${\cal E} = 10^{-4}~{\rm g}\cdot {\rm yr}$, while $\sqrt{N^{}_{\rm int}}/N^{}_{\rm int} \approx 10^{-7}$ for PTOLEMY with $\Delta K^{}_e = 5~{\rm eV}$ and ${\cal E} = 100~{\rm g}\cdot {\rm yr}$. Both values are much larger than the corresponding distortion amplitudes. It is thus evident that the uncertainty in $A^{}_{\beta}$ will be dominated by the intrinsic statistical fluctuation of the observed beta-decay events in future experiments. As the data fluctuation near the endpoint is most significant among the entire spectrum, the influence of the spectral distortion as indicated in Eq.~(\ref{eq:spectrumratio}) is not important. Hence we conclude that the classical spectrum with the neutrino masses corrected by $m^{}_{i} \rightarrow m^{}_{i}\cdot\left({m^{}_{^3 {\rm  He}}}/{m^{}_{^3 {\rm H}}}\right)$ in Eq.~(\ref{eq:Gammaprime}) works as well as the exact relativistic spectrum in Eq.~(\ref{eq:betaExact}) for future beta-decay experiments.

It is worthwhile to emphasize that because of a finite energy resolution $\Delta$, which is normally much larger than the absolute neutrino mass $m^{}_i$, it is difficult to resolve the true endpoint. Hence, the experimental sensitivity to neutrino masses is in fact governed by the integrated number of beta-decay events within a specified energy window below the endpoint. 
Taking this energy window to be the experimental energy resolution, $\Delta K^{}_e = \Delta$, we can figure out the expected statistics around the endpoint according to Eq.~(\ref{eq:Nint}). For a conservative experimental setup, e.g. that is achievable in KATRIN with $\Delta = 1~{\rm eV}$ and ${\cal E} = 10^{-4}~{\rm g}\cdot {\rm yr}$, the expected event number within the endpoint bin is $3.2 \times 10^5$. In the limit of $ m^{}_{\beta}  \ll  \Delta $, a finite neutrino mass will induce a relative deviation of events within the window $\Delta$ by $3/2 \cdot m^2_{\beta}/ \Delta^2 $. Therefore, the sensitivity to $m^{}_{\beta}$ is roughly scaled as $\left(\Delta/{\cal E}\right)^{1/4}$. 
The choice of energy window is limited by the smearing effect of finite energy resolution, and the sensitivity to neutrino masses will drop with a larger energy resolution.
This can be compensated by increasing the exposure, such that an efficient event number can be acquired to resolve the overall shift due to finite neutrino masses.

Unless stated otherwise, we will refer from now on to ${{\rm d}\Gamma^\prime_{\rm cl}}/{{\rm d}K^{}_e}$ as the
{\it exact spectrum} in the remaining discussion.

\subsection{Validity of the effective mass}

In principle, it is the exact relativistic spectrum that should be confronted with the experimental observation in order to extract the absolute neutrino masses $m^{}_i$ (for $i = 1, 2, 3$), since $|U^{}_{ei}|^2$ (for $i = 1, 2, 3$) can be precisely measured in neutrino oscillation experiments. However, often the effective electron spectrum with only one mass parameter is considered (see e.g., \cite{Vissani:2000ci})
\begin{eqnarray}\label{eq:Gammaeff}
   \frac{{\rm d}\Gamma^{}_{\rm eff}}{{\rm d}K^{}_e} = N^{}_{\rm T} \frac{\overline{\sigma}^{}_{\rm cl}(E^{}_e)}{\pi^2} \sqrt{(K^{}_{\rm end,0} - K^{}_{e})^2 - \left(m^{}_{\beta} \frac{m^{}_{^3 {\rm  He}}}{m^{}_{^3 {\rm H}}}\right)^2} \left(K^{}_{\rm end,0}-K^{}_{e}\right) ,
%     (12)
\end{eqnarray}
where the effective neutrino mass $m^{}_\beta$ is usually defined as in Eq.~(\ref{eq:mbeta}).
Note that for consistency we have kept the near-unity factor ${m^{}_{^3 {\rm  He}}}/{m^{}_{^3 {\rm H}}}$ as in ${{\rm d}\Gamma^\prime_{\rm cl}}/{{\rm d}K^{}_e}$ of Eq.~(\ref{eq:Gammaprime}), which is necessary when it comes to experiments beyond KATRIN and Project 8, i.e.\ PTOLEMY.
However, ${m^{}_{^3 {\rm He}}}/{m^{}_{^3 {\rm H}}}$ appears as an overall factor to all neutrino mass parameters, so the quantitative impact on our discussion of the validity of the effective mass is actually negligible\footnote{One can easily check that the relation $m^{2}_\beta \equiv |U^{}_{e1}|^2 m^2_1 + |U^{}_{e2}|^2 m^2_2 + |U^{}_{e3}|^2 m^2_3$ is stable under $m^{}_{i} \rightarrow m^{}_{i}\cdot\left({m^{}_{^3 {\rm  He}}}/{m^{}_{^3 {\rm H}}}\right)$ and $m^{}_{\beta} \rightarrow m^{}_{\beta}\cdot\left({m^{}_{^3 {\rm  He}}}/{m^{}_{^3 {\rm H}}}\right)$.
Thus any quantitative conclusion made by considering ${m^{}_{^3 {\rm  He}}}/{m^{}_{^3 {\rm H}}}$ corrections can be directly applied to the case without correction of ${m^{}_{^3 {\rm  He}}}/{m^{}_{^3 {\rm H}}}$ by shifting all neutrino masses with a relative fraction as small as $10^{-4}$, and vice versa.},
but we keep it nevertheless in our numerical calculations.
Another important point is that the endpoint energy in Eq.~(\ref{eq:K0end}) should be corrected if we take account of  binding energies as well as excitations of the daughter system in an actual experiment. For instance, in KATRIN or Project 8 with the molecular tritium target, a correction of $16.29~{\rm eV}$ to the endpoint energy should be considered owing to the binding energies of the mother tritium pair, the  daughter tritium-helium molecule, and the combination of the ionized electron \cite{Otten:2008zz}. Compared to the atomic case, the recoil energy of the molecular state will also be reduced by a factor of two due to the doubled mass, which will boost  the endpoint energy of electrons. For PTOLEMY, with a foreseen possibility of atomic tritium weakly bounded to a graphene layer \cite{Betts:2013uya}, the ionized electron will inevitably interact with the complex graphene binding system. Furthermore, the recoiled helium (with a kinetic energy of 3.4 eV) will escape the graphene binding structure with a sub-eV binding energy. All these effects need to be systemically considered in the experiment to evaluate the final endpoint energy $K^{}_{\rm end,0}$. 
 However, in our case, the spectrum in Eq.~(\ref{eq:Gammaeff}) mostly depends on the relative deviation from the endpoint energy $K^{}_{\rm end,0}-K^{}_{e}$, and $\overline{\sigma}^{}_{\rm cl}(E^{}_e)$ changes very slowly as a function of $K^{}_{\rm end,0}-K^{}_{e}$ near the endpoint. Hence, our results which will be presented in terms of $K^{}_{\rm end,0}-K^{}_{e}$ are still valid if a different $K^{}_{\rm end,0}$ is considered.
 On the other hand, the  final-state excitations of molecules will smear the energy of outgoing electrons \cite{Saenz:2000dul,Fackler:1985nm,Doss:2006zv,Bodine:2015sma}, and this effect will be taken into account as the irreducible energy resolution of the experiment.

Let us summarize the existing expressions of the electron spectra defined in this work:
(i) the exact relativistic beta spectrum ${\mathrm{d} \Gamma^{}_{\rm rel}}/{\mathrm{d} K^{}_{e}}$
without making approximations, see Eq.\ (\ref{eq:betaExact}); (ii) the classical spectrum
${{\rm d}\Gamma^{}_{\rm cl}}/{{\rm d}K^{}_e}$ in the limit of
${m^{}_{^3 {\rm He}}}/{m^{}_{^3 {\rm H}}} \rightarrow 1$ and ${m^{}_{e}}/{m^{}_{^3 {\rm H}}} \rightarrow 0$,
see Eq.\ (\ref{eq:betaclassic}); (iii) the modified classical spectrum
${{\rm d}\Gamma^{\prime}_{\rm cl}}/{{\rm d}K^{}_e}$ by making the replacement
$m^{}_{i} \rightarrow m^{}_{i}\cdot\left({m^{}_{^3 {\rm  He}}}/{m^{}_{^3 {\rm H}}}\right)$
in ${{\rm d}\Gamma^{}_{\rm cl}}/{{\rm d}K^{}_e}$, see Eq.\ (\ref{eq:Gammaprime});
(iv) the effective electron spectrum ${{\rm d}\Gamma^{}_{\rm eff}}/{{\rm d}K^{}_e}$
defined in Eq.~(\ref{eq:Gammaeff}).
We have seen that the difference between ${{\rm d}\Gamma^{\prime}_{\rm cl}}/{{\rm d}K^{}_e}$
in Eq.\ (\ref{eq:Gammaprime}) and the classical spectrum ${{\rm d}\Gamma^{}_{\rm cl}}/{{\rm d}K^{}_e}$ in Eq.\  (\ref{eq:betaclassic}) plays only a role when PTOLEMY is considered. In addition, the difference to the relativistic spectrum ${\mathrm{d} \Gamma^{}_{\rm rel}}/{\mathrm{d} K^{}_{e}}$ in
Eq.\ (\ref{eq:betaExact}) is minuscule and the classical spectra can be considered as the exact ones.
It remains to compare the so-defined exact spectrum
${{\rm d}\Gamma^{\prime}_{\rm cl}}/{{\rm d}K^{}_e}$ in Eq.\ (\ref{eq:Gammaprime}) to the effective one
${{\rm d}\Gamma^{}_{\rm eff}}/{{\rm d}K^{}_e}$ in Eq.~(\ref{eq:Gammaeff}).

%%%%%%%%%%%%%%%%%%%%%%%%%%%%%%%% Fig. 1 %%%%%%%%%%%%%%%%%%%%%%%%%%%%%%%%%
\begin{figure}[t!]
\begin{center}
\hspace{-0.4cm}
\includegraphics[width=0.49\textwidth]{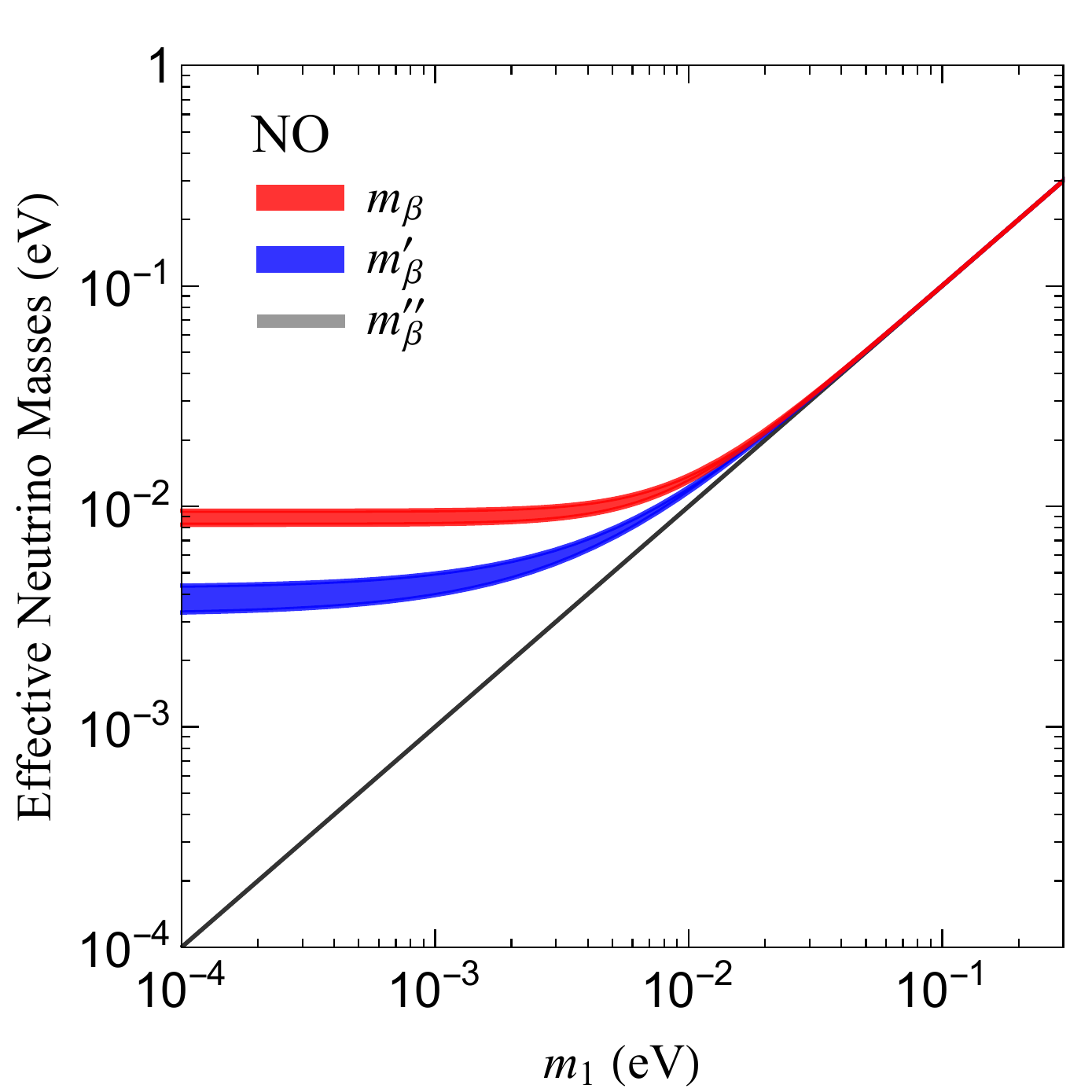}
\includegraphics[width=0.49\textwidth]{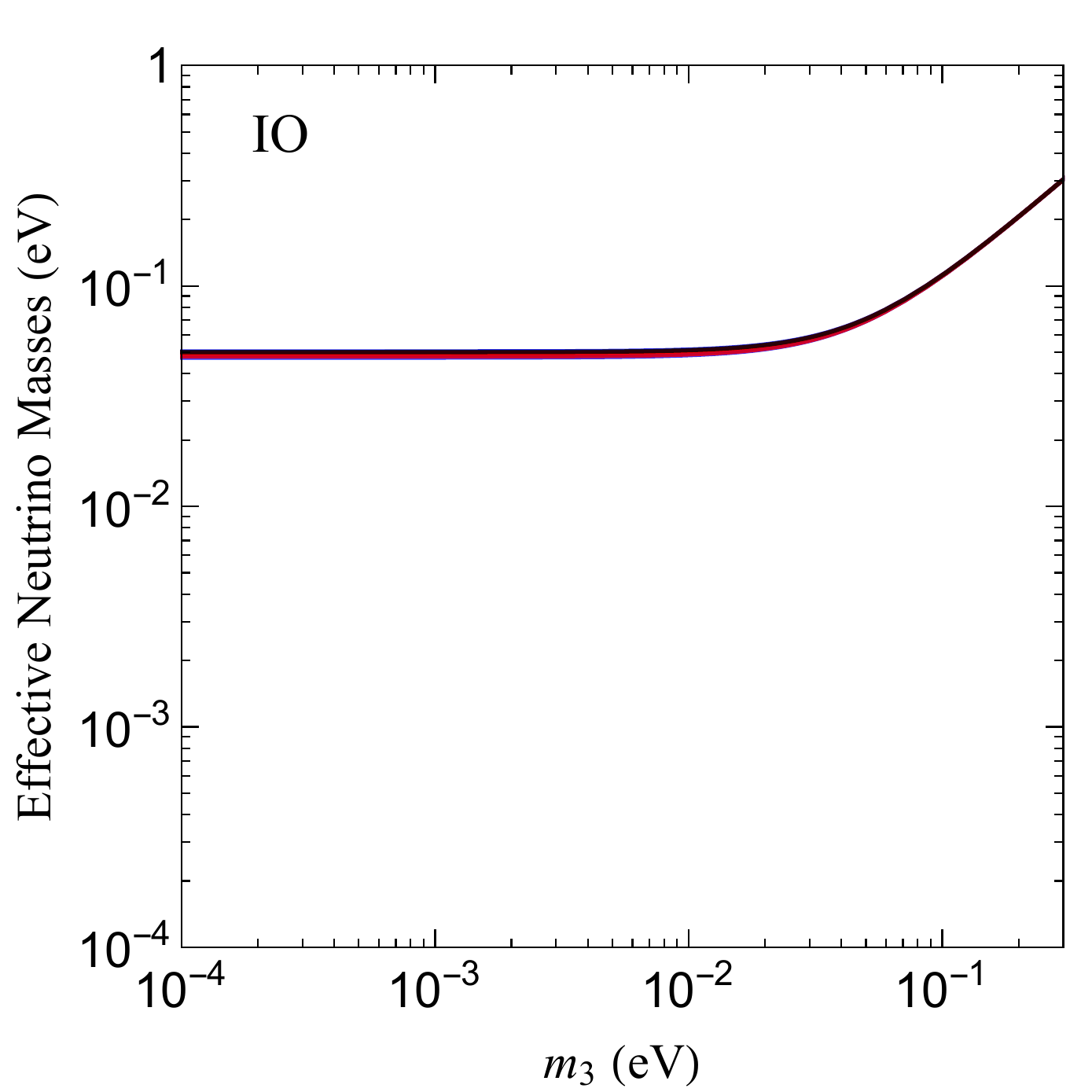}
\end{center}
\vspace{-.50cm}
\caption{The effective neutrino masses versus $m^{}_{1}$ for NO (left panel) and $m^{}_{3}$ for IO (right panel). The $3\sigma$ CL uncertainties of oscillation parameters have been considered. A similar plot for absolute neutrino masses was given in Ref.~\cite{Beacom:2002cb}.}
\label{fig:mbetaVSm1}
\end{figure}
%%%%%%%%%%%%%%%%%%%%%%%%%%%%%%%%%%%%%%%%%%%%%%%%%%%%%%%%%%%%%%%%%%%%%%%%%%%%%
Moreover, in the literature  two different definitions of the effective neutrino mass have also been introduced~\cite{Farzan:2001cj, Studnik:2001hs,Farzan:2002zq}, namely
\begin{eqnarray}\label{eq:mbeta2}
m^{\prime}_{\beta} \equiv \sum^{3}_{i=1} m^{}_{i} |U^{}_{ei}|^2 \;, \quad
m^{\prime\prime}_{\beta} \equiv m^{}_{1} \; .
%     (13)
\end{eqnarray}
The effective electron spectrum can then be obtained by replacing $m^{}_\beta$ in Eq.~(\ref{eq:Gammaeff}) with $m^\prime_\beta$ or $m^{\prime \prime}_\beta$. In this subsection, we discuss the difference among those three effective neutrino masses and clarify their validity  with future beta-decay experiments in mind.

As has been observed in Ref.~\cite{Farzan:2002zq}, the three effective neutrino masses have different accuracies in fitting the exact spectrum. If neutrino masses are quasi-degenerate, all three effective masses provide very good fits and their relative differences are very small. For example, it is easy to verify that $(m^{\prime 2}_{\beta} - m^{\prime\prime 2}_{\beta})/m^{\prime\prime}_{\beta} \lesssim 10^{-3}$. If the chosen energy window satisfies $\Delta K^{}_e < 2\, m^{}_{\beta}$, then $m^{\prime}_{\beta}$ can give a better fit than $m^{\prime\prime}_\beta$, whereas $m^{}_{\beta}$ is still an excellent parameter in fitting the spectrum with an almost negligible difference $(m^{2}_{\beta} - m^{\prime 2}_{\beta })/m^{\prime\prime 2}_{\beta} \lesssim 10^{-5}$. If neutrino masses are hierarchical, $m^{}_{\beta}$ always fits better to the spectrum than the other two variants. In case of an extremely small value of the  lightest neutrino mass, both $m^{\prime}_{\beta}$ and $m^{\prime\prime}_{\beta}$ are unable to offer a good fit to the true spectrum.
In Fig.~\ref{fig:mbetaVSm1} we plot three effective neutrino masses in terms of the lightest neutrino mass which is $m^{}_{1}$ for NO and $m^{}_{3}$ for IO. One can observe that their differences are significant in NO when $m^{}_{1}$ is small, but in IO the differences are always unnoticeable. The situation of IO can be attributed to the fact that the contribution of $m^{}_{3}$ is suppressed by $|U^{}_{e3}|^2$ while the remaining two neutrino masses $m^{}_{1}$ and $m^{}_{2}$ are always nearly degenerate due to the relation $\Delta m^2_{21} \ll |\Delta m^2_{31}|$.

To be more explicit, we look carefully at the main difference between the exact spectrum ${{\rm d}\Gamma^{\prime}_{\rm cl}}/{{\rm d}K^{}_e}$ and the effective one ${{\rm d}\Gamma^{}_{\rm eff}}/{{\rm d}K^{}_e}$.
The difference stems from the kinematical functions, namely \cite{Farzan:2002zq}
\begin{eqnarray}\label{eq:betacompare1}
{\mathrm{d} \Gamma^{\prime}_{\rm cl}}/{\mathrm{d} K^{}_{e}} & \propto & \sum^{3}_{i=1} |U^{}_{ei}|^2\sqrt{(K^{}_{\rm end,0}-K^{}_{e})^2-\left(m^{}_{i} \cdot \frac{m^{}_{^3 {\rm  He}}}{m^{}_{^3 {\rm H}}}\right)^2} \left(K^{}_{\rm end,0}-K^{}_{e}\right), \\ \label{eq:betacompare2}
{\mathrm{d} \Gamma^{}_{\rm eff}}/{\mathrm{d} K^{}_{e}} & \propto  &\sqrt{(K^{}_{\rm end,0}-K^{}_{e})^2-\left(m^{(\prime,\prime\prime)}_{\beta}\cdot \frac{m^{}_{^3 {\rm  He}}}{m^{}_{^3 {\rm H}}}\right)^2} \left(K^{}_{\rm end,0}-K^{}_{e}\right) ,
%     (14-15)
\end{eqnarray}
where the spectra involving our three effective neutrino masses $m^{}_\beta$, $m^\prime_\beta$ and $m^{\prime\prime}_\beta$ are collectively given. To analyze the difference we take $m^{}_\beta$ in the NO case for example, but the other effective masses and the IO case can be studied in a similar way. Let us start with the endpoint of the electron spectrum and then go to lower energies.
For convenience the factor of ${m^{}_{^3 {\rm  He}}}/{m^{}_{^3 {\rm H}}}$ is omitted  in the following qualitative discussion, which of course will not affect the main feature of the result as we noted above.
\begin{enumerate}
\item For the exact spectrum, the endpoint energy $K^{}_{\rm end}$ is set by the smallest neutrino mass, i.e., $K^{}_{\rm end,0} - K^{}_{\rm end} = m^{}_{1}$, while it is $m^{}_{\beta}$ for the effective spectrum ${\mathrm{d} \Gamma^{}_{\rm eff}}/{\mathrm{d} K^{}_{e}}$. Since $m^2_{\beta} = m^2_1 + \Delta m^2_{21} |U^{}_{e2}|^2 + \Delta m^2_{31} |U^{}_{e3}|^2 > m^2_{1}$, the endpoint energy of the effective spectrum ${\mathrm{d} \Gamma^{}_{\rm eff}}/{\mathrm{d} K^{}_{e}}$ is smaller than that of the exact one ${\mathrm{d} \Gamma^{\prime}_{\rm cl}}/{\mathrm{d} K^{}_{e}}$. Therefore, starting from the electron kinetic energy of $K^{}_{e} = K^{}_{\rm end,0} - m^{}_{1}$ and going to smaller values, the effective spectrum ${\mathrm{d} \Gamma}^{}_{\rm eff}/{\mathrm{d} K^{}_{e}}$ is always vanishing and thus should be lying below the exact one ${\mathrm{d} \Gamma^{\prime}_{\rm cl}}/{\mathrm{d} K^{}_{e}}$.

\item As $K^{}_e$ is decreasing further, we come to the point at which $K^{}_{\rm end,0} - K^{}_{e} = m^{}_\beta$ is satisfied. Note that $m^2_\beta = m^2_3 - \Delta m^2_{31} |U^{}_{e1}|^2 - \Delta m^2_{32} |U^{}_{e2}|^2 < m^2_3$ holds. Therefore, for $m^{}_\beta < K^{}_{\rm end,0} - K^{}_{e} < m^{}_3$, ${\mathrm{d} \Gamma}^{}_{\rm eff}/{\mathrm{d} K^{}_{e}}$ becomes nonzero. As indicated in Eqs.~(\ref{eq:betacompare1}) and (\ref{eq:betacompare2}), before the decay channel corresponding to $m^{}_{3}$ is switched on, ${\mathrm{d} \Gamma}^{}_{\rm eff}/{\mathrm{d} K^{}_{e}}$ is about to exceed ${\mathrm{d} \Gamma}^{\prime}_{\rm cl}/{\mathrm{d} K^{}_{e}}$. At $K^{}_{e} = K^{}_{\rm end,0} - m^{}_{3}$, we have ${\mathrm{d} \Gamma}^{}_{\rm eff}/{\mathrm{d} K^{}_{e}} \propto \sqrt{|U^{}_{e1}|^2 \Delta m^2_{31} + |U^{}_{e2}|^2 \Delta m^2_{32}} \approx \sqrt{|U^{}_{e1}|^2 + |U^{}_{e2}|^2}\sqrt{\Delta m^2_{31}}$ and ${\mathrm{d} \Gamma}^{\prime}_{\rm cl}/{\mathrm{d} K^{}_{e}} \propto |U^{}_{e1}|^2 \sqrt{\Delta m^2_{31}} + |U^{}_{e2}|^2 \sqrt{\Delta m^2_{32}} \approx (|U^{}_{e1}|^2 + |U^{}_{e2}|^2)\sqrt{\Delta m^2_{31}}$, where $\Delta m^2_{21} \ll \Delta m^2_{31}$ has been taken into account, leading to ${\mathrm{d} \Gamma}^{}_{\rm eff}/{\mathrm{d} K^{}_{e}} > {\mathrm{d} \Gamma}^{\prime}_{\rm cl}/{\mathrm{d} K^{}_{e}}$.

\item When we go far below the endpoint, e.g., $K^{}_{e} \ll K^{}_{\rm end,0} - m^{}_{i}$ or equivalently $K^{}_{\rm end,0} - K^{}_e \gg m^{}_i$, the neutrino masses can be neglected and thus these two spectra coincide with each other. Therefore, for $m^{}_\beta$ under consideration, the difference between ${\mathrm{d} \Gamma}^{}_{\rm eff}/{\mathrm{d} K^{}_{e}}$ and ${\mathrm{d} \Gamma}^{\prime}_{\rm cl}/{\mathrm{d} K^{}_{e}}$ could change its sign in the narrow range below the endpoint but finally converges to zero.
\end{enumerate}
%%%%%%%%%%%%%%%%%%%%%%%%%%%%%%%% Fig. 2 %%%%%%%%%%%%%%%%%%%%%%%%%%%%%%%%%
\begin{figure}[t!]
\begin{center}
\hspace{-0.4cm}
\includegraphics[width=0.49\textwidth]{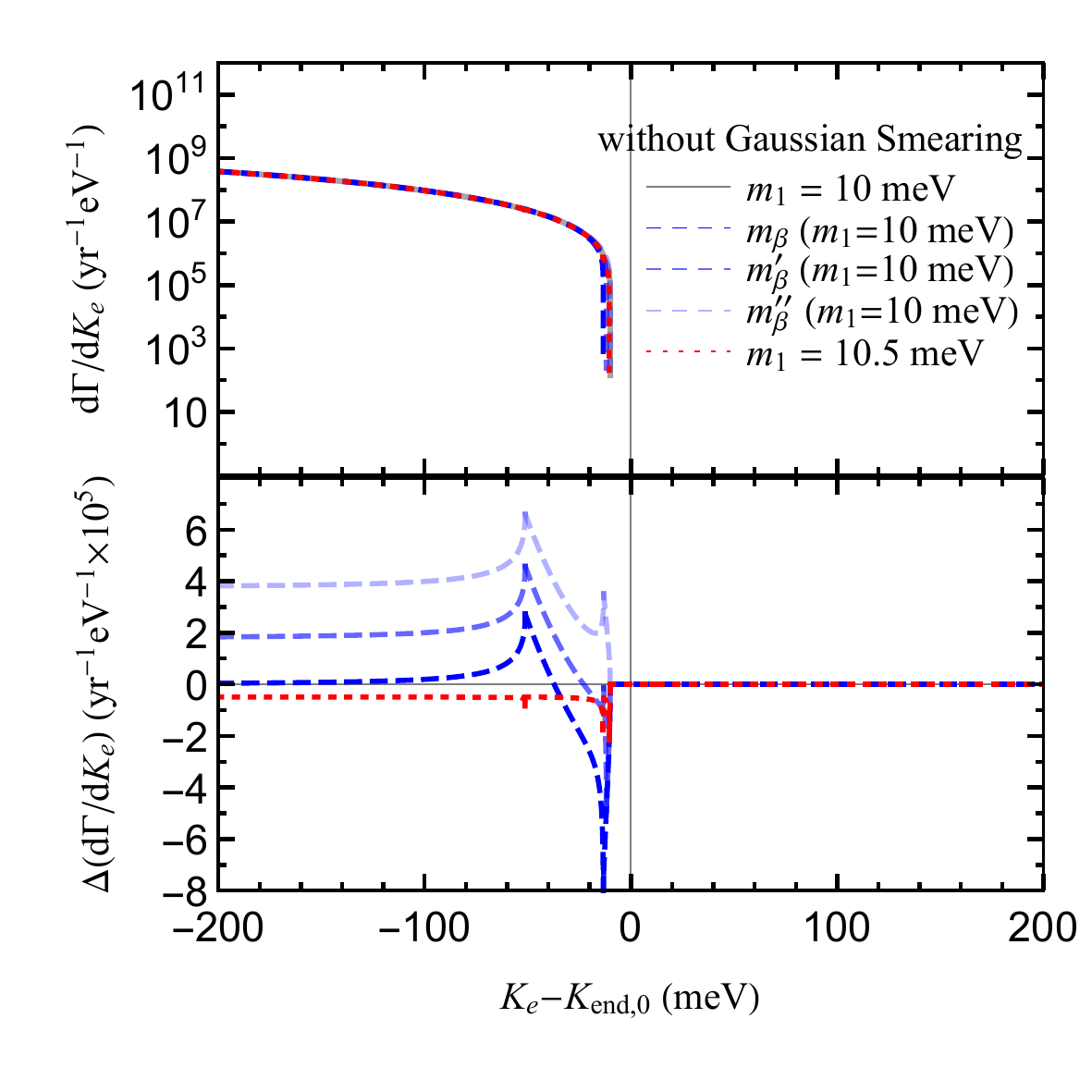}
\includegraphics[width=0.49\textwidth]{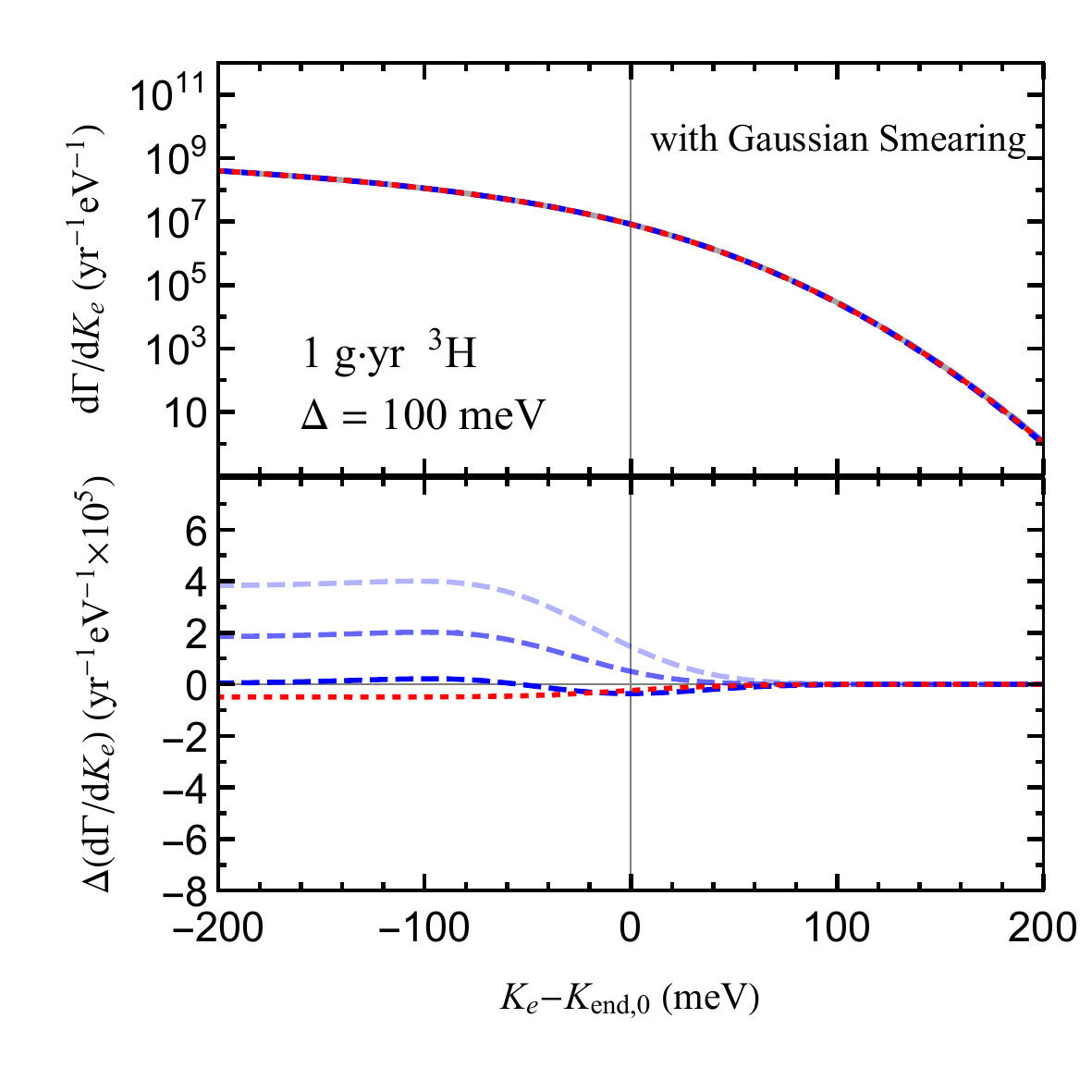}
\end{center}
\vspace{-1.0cm}
\caption{Illustration of the electron spectrum from tritium beta decays, where the total exposure ${\cal E} = 1~{\rm g}\cdot {\rm yr}$ and the best-fit values of neutrino mixing angles and mass-squared differences are assumed. The exact spectra ${\mathrm{d} \Gamma^{\prime}_{\rm cl}}/{\mathrm{d} K^{}_{e}}$ with $m^{}_1 = 10~{\rm meV}$ and $m^{}_1 = 10.5~{\rm meV}$ are shown as the gray solid and red dotted curve, respectively. The effective spectra ${\mathrm{d} \Gamma^{}_{\rm eff}}/{\mathrm{d} K^{}_{e}}$ with $m^{}_\beta = 13.4~{\rm meV}$, $m^\prime_\beta = 11.9~{\rm meV}$ and $m^{\prime\prime}_\beta = 10~{\rm meV}$, corresponding to $m^{}_1 = 10~{\rm meV}$, are represented by the dark, medium and light blue dashed curves. In the left panel, the energy smearing is ignored, while an energy resolution of $\Delta = 100~{\rm meV}$ is taken into account in the right panel. In both panels, the real spectra are depicted in the upper subgraph, whereas their deviations from the exact spectrum with $m^{}_1 = 10~{\rm meV}$ are given in the lower subgraph.}
\label{fig:spectrumVSKe}
\end{figure}
%%%%%%%%%%%%%%%%%%%%%%%%%%%%%%%%%%%%%%%%%%%%%%%%%%%%%%%%%%%%%%%%%%%%%%%%%%%%%

For illustration, we show in Fig.~\ref{fig:spectrumVSKe} the electron spectra in the narrow energy region $-200~{\rm meV} \leq K^{}_e - K^{}_{\rm end, 0} \leq 200~{\rm meV}$ around the endpoint, where possible background events are ignored. In addition, the total exposure for the tritium beta-decay experiment is taken to be ${\cal E} = 1~{\rm g}\cdot {\rm yr}$. In the left panel of Fig.~\ref{fig:spectrumVSKe}, the exact spectrum ${\mathrm{d} \Gamma}^{\prime}_{\rm cl}/{\mathrm{d} K^{}_{e}}$ with $m^{}_{1} = 10~{\rm meV}$ is plotted as the gray solid curve, while that with $m^{}_{1}=10.5~{\rm meV}$ is represented by the red dotted curve for comparison. The effective spectra ${\mathrm{d} \Gamma}^{}_{\rm eff}/{\mathrm{d} K^{}_{e}}$ for $m^{}_\beta = 13.4~{\rm meV}$, $m^\prime_\beta = 11.9~{\rm meV}$ and $m^{\prime\prime}_\beta = 10~{\rm meV}$ are given by the dark, medium and light blue dashed curves, respectively. Those values are obtained for a smallest mass of $m_1 = 10$ meV and the current best-fit values of the oscillation parameters \cite{Esteban:2018azc}. Since it is hard to distinguish these spectra, as can be seen in the upper subgraph in the left panel, we depict their deviations from the exact spectrum,
$$\Delta(\mathrm{d} \Gamma / \mathrm{d} K^{}_{e})\equiv \mathrm{d} \Gamma^{}_{\rm eff} / \mathrm{d} K^{}_{e} - \mathrm{d} \Gamma^{\prime}_{\rm cl} / \mathrm{d} K^{}_{e}\,,
$$
with $m^{}_1 = 10~{\rm meV}$ in the lower subgraph. The behavior of these deviations can be well understood analytically, as we have already explained by using Eqs.~(\ref{eq:betacompare1}) and (\ref{eq:betacompare2}). As for the exact spectrum $\mathrm{d} \Gamma^{\prime}_{\rm cl} / \mathrm{d} K^{}_{e}$ with $m^{}_1 = 10.5~{\rm meV}$, it is always lying below that with $m^{}_1 = 10~{\rm meV}$. The reason is simply that the kinematical function $[(K^{}_{\rm end,0} - K^{}_e)^2 -(m^{}_{i} \cdot {m^{}_{^3 {\rm  He}}}/{m^{}_{^3 {\rm H}}})^2 ]^{1/2}$ in the exact spectrum $\mathrm{d} \Gamma^{\prime}_{\rm cl} / \mathrm{d} K^{}_{e}$ becomes smaller for larger values of $m^{}_i$.

The finite energy resolution of the detector has been ignored in the left panel of Fig.~\ref{fig:spectrumVSKe}, but is taken into account in the calculations of the energy spectra and their deviations from $\mathrm{d} \Gamma^{\prime}_{\rm cl} / \mathrm{d} K^{}_{e}$ with $m^{}_1 = 10~{\rm meV}$ in the right panel.  Assuming the energy resolution \footnote{The values of energy resolution in this work are all referred to as the $1\sigma$ deviation of the energy reconstruction.} of the detector to be $\Delta = 100~{\rm meV}$ and taking the Gaussian form, we can derive the energy spectrum with  smearing effects as follows
\begin{eqnarray}\label{eq:spectrum}
\frac{\mathrm{d} {\Gamma}^{}_{\Delta}}{\mathrm{d} K^{}_{e}} (K^{}_{e})& = & \frac{1}{\sqrt{2\pi}(\Delta/\sqrt{8 \mathrm{ln}2})} \int^{+\infty}_{-\infty} \mathrm{d}K^{\prime}_{e} \frac{\mathrm{d} {\Gamma}}{\mathrm{d} K^{\prime}_{e}}(K^{\prime}_{e})\, \mathrm{exp}\left[ -\frac{(K^{}_{e}-K^{\prime}_{e})^2}{2(\Delta\sqrt{8\mathrm{ln}2})^2}\right]  ,
%     (16)
\end{eqnarray}
which has been plotted in the right panel for both $\mathrm{d} \Gamma^{\prime}_{\rm cl} / \mathrm{d} K^{}_{e}$ and $\mathrm{d} \Gamma^{}_{\rm eff} / \mathrm{d} K^{}_{e}$.
Note that we have not yet specified any planned experimental configuration so far, because the main purpose here is to understand the behavior of deviations caused by using different effective neutrino masses.
Two interesting observations can be made and deserve further discussions:
\begin{itemize}
\item
First, when energy smearing effects are included, the difference between $\mathrm{d} \Gamma^{}_{\rm eff} / \mathrm{d} K^{}_{e}$ and $\mathrm{d} \Gamma^{\prime}_{\rm cl} / \mathrm{d} K^{}_{e}$ will be averaged over the electron kinetic energy, reducing the discrepancy between them. This effect is more significant for the electron kinetic energy closer to the endpoint. Therefore, if the energy resolution is extremely good, the error caused by using the effective spectrum becomes larger. In this case, one needs to fit the experimental data by implementing $\mathrm{d} \Gamma^{\prime}_{\rm cl} / \mathrm{d} K^{}_{e}$ with the lightest neutrino mass as the fundamental parameter.

\item
Second, the effective spectrum $\mathrm{d} \Gamma^{}_{\rm eff} / \mathrm{d} K^{}_{e}$ with $m^{}_{\beta}$ converges to the exact one in the energy region far below the endpoint. Moreover, it is interesting to notice that even though the difference $\Delta(\mathrm{d} \Gamma / \mathrm{d} K^{}_{e})$ between ${\mathrm{d} \Gamma}^{}_{\rm eff}/{\mathrm{d} K^{}_{e}}$ with $m^{}_\beta$ and ${\mathrm{d} \Gamma}^{\prime}_{\rm cl}/{\mathrm{d} K^{}_{e}}$ can be either positive or negative, the total number of beta-decay events within a very wide energy window is approximately vanishing.
To be more concrete, the integration of $\Delta(\mathrm{d} \Gamma / \mathrm{d} K^{}_{e})$ over an energy window $\Delta K^{}_{e}$ scales as $\Delta(\mathrm{d} \Gamma / \mathrm{d} K^{}_{e}) \propto m^{}_{\beta}/\Delta K^{}_{e}$\cite{Farzan:2002zq}, which will be vanishing when $\Delta K^{}_{e} \gg m^{}_{\beta}$.
If the energy resolution $\Delta = 100~{\rm meV}$ is larger than absolute neutrino masses, we can evaluate the difference between the effective spectra in the region of $K^{}_e < K^{}_{\rm end, 0} - \Delta$ via series expansion in terms of $m^2_i/\Delta^2$, namely,
\begin{eqnarray}\label{eq:difference}
\frac{{\rm d}\Gamma^{}_{\rm eff}}{{\rm d}K^{}_e} - \frac{{\rm d}\Gamma^{\prime}_{\rm cl}}{{\rm d}K^{}_e} \propto
\left\{
\begin{array}{lcl}
0, & ~ & \hbox{for $m^{}_\beta$ \;;} \\
m^{\prime 2}_\beta - m^2_\beta, & ~ & \hbox{for $m^\prime_\beta$ \;;} \\
m^{\prime\prime 2}_\beta - m^2_\beta, & ~ & \hbox{for $m^{\prime\prime}_\beta$ \;,}
\end{array}
\right.
%     (17)
\end{eqnarray}
where all higher-order terms of ${\cal O}(m^4_i/\Delta^4)$ have been omitted. Consequently, the effective spectrum with $m^{}_\beta$ can fit perfectly the experimental observation, whereas a sizable overall shift is left for $m^{\prime}_{\beta}$ as well as for $m^{\prime\prime}_{\beta}$. As we have mentioned before, although the energy resolution is not good enough to completely pin down the endpoint, the experimental sensitivity to absolute neutrino masses can be obtained by observing the total number of beta-decay events within the energy window around the endpoint. 
\end{itemize}

An immediate question is whether the effective neutrino mass is still a useful parameter for future beta-decay experiments. Put alternatively, does $\mathrm{d} \Gamma^{\prime}_{\rm cl} / \mathrm{d} K^{}_{e}$
in Eq.\ (\ref{eq:Gammaprime}) provide a good description of
the effective spectrum ${\mathrm{d} \Gamma}^{}_{\rm eff}/{\mathrm{d} K^{}_{e}}$ in Eq.\ (\ref{eq:Gammaeff})? We will now investigate the validity of the effective neutrino mass by following a simple statistical approach. The strategy of our numerical analysis is summarized in the following.

Given the target mass and the operation time $T$ (i.e., the total exposure ${\cal E}$), we simulate the experimental data by using the exact spectrum $\mathrm{d} \Gamma^{\prime}_{\rm cl} / \mathrm{d} K^{}_{e}$ and divide the simulated data into a number of energy bins with bin width $\Delta$, which is taken to be the energy resolution of the detector. In general, the event number in the $i$th energy bin $[E^{}_i - \Delta/2, E^{}_i + \Delta/2]$ is given by the integration of the spectrum over the bin width
\begin{eqnarray}\label{eq:binnumber}
N^{}_{i} = T \int^{E^{}_{i} + \Delta/2}_{E^{}_{i}-\Delta/2} \frac{\mathrm{d} {\Gamma}^{}_{\Delta}}{\mathrm{d} K^{}_{e}} \mathrm{d} K^{}_{e} \; ,
\end{eqnarray}
where $E^{}_i$ denotes the mean value of the electron kinetic energy in the $i$th bin and ${\rm d}\Gamma^{}_\Delta/{\rm d}K^{}_e$ is the convolution of a spectrum
%the exact spectrum ${\rm d}\Gamma^{}_{\rm cl}/{\rm d}K^{}_e$
with a Gaussian smearing function as in Eq.~(\ref{eq:spectrum}). The simulated event number $N^{\rm cl}_i$ in each energy bin is calculated by using ${\rm d}\Gamma^{\prime}_{\rm cl}/{\rm d}K^{}_e$ with a specified value of $m^{}_1$ (i.e., the lightest neutrino mass in the NO case). On the other hand, to clarify how good ${\rm d}\Gamma^{}_{\rm eff}/{\rm d}K^{}_e$ can describe the true data, the predicted event number $N^{\rm eff}_i$ in each energy bin is calculated in the same way but with the effective spectrum ${\rm d}\Gamma^{}_{\rm eff}/{\rm d}K^{}_e$, which will be subsequently sent to fit  the simulated true data $N^{\rm cl}_{i}$.

%%%%%%%%%%%%%%%%%%%%%%%%%%%%%%%% Fig. 2 %%%%%%%%%%%%%%%%%%%%%%%%%%%%%%%%%
\begin{figure}[t!]
	\begin{center}
		\includegraphics[width=0.43\textwidth]{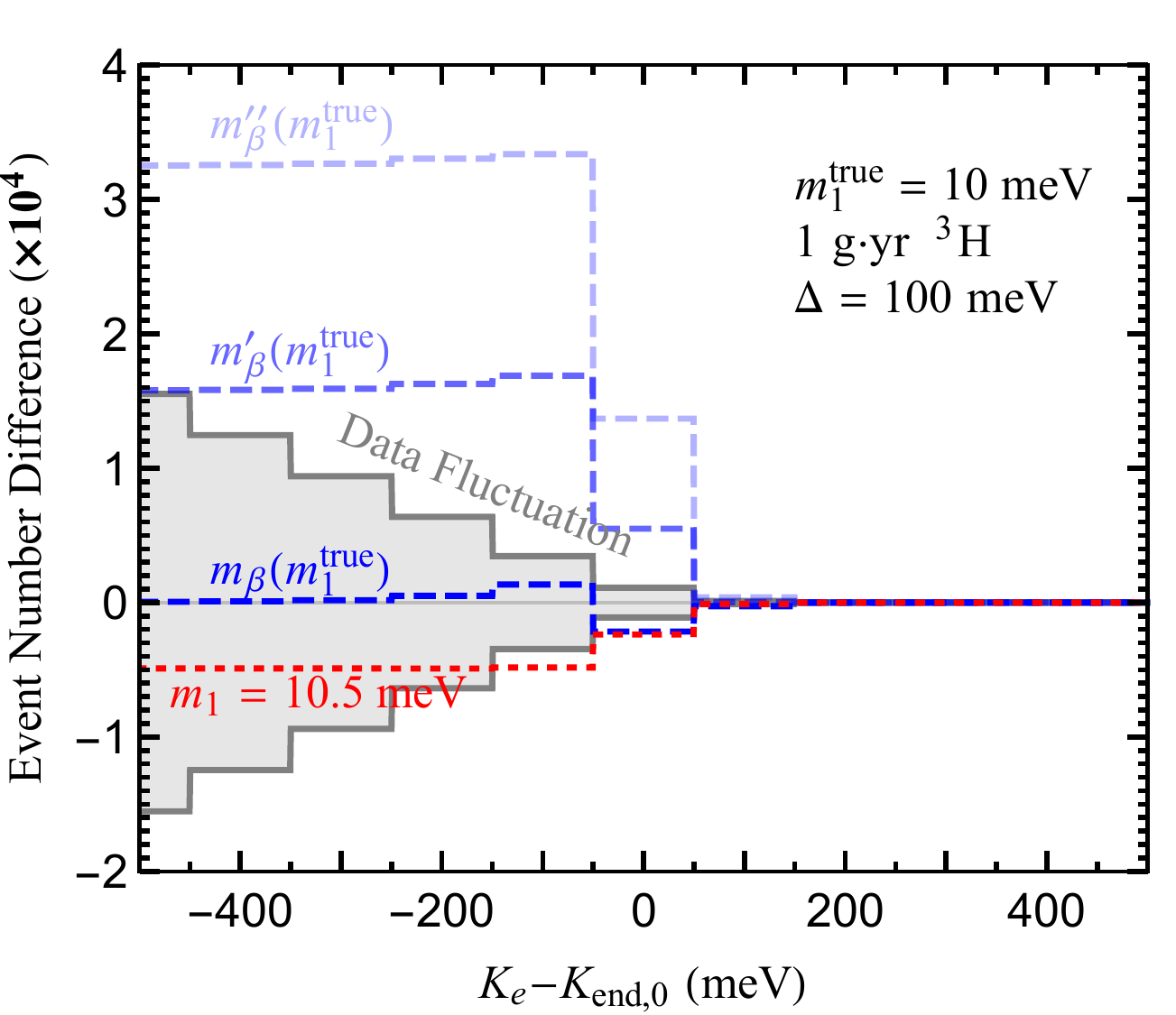}
		\includegraphics[width=0.43\textwidth]{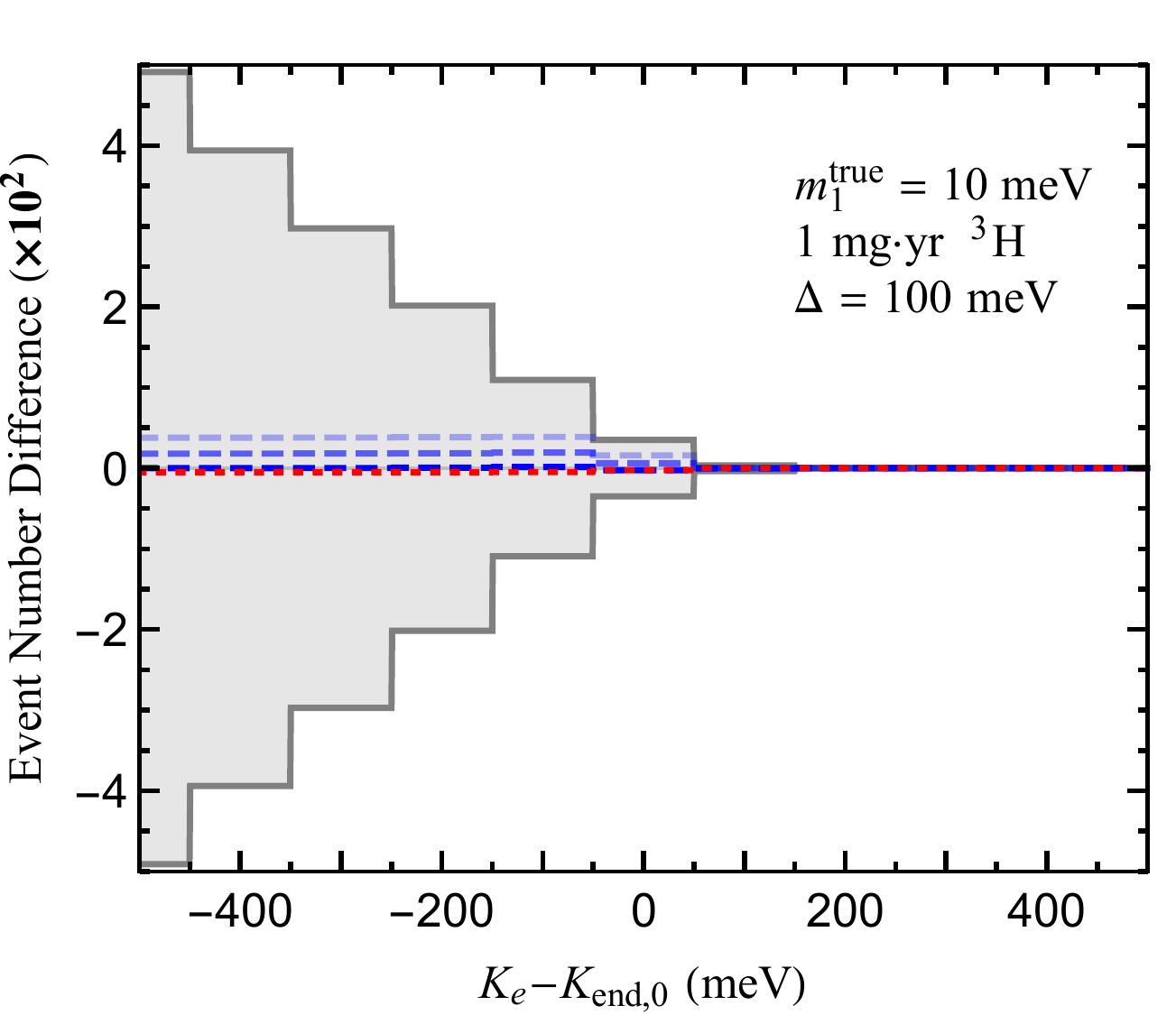}
		\includegraphics[width=0.43\textwidth]{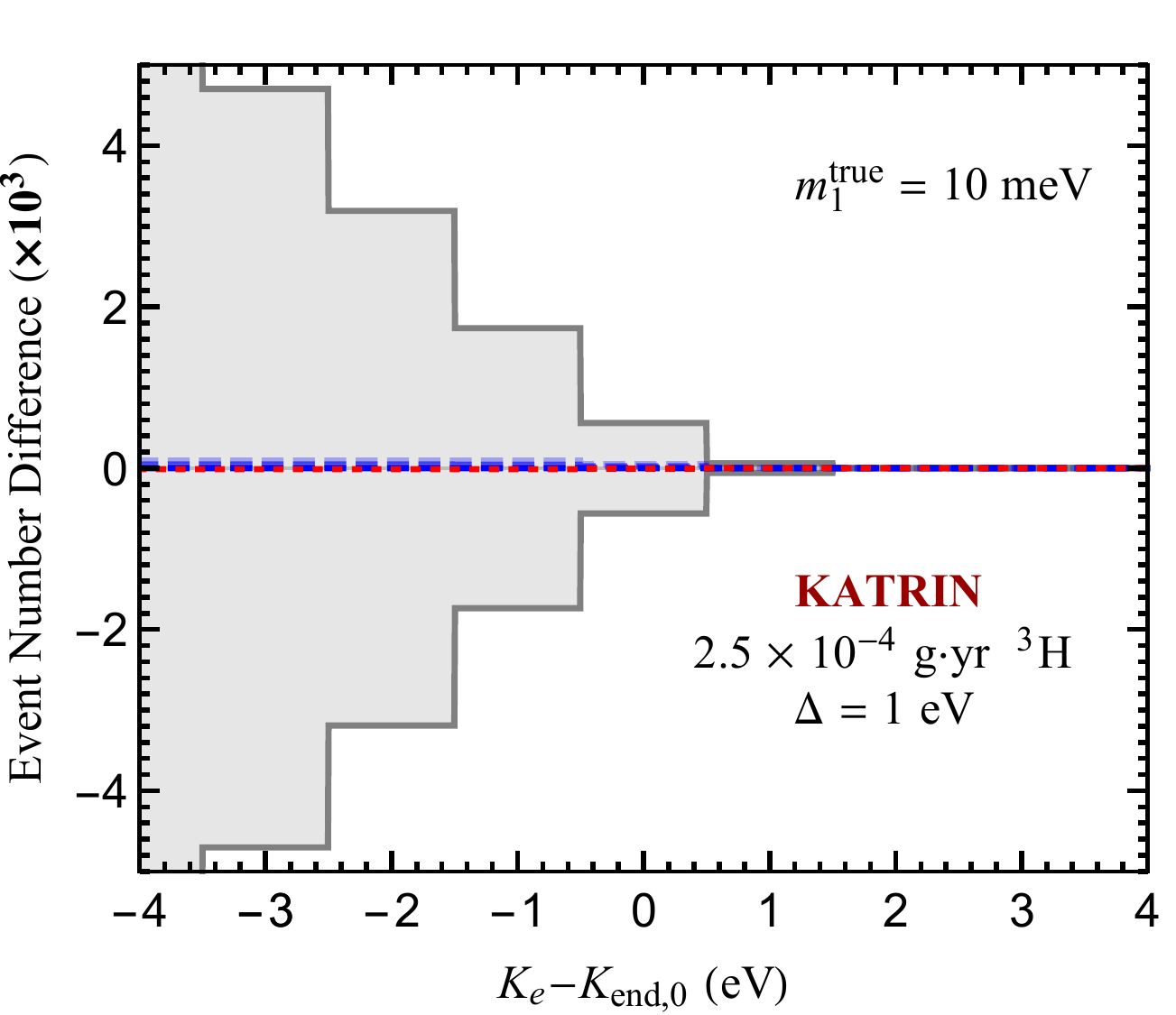}
		\includegraphics[width=0.43\textwidth]{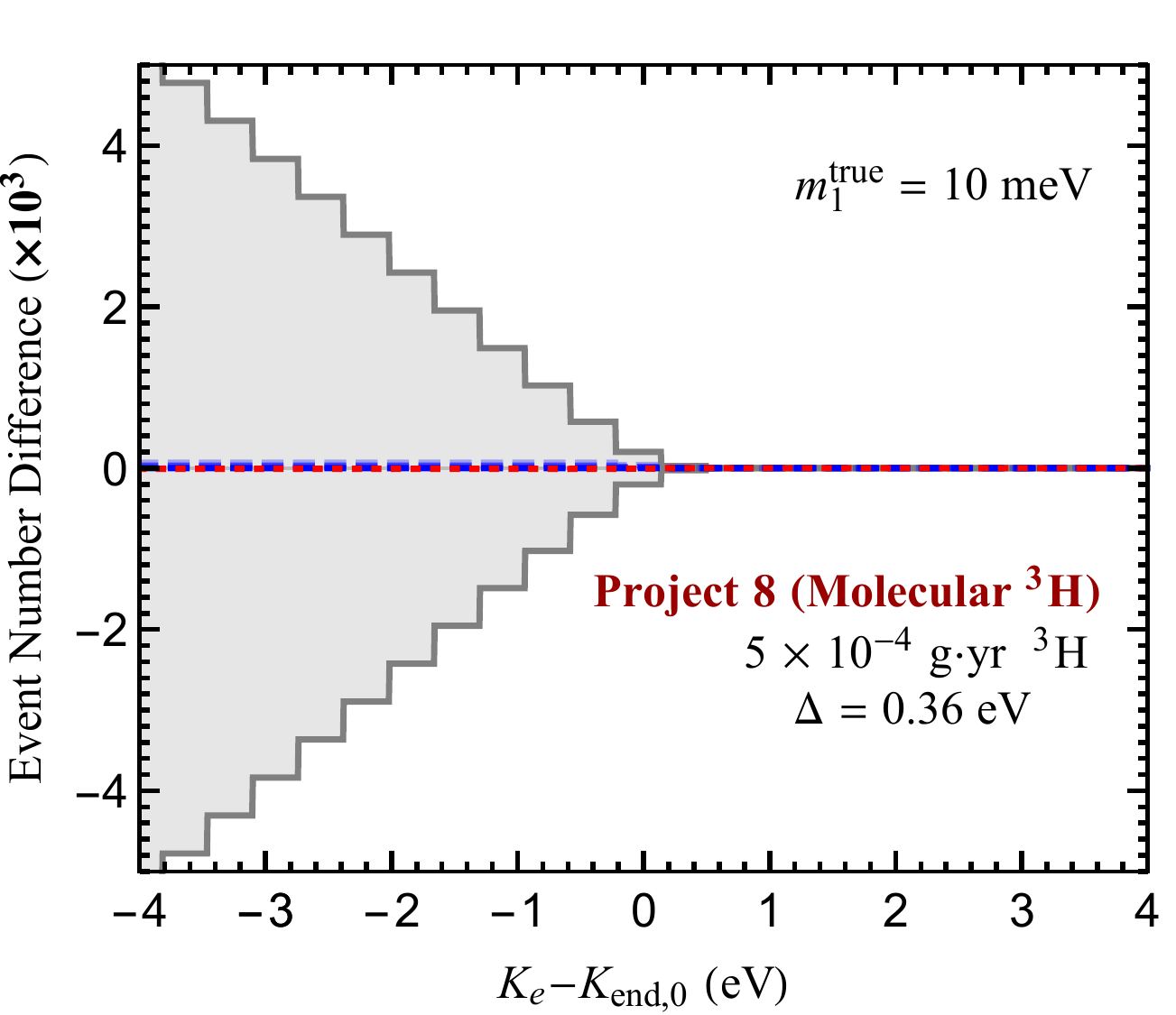}
		\includegraphics[width=0.43\textwidth]{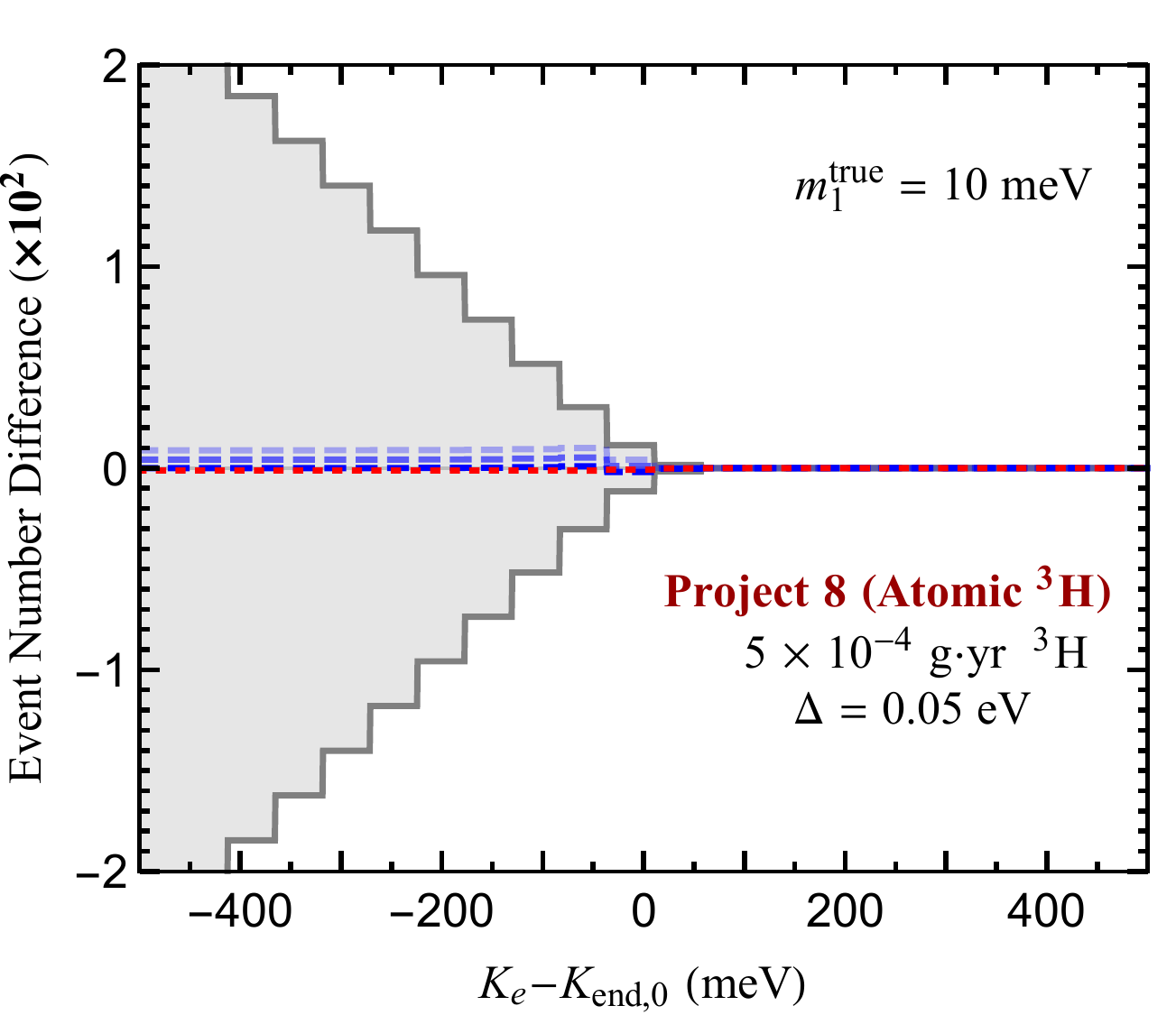}
		\includegraphics[width=0.43\textwidth]{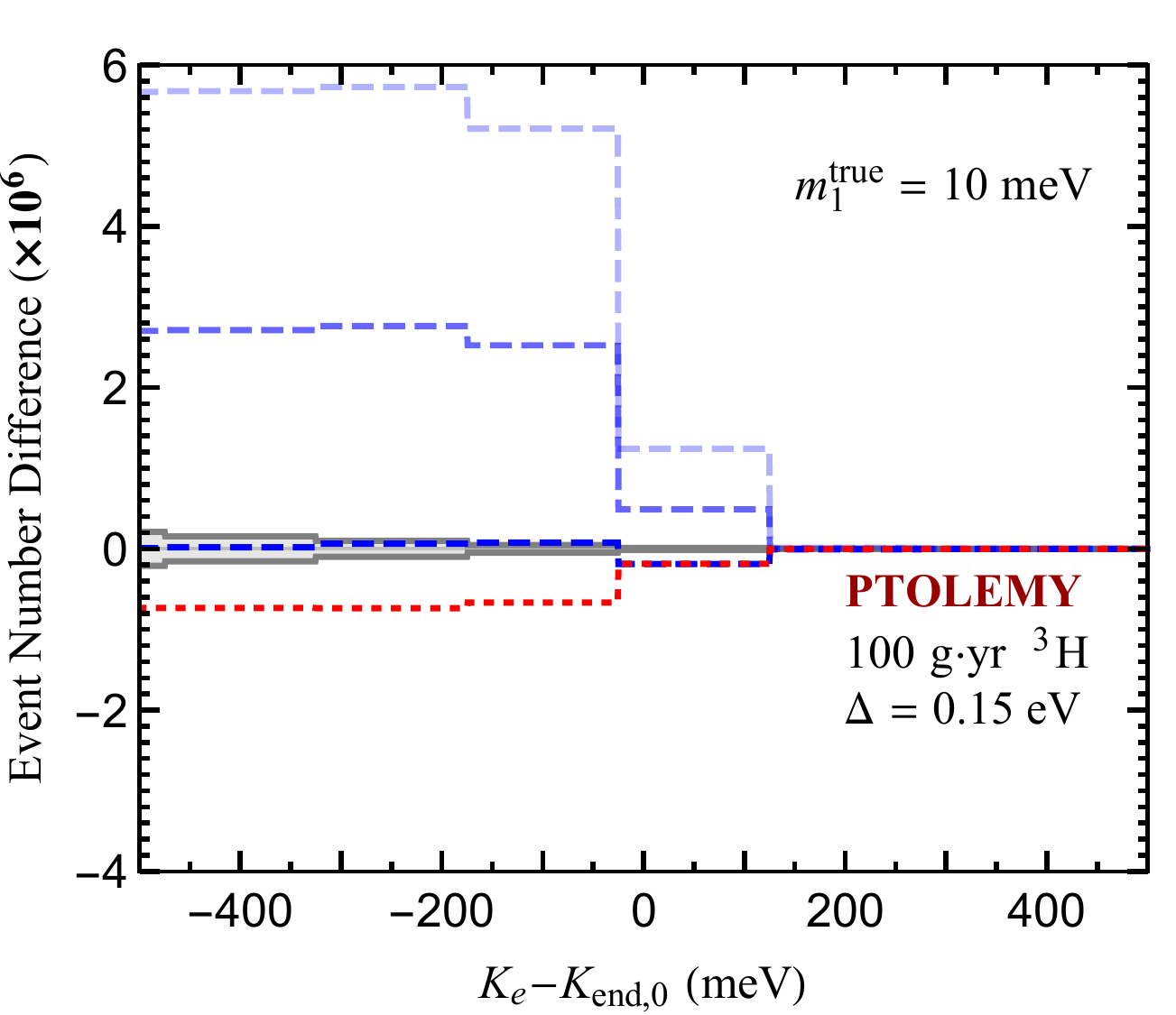}
	\end{center}
	\vspace{-0.5cm}
	\caption{The difference of event numbers for the effective spectrum $\mathrm{d} \Gamma^{}_{\rm eff} / \mathrm{d} K^{}_{e}$ and the exact one $\mathrm{d} \Gamma^{\prime}_{\rm cl} / \mathrm{d} K^{}_{e}$. The data are simulated by taking $m^{\rm true}_{1} = 10~{\rm meV}$. The blue histograms signify the event number deviations of the effective spectra from the exact one, whereas the gray filled histograms stand for the statistical fluctuations. Two nominal experimental setups have been assumed in the upper two panels, and in the remaining four panels we illustrate the cases of realistic experiments including KATRIN, Project 8 with molecular tritium and with atomic one, and PTOLEMY. Note that different scales on the axes have been adopted for each plot.}
	\label{fig:spectrumBinned}
\end{figure}
%%%%%%%%%%%%%%%%%%%%%%%%%%%%%%%%%%%%%%%%%%%%%%%%%%%%%%%%%%%%%%%%%%%%%%%%%%%%%

It should be noted that KATRIN operating in the ordinary mode with the MAC-E-Filter observes actually the integrated number of beta-decay events and has to reconstruct the differential spectrum by adjusting the retarding potential to scan over a certain energy window containing the endpoint. The number of events for the differential spectrum in each energy bin turns out to be $N^{}_{i} = N^{\rm int}_{i} - N^{\rm int}_{i-1}$, where $N^{\rm int}_{i}$ is the event number of the integrated spectrum for the scanning point corresponding to $E^{}_i$. For this reason, the statistical fluctuation of the event number for the reconstructed differential spectrum can be estimated as $\sqrt{N^{\rm int}_{i} + N^{\rm int}_{i-1}}$, which should be compared with that of $\sqrt{N^{}_{i}}$ for the direct measurement. Meanwhile, a longer time of data taking is also expected. Therefore our result should be taken to be conservative when considering KATRIN-like experiments operating in integrated mode.

KATRIN can also directly measure the non-integrated beta spectrum in a  possible MAC-E-TOF mode, as described in Appendix~\ref{sec:appA}.
Since tritium experiments in the future tend to adopt non-integrated modes to maximize the neutrino mass sensitivity, we shall focus on this scenario.
For those tritium experiments operating in the non-integrated mode, we will use the following experimental configurations: (i) KATRIN with the target mass $m^{}_{\rm KATRIN} = 2.5 \times 10^{-4}~{\rm g}$ and energy resolution $\Delta^{}_{\rm KATRIN} = 1~{\rm eV}$; (ii) Project 8 loaded with molecular tritium gas with $m^{}_{\rm P8} = 5 \times 10^{-4}~{\rm g}$ and $\Delta^{}_{\rm P8m} = 0.36~{\rm eV}$; (iii) Project 8 loaded with atomic tritium gas with $m^{}_{\rm P8} = 5 \times 10^{-4}~{\rm g}$ and $\Delta^{}_{\rm P8a} = 0.05~{\rm eV}$;
(iv) PTOLEMY with $m^{}_{\rm PTOLEMY} = 100~{\rm g}$ and $\Delta^{}_{\rm PTOLEMY} = 0.15~{\rm eV}$. The details can be found in Appendix~\ref{sec:appA}. 
For Project 8 loaded with molecular tritium, the energy resolution is limited by the irreducible width of the final-state molecular excitations \cite{Monreal:2009za}.
This limitation can be overcome by switching the target to atomic tritium.
Note that for KATRIN  in the MAC-E-TOF mode, which is still under development, the penalties of the tritium decay rate and energy resolution due to the chopping procedure (see Appendix~\ref{sec:appA}) are ignored, so the configuration here is somewhat idealized for KATRIN.
Nevertheless, we will find the effect of using $m^{}_{\beta}$ even in this ideal KATRIN setup is negligible.
We adopt Gaussian distributions as in Eq.~(\ref{eq:spectrum}) for the uncertainties caused by finite energy resolutions in all experiments. In a more realistic analysis with all experimental details taken into account, one should consider a strict shape for the energy resolution function, e.g., a triangle-like shape for KATRIN in the  developing MAC-E-TOF mode. 
The actual shape for Project 8 with molecular tritium should  also be calculated with detailed consideration of  final-state excitations.
However, a different shape of the energy resolution from Gaussian should not affect our results by orders of magnitude.

%%%%%%%%%%%%%%%%%%%%%%%%%%%%%%%% Fig. 3 %%%%%%%%%%%%%%%%%%%%%%%%%%%%%%%%%
\begin{figure}[t!]
	\begin{center}
		\hspace{-0.2cm}
		\includegraphics[width=0.48\textwidth]{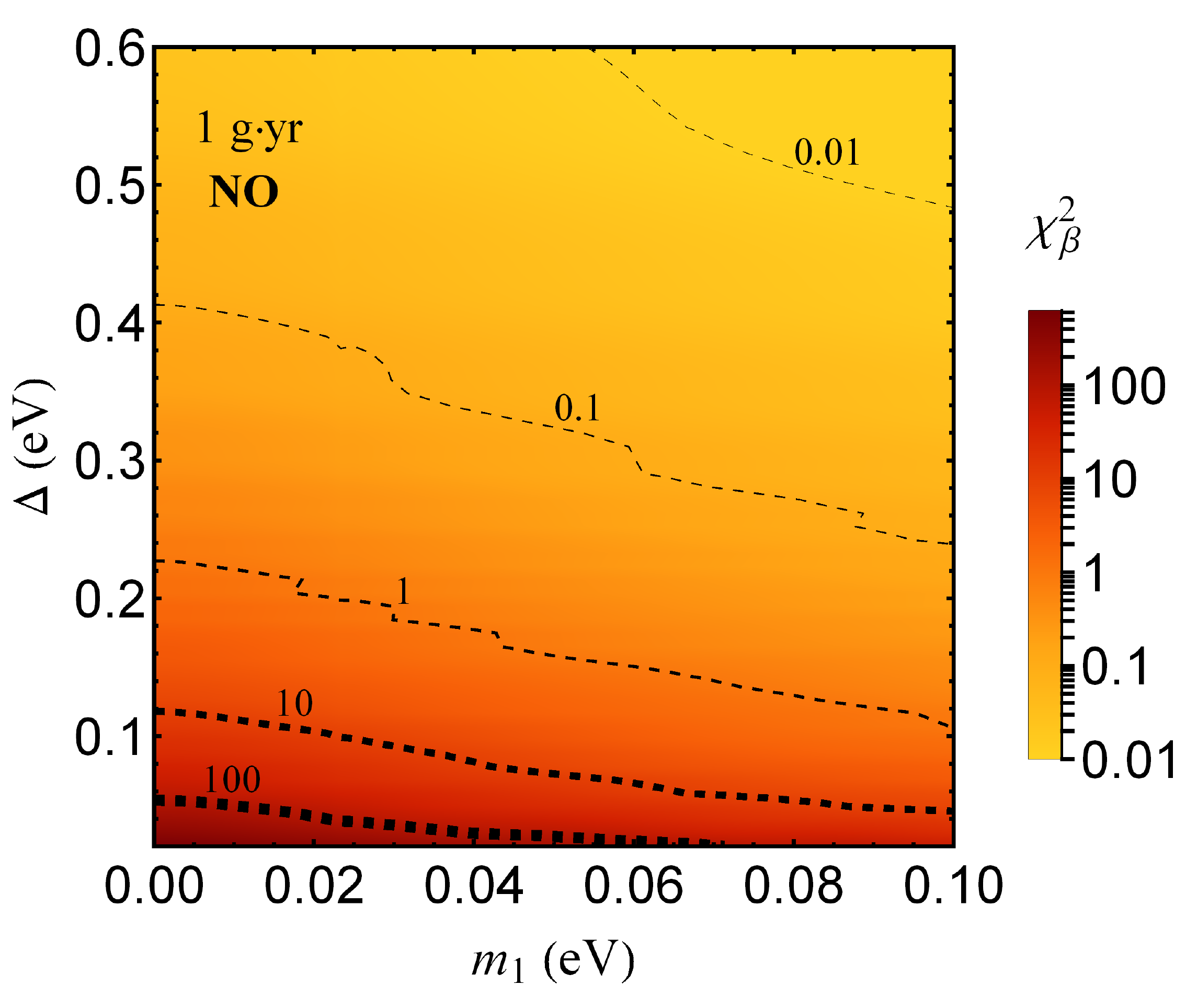}
		\includegraphics[width=0.48\textwidth]{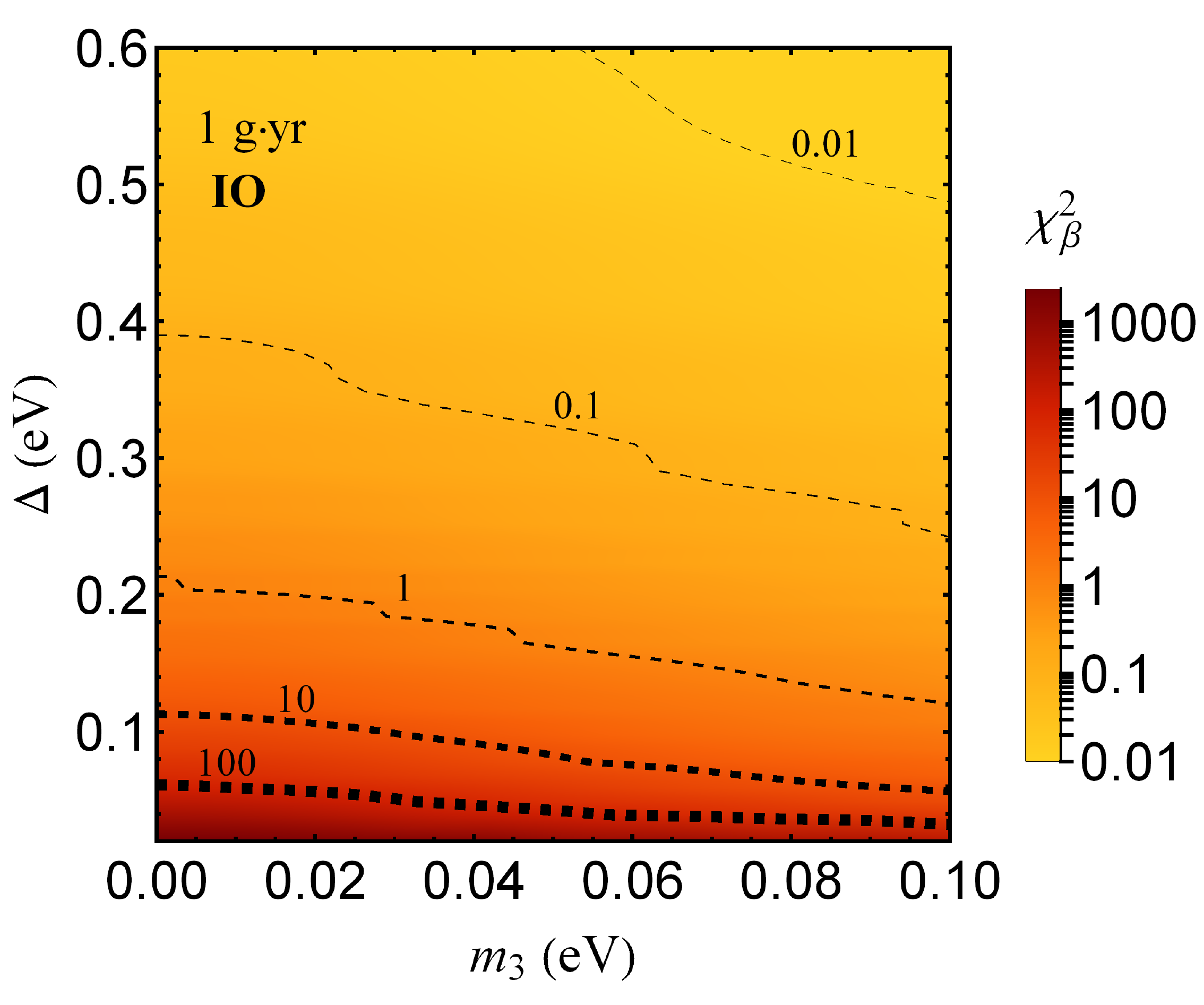}
	\end{center}
	\vspace{-0.5cm}
	\caption{The contours of $\chi^2_\beta$, see Eq.\ (\ref{eq:chi2}),
	  arising from the description of the electron spectrum by using the effective neutrino mass $m^{}_{\beta}$, are displayed in the plane of smallest mass versus resolution, i.e.\ the
          $m^{}_{1}$-$\Delta$ plane for the NO case (left panel), and the $m^{}_{3}$-$\Delta$ plane for the IO case (right panel).}
	\label{fig:chi2VSDeltaVSm}
\end{figure}
%%%%%%%%%%%%%%%%%%%%%%%%%%%%%%%%%%%%%%%%%%%%%%%%%%%%%%%%%%%%%%%%%%%%%%%%%%%%%
In Fig.~\ref{fig:spectrumBinned}, we show the difference in the event numbers of ${{\rm d}\Gamma^{}_{\rm eff}}/{{\rm d}K^{}_e}$ and ${{\rm d}\Gamma^{\prime}_{\rm cl}}/{{\rm d}K^{}_e}$, together with the statistical fluctuation of the events.
In the upper two panels two nominal experimental setups have been chosen for demonstration, and in the remaining four panels we illustrate the cases of realistic experiments.
In all panels, the data are simulated with ${{\rm d}\Gamma^{\prime}_{\rm cl}}/{{\rm d}K^{}_e}$, for which a true value of the lightest neutrino mass $m^{\rm true}_1 = 10~{\rm meV}$ has been input, and the data fluctuations are represented by the filled gray histograms. For comparison, the event number difference in each energy bin has been calculated for three effective spectra with $m^{}_\beta = 13.4~{\rm meV}$, $m^\prime_\beta = 11.9~{\rm meV}$ and $m^{\prime \prime}_\beta = 10~{\rm meV}$, which is denoted as the blue dashed curves. In addition, the gray solid curve denotes the exact spectrum with $m^{\rm true}_1 = 10~{\rm meV}$ as in Fig.~\ref{fig:spectrumVSKe}, while the red dotted curve is for
$m^{\rm true}_1 = 10.5~{\rm meV}$.
From Fig.~\ref{fig:spectrumBinned}, two important observations can be made. First, for a smaller exposure such as in KATRIN and Project 8, the statistical fluctuation can easily overwhelm the deviations, rendering the effective description of the beta spectrum more reliable. Second, for $m^{}_{\beta}$, the error caused by using the effective spectrum is most significant in the energy bin containing the endpoint. The reason is obvious, namely that the data  fluctuation increases and the deviation decreases, as the energy moves away from the endpoint. For the other two effective masses $m^{\prime}_{\beta}$ and $m^{\prime\prime}_{\beta}$, the deviations are even more significant.

%%%%%%%%%%%%%%%%%%%%%%%%%%%%%%%% Fig. 4 %%%%%%%%%%%%%%%%%%%%%%%%%%%%%%%%%
\begin{figure}[t!]
	\begin{center}
		\hspace{-0.2cm}
		\includegraphics[width=0.48\textwidth]{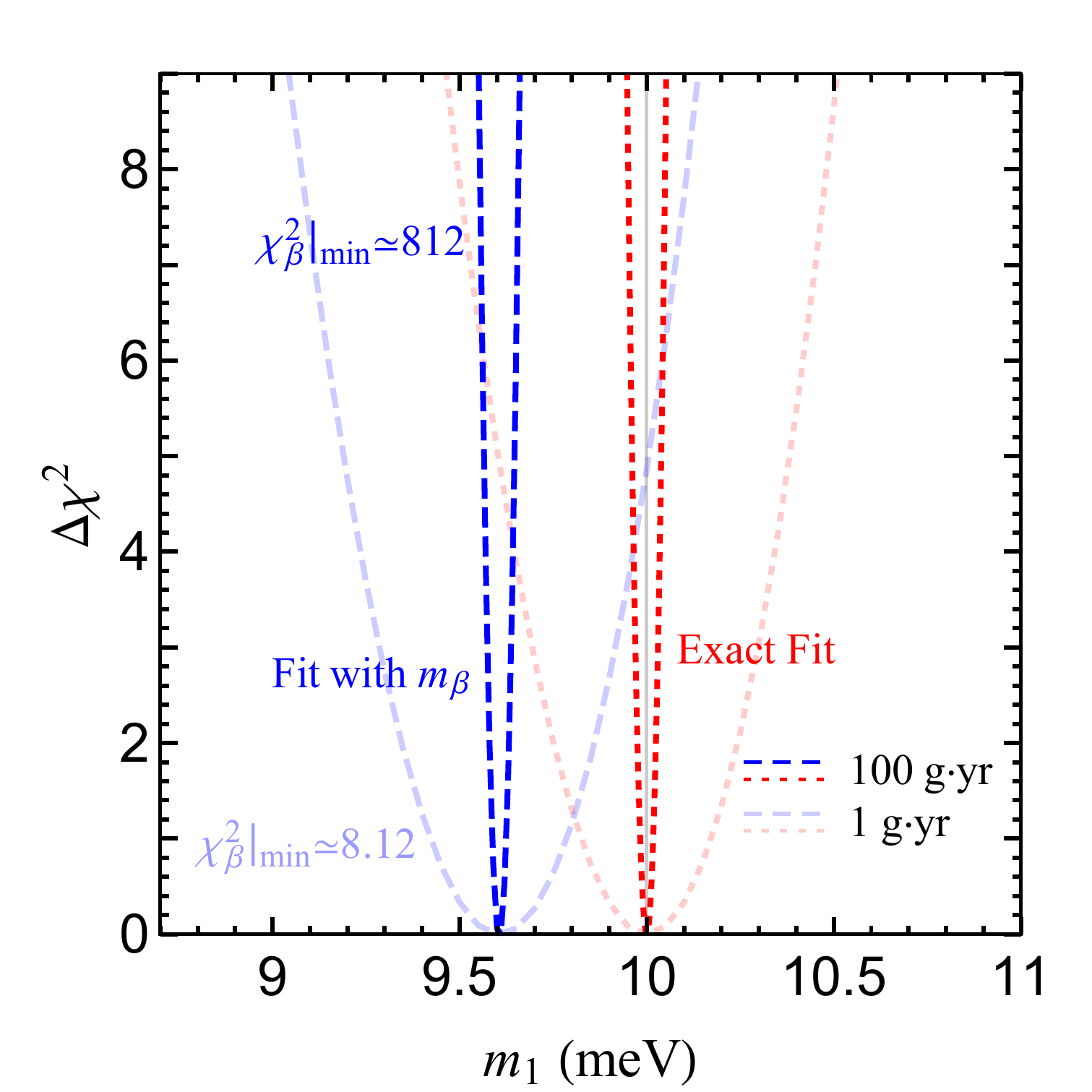}
	\end{center}
	\vspace{-0.5cm}
	\caption{The function $\Delta \chi^2 \equiv \chi^2_\beta - \chi^2_\beta|^{}_{\rm min}$ is shown with respect to the lightest neutrino mass $m^{}_{1}$ in the NO case. The dark red curve is generated by fitting with the exact spectrum $\mathrm{d} \Gamma^{\prime}_{\rm cl} / \mathrm{d} K^{}_{e}$, while the dark blue one is by using the effective spectrum   $\mathrm{d} \Gamma^{}_{\rm eff} / \mathrm{d} K^{}_{e}$ with $m^{}_{\beta}$, for the exposure of ${\cal E} = 100~{\rm g}\cdot {\rm yr}$. The light curves are for ${\cal E} = 1~{\rm g}\cdot {\rm yr}$ with all else being the same. For illustration, we use an  energy resolution of $\Delta = 0.1~{\rm eV}$.}
	\label{fig:chi2VSm1}
\end{figure}
%%%%%%%%%%%%%%%%%%%%%%%%%%%%%%%%%%%%%%%%%%%%%%%%%%%%%%%%%%%%%%%%%%%%%%%%%%%%%

\begin{table}[b]
	\small
	\centering
	\begin{tabular*}{\textwidth}{l @{\extracolsep{\fill}} c c| c c c c }
	\hline
    \hline
	$m^{}_{\rm L}=0~{\rm eV}$&  Target Mass & $\Delta$ & $\chi^2_{\beta}$, NO & $\chi^2_{\beta}$, IO  & $\Delta \chi^2_{\rm true}$, NO & $\Delta \chi^2_{\rm true}$, IO
	  \\
	\hline
	 KATRIN & $2.5 \times 10^{-4}~{\rm g}$ & $1~{\rm eV}$ & $7.4{\times} 10^{-7}$ & $5.6{\times} 10^{-7}$ &  $1.3{\times} 10^{-7}$ & $1.1 {\times} 10^{-7}$\\
	  Project 8 (Molecular $^3{\rm H}$)& $5\times 10^{-4}~{\rm g}$  & $0.36~{\rm eV}$ & $8.9{\times} 10^{-5}$ & $6.1{\times} 10^{-5}$ &  $2.0{\times} 10^{-5}$ & $1.4 {\times} 10^{-5}$\\
	  Project 8 (Atomic $^3{\rm H}$)& $5\times 10^{-4}~{\rm g}$  & $0.05~{\rm eV}$ & $0.064$ & $0.13$ &  $0.032$ & $0.015$\\
	 PTOLEMY & $100~{\rm g}$ & $0.15~{\rm eV}$ & $428$ & $331$ & $141$ & $81$ \\
	 \hline
	 \hline
	\end{tabular*}
	\caption{The configurations of tritium beta decay experiments and the resulting $\chi^2_{\beta}$ and $\Delta \chi^2_{\rm true}$ defined in Eqs.~(\ref{eq:chi2}) and (\ref{eq:Dchi2}) arising from the description of the electron spectrum by using the effective neutrino mass $m^{}_{\beta}$. One year of data taking has been assumed. No background is assumed, and the $\chi^2$-values can be further reduced taking into account possible background contributions.}
	\label{table:betaDecaysChi2}
\end{table}	

To quantify the difference between the effective and exact spectra in a statistical approach, we define $\Delta N^{}_{i} \equiv N^{\rm eff}_{i}-N^{\rm cl}_{i}$ in each energy bin and take $\sqrt{N^{\rm cl}_i}$ to be the corresponding statistical uncertainty.\footnote{This is true when the event number in each energy bin is large, such that the fluctuation follows approximately a Gaussian distribution, which turns out to be true for all tritium experiments in our consideration as can be noticed in Fig.~\ref{fig:spectrumBinned}.} In this way, if $\Delta N^{}_{i}$ is negligible compared to $\sqrt{N^{\rm cl}_{i}}$, one can claim that the error due to the use of the effective spectrum is unimportant in that energy bin. For the whole energy spectrum, the $\chi^2$-function can be constructed as
\begin{eqnarray}\label{eq:chi2}
\chi^2_{\beta} = \sum^{}_{i} \frac{\left(\Delta N^{}_{i}\right)^2}{N^{\rm cl}_{i}} \; ,
%     (19)
\end{eqnarray}
where $i$ runs over the number of energy bins.
This $\chi^2$-function measures to what degree the effective spectrum ${{\rm d}\Gamma^{}_{\rm eff}}/{{\rm d}K^{}_e}$ deviates from the exact one ${{\rm d}\Gamma^{\prime}_{\rm cl}}/{{\rm d}K^{}_e}$. Because we have used ${{\rm d}\Gamma^{\prime}_{\rm cl}}/{{\rm d}K^{}_e}$ to generate the true data, from the model selection perspective (i.e., fitting two different models ${{\rm d}\Gamma^{\prime}_{\rm cl}}/{{\rm d}K^{}_e}$ and ${{\rm d}\Gamma^{}_{\rm eff}}/{{\rm d}K^{}_e}$ with the same data, respectively), $\chi^2_{\beta}$ defines the statistical significance with which one can favor ${{\rm d}\Gamma^{\prime}_{\rm cl}}/{{\rm d}K^{}_e}$ over ${{\rm d}\Gamma^{}_{\rm eff}}/{{\rm d}K^{}_e}$.
If one insists in using the effective spectrum ${{\rm d}\Gamma^{}_{\rm eff}}/{{\rm d}K^{}_e}$ to fit the data, $\chi^2_{\beta}$ also measures the goodness of fit $\chi^2_{\beta}/v$ of ${{\rm d}\Gamma^{}_{\rm eff}}/{{\rm d}K^{}_e}$ given the degree of freedom $v$ in fitting. Since most deviations of ${{\rm d}\Gamma^{}_{\rm eff}}/{{\rm d}K^{}_e}$ from ${{\rm d}\Gamma^{\prime}_{\rm cl}}/{{\rm d}K^{}_e}$ distribute only in a few energy bins around the endpoint, the degree of freedom can be $v= \mathcal{O}(1)$ depending on the number of bins we use in the actual fit.
As has been mentioned previously, we have fixed the bin size to be the energy resolution $\Delta$. In principle, the bin width can be chosen freely. The smaller the bin width is, the more information one can acquire in the fit. However, this is limited by the energy resolution of an experiment, which will smooth out the information within a comparable bin size, such that further decreasing the bin size will not improve the result anymore. We have numerically checked that by choosing a bin width smaller than the energy resolution, e.g. $\Delta/8$, the $\chi^2$-function defined in Eq.~(\ref{eq:chi2}) will  increase only by a factor of $\sim 70\%$. Further reducing the bin width will not alter this result.

Let us make some remarks on the other input in our numerical calculations. First, the best-fit values of neutrino oscillation parameters from Ref.~\cite{Esteban:2018azc} are adopted. Second, the energy window for the analysis has been taken to be $K^{}_{e} - K^{}_{\rm end,0} \in (-4\cdots 4)~{\rm eV}$. Third, we have assumed no background contributions. The inclusion of possible background events will reduce the value of $\chi^2_\beta$, leading to a smaller statistical deviation of the effective spectrum ${{\rm d}\Gamma^{}_{\rm eff}}/{{\rm d}K^{}_e}$ from the classical one ${{\rm d}\Gamma^{\prime}_{\rm cl}}/{{\rm d}K^{}_e}$. Four, we take the normalization factor to be one, as it can be precisely determined by choosing a wider energy window in realistic experiments.

In the left panel of Fig.~\ref{fig:chi2VSDeltaVSm}, for each pair of $m^{}_{1}$ in the range of $(0\cdots 0.1)~{\rm eV}$ and $\Delta$ in the range of $(0.02\cdots 0.6)~{\rm eV}$, we present the value of $\chi^2_\beta$ for the total exposure of ${\cal E} = 1~{\rm g\cdot yr}$ in the NO case, where the effective spectrum with $m^{}_\beta$ is adopted for illustration. Similar calculations have also been carried out in the IO case and the results are given in the $m^{}_3$-$\Delta$ plane in the right panel. Roughly speaking, for those values of $\Delta$ and $m^{}_1$ in the NO case (or $m^{}_3$ in the IO case) corresponding to $\chi^2_\beta \lesssim 0.1$, the effective spectrum ${{\rm d}\Gamma^{}_{\rm eff}}/{{\rm d}K^{}_e}$ with $m^{}_\beta$ is reasonably good to describe the data,
i.e., with negligible and fragile statistical significance to discriminate ${{\rm d}\Gamma^{\prime}_{\rm cl}}/{{\rm d}K^{}_e}$ from ${{\rm d}\Gamma^{}_{\rm eff}}/{{\rm d}K^{}_e}$ and no noticeable impact on the goodness-of-fit.
In the same sense, we can also conclude that $m^{}_\beta$ is no longer a safe parameter for those values of $\Delta$ and $m^{}_1$ (or $m^{}_3$) corresponding to $\chi^2_\beta \gtrsim 10$,
i.e., the statistical power to favor ${{\rm d}\Gamma^{\prime}_{\rm cl}}/{{\rm d}K^{}_e}$ over ${{\rm d}\Gamma^{}_{\rm eff}}/{{\rm d}K^{}_e}$ is more than $3\sigma$ and a considerable impact on the goodness of fit arises (the $p$-value of fit is 0.001565 for $\chi^2_\beta = 10$ and $v=1$, and the model ${{\rm d}\Gamma^{}_{\rm eff}}/{{\rm d}K^{}_e}$ is almost ruled out by the data).

Although we have fixed the exposure at ${\cal E} = 1~{\rm g}\cdot {\rm yr}$, it is straightforward to derive the values of $\chi^2_\beta$ for a different exposure by noting the fact that $(\Delta N^{}_{i})^2/{N^{\rm cl}_{i}}$ is linearly proportional to $\mathcal{E}$. As a consequence, for a different exposure $\widetilde{\cal E}$, the original values of $\chi^2_\beta$ for ${\cal E}$ will be modified to be $\chi^2_\beta \cdot \widetilde{\cal E}/{\cal E}$. For instance, the original value of the contour $ \chi^2_{\beta} = 10$ for ${\cal E} = 1~{\rm g\cdot yr}$ should be changed to $\chi^2_\beta = 0.01$ for $\mathcal{E} = 1~{\rm mg\cdot yr}$.
Nevertheless, if we insist in using ${{\rm d}\Gamma^{}_{\rm eff}}/{{\rm d}K^{}_e}$ to fit the data regardless of the statistical preference for the true model ${{\rm d}\Gamma^{\prime}_{\rm cl}}/{{\rm d}K^{}_e}$ and a poor goodness-of-fit, the parameter estimation of $m^{}_{\beta}$ can always be performed based on using ${{\rm d}\Gamma^{}_{\rm eff}}/{{\rm d}K^{}_e}$.
In this case, a large value of $\chi^2_\beta$ does not necessarily mean a large value of $\Delta \chi^2 \equiv \chi^2_\beta - \chi^2_\beta|^{}_{\rm min}$ in parameter estimations, where $\chi^2_\beta|^{}_{\rm min}$ denotes the minimum of $\chi^2_\beta$ by freely adjusting $m^{}_\beta$ or $m^{}_1$.

%%%%%%%%%%%%%%%%%%%%%%%%%%%%%%%%%
\begin{figure}[t]
	\begin{center}
		\hspace{-0.2cm}
		\includegraphics[width=0.48\textwidth]{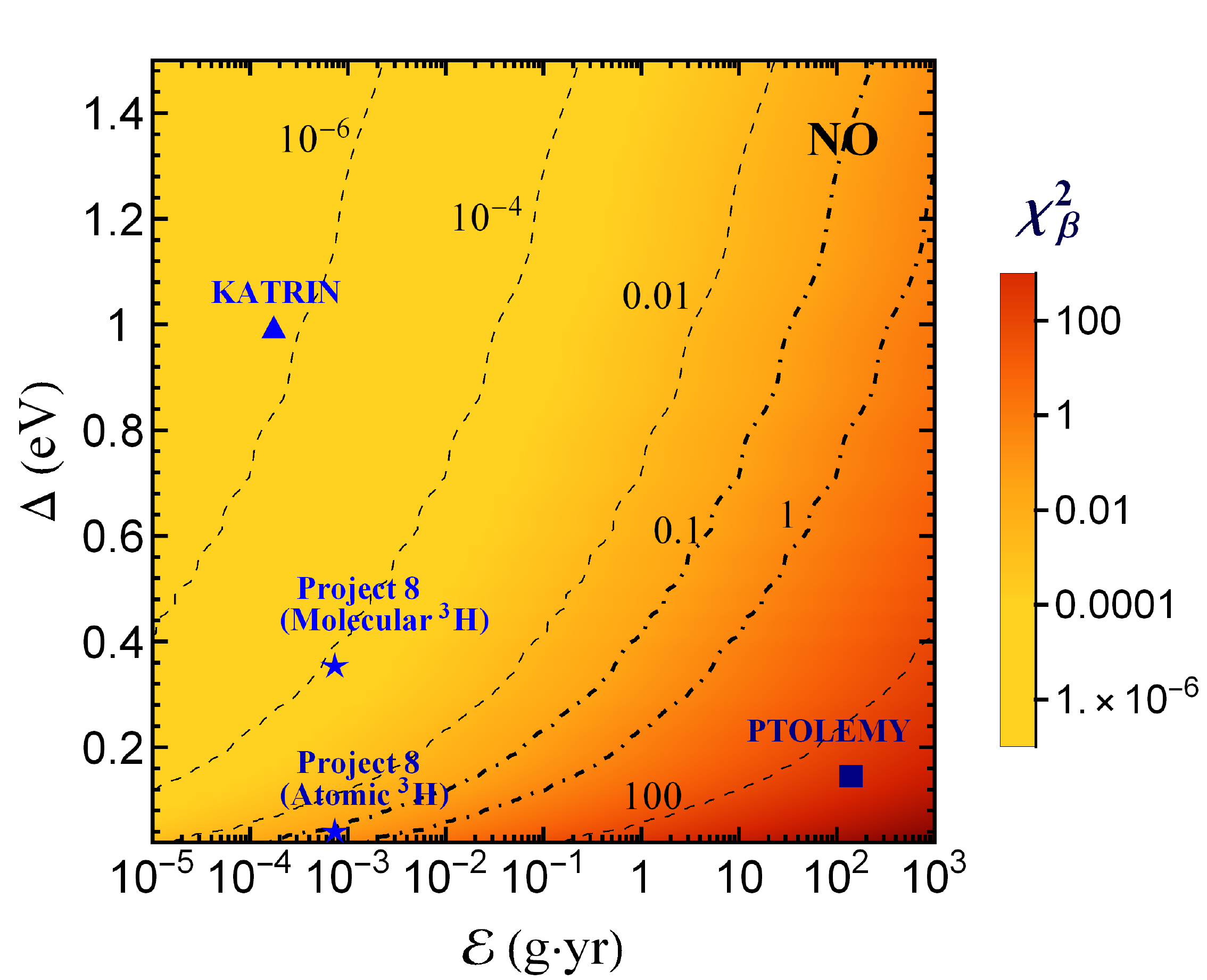}
		\includegraphics[width=0.48\textwidth]{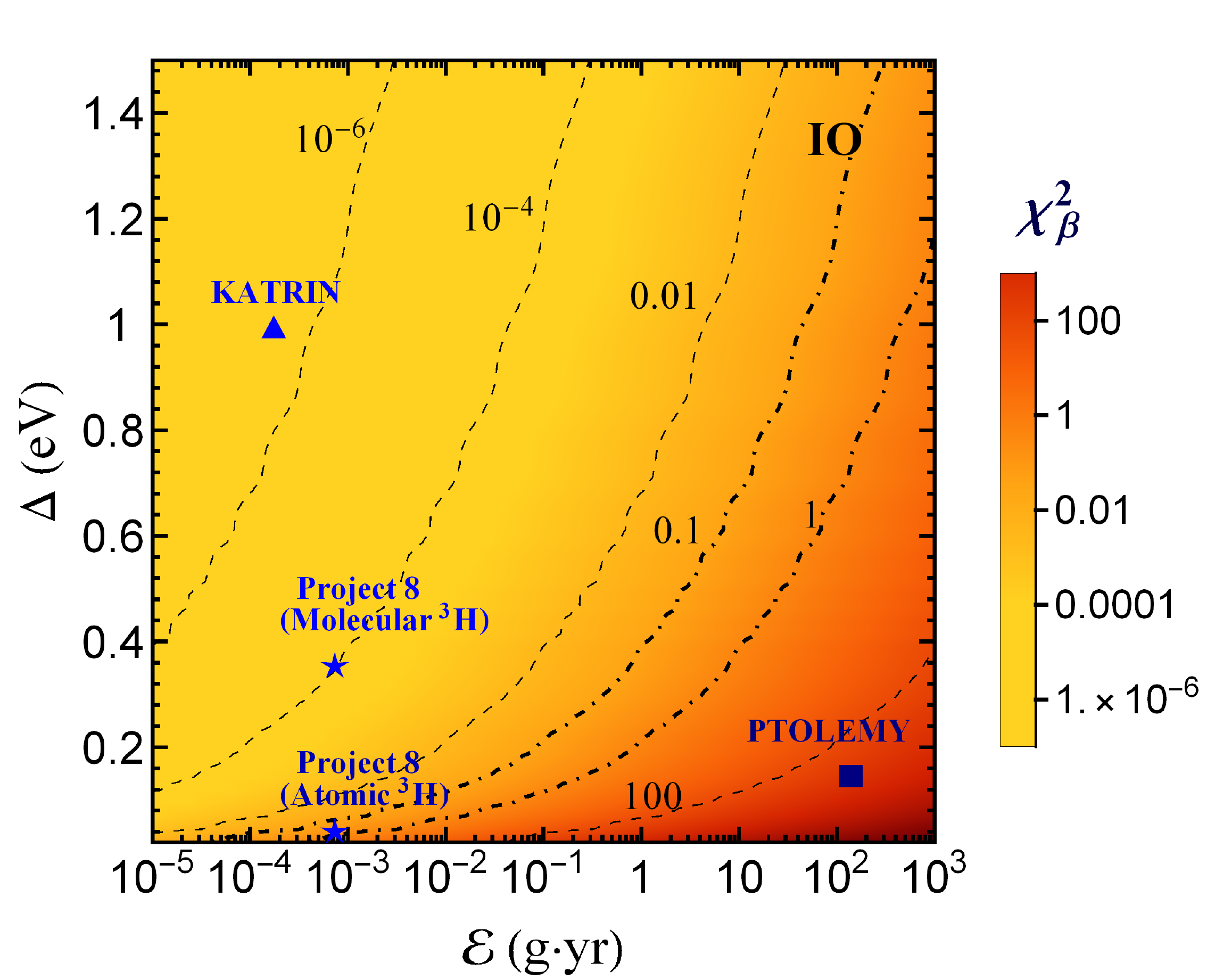}
		\includegraphics[width=0.48\textwidth]{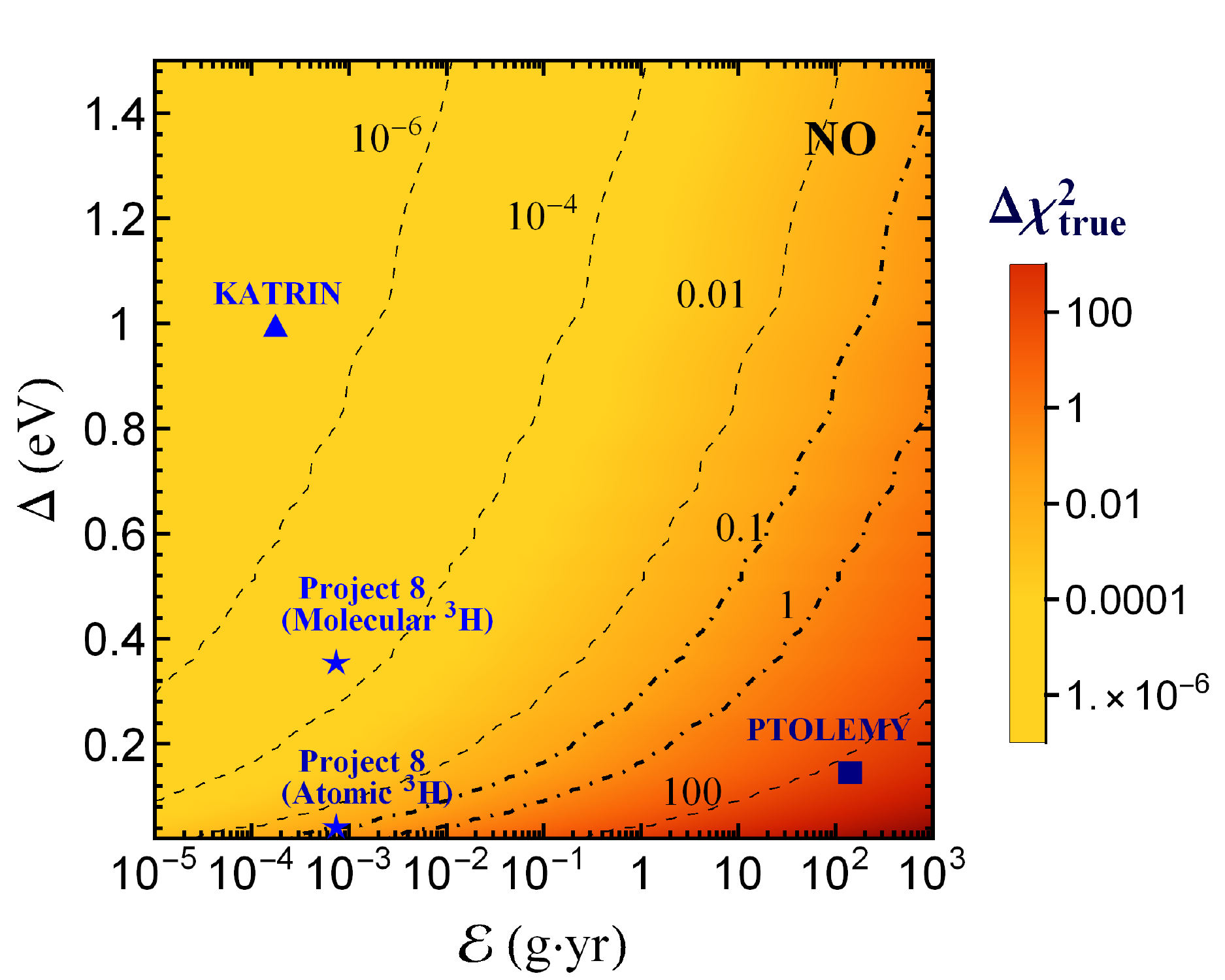}
		\includegraphics[width=0.48\textwidth]{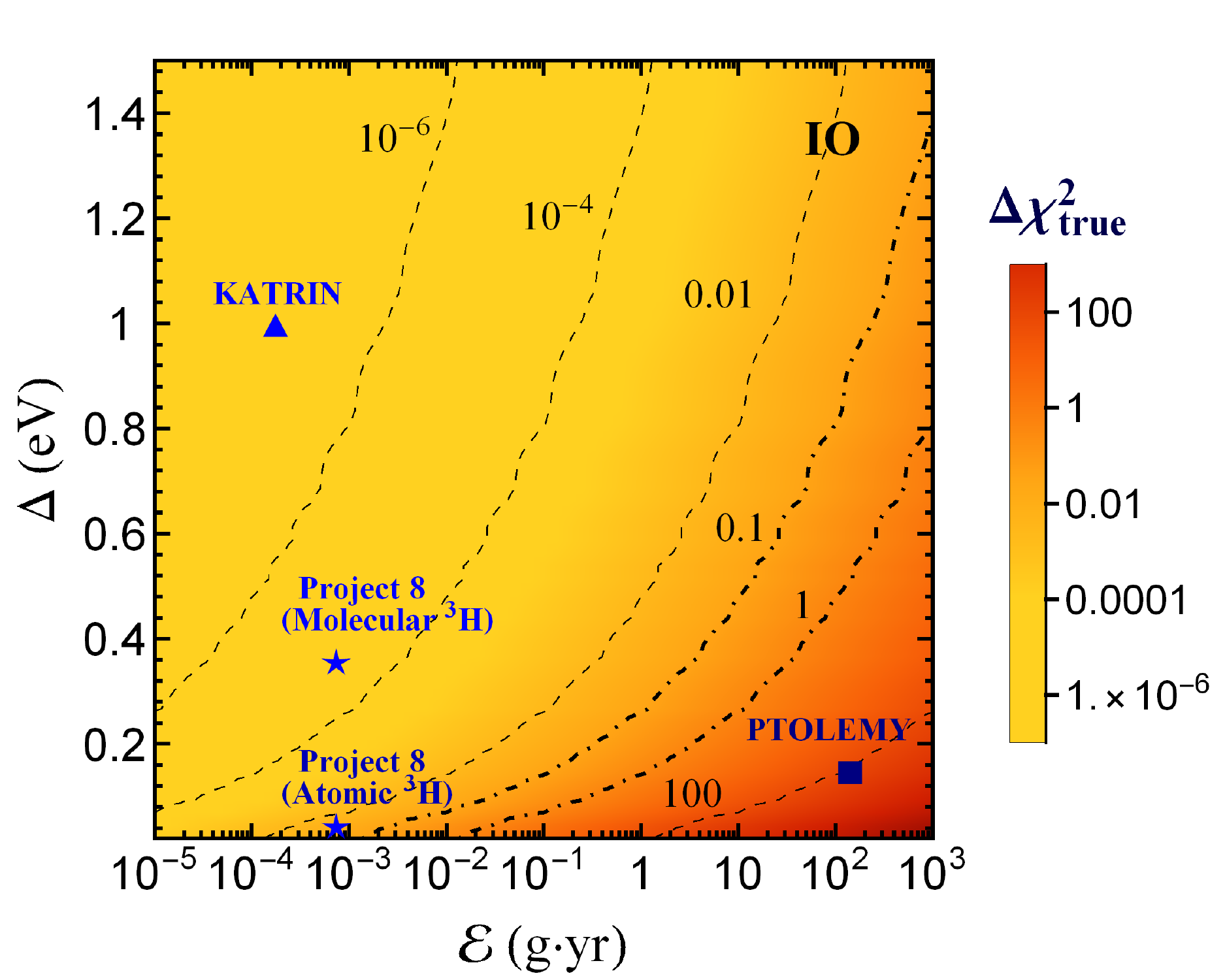}
	\end{center}
	\vspace{-0.5cm}
	\caption{The contours of $\chi^2_\beta$ and $\Delta \chi^2_{\rm true}$, see Eqs.~(\ref{eq:chi2}) and (\ref{eq:Dchi2}),
	  arising from the description of the electron spectrum by using the effective neutrino mass $m^{}_{\beta}$, are displayed in the exposure-resolution
          ($\mathcal{E}$-$\Delta$) plane for the NO case (left two panels) and the IO case (right two panels). The lightest neutrino mass is fixed to $0~{\rm eV}$.}
	\label{fig:chi2VSExposureVSDelta}
\end{figure}
%%%%%%%%%%%%%%%%%%%%%%%%%%%%%%%%%%%%%%%%%%%%%%%%%%%%%%%%%%%%%%%%%%%%%%%%%%%%%
	
To explicitly show the error of fitting the neutrino mass $m^{}_1$ with the effective spectrum, we calculate $\Delta \chi^2$ and present the final result with respect to $m^{}_1$ in Fig.~\ref{fig:chi2VSm1}.
In our calculations, we assume the true value of $m^{}_{1}$ to be $10~{\rm meV}$, corresponding to $m^{}_{\beta} = 13.4~{\rm meV}$. The energy resolution is fixed to $0.1~{\rm eV}$. The dark red curve represents the result obtained by fitting with the exact spectrum ${{\rm d}\Gamma^{\prime}_{\rm cl}}/{{\rm d}K^{}_e}$, while the dark blue one corresponds to the fit by using the effective spectrum ${{\rm d}\Gamma^{}_{\rm eff}}/{{\rm d}K^{}_e}$, given the exposure of ${\cal E} = 100~{\rm g}\cdot {\rm yr}$. The light curves stand for the case with ${\cal E} = 1~{\rm g}\cdot {\rm yr}$. One can observe that if the effective spectrum with $m^{}_\beta$ is used, the best-fit value of $m^{}_1$ is found to be $m^{\rm bf}_{1} = 9.6~{\rm meV}$, which deviates notably from $m^{\rm bf}_{1} = 10~{\rm meV}$ obtained by using the exact spectrum. Even for the exposure of ${\cal E} = 1~{\rm g\cdot yr}$ the true value is outside of $\Delta \chi^2 \lesssim 4$ when fitting with the effective spectrum. The situation becomes worse if we take a larger exposure.

To systematically study how far the parameter value fitted by using ${{\rm d}\Gamma^{}_{\rm eff}}/{{\rm d}K^{}_e}$ can deviate from the true one, we define the following difference of $\chi^2$
\begin{eqnarray}\label{eq:Dchi2}
\Delta \chi^2_{\rm true} \equiv \chi^{2}_{\beta}(m^{}_{\beta}=m^{\rm true}_{\beta}) - \chi^{2}_{\beta}(m^{}_{\beta} = m^{\rm bf}_{\beta})|^{}_{\rm min} \; ,
%     (19)
\end{eqnarray}
where $\chi^{2}_{\beta}(m^{}_{\beta}=m^{\rm true}_{\beta})$ is the $\chi^2$ value when $m^{}_{\beta}$ is set to $m^{\rm true}_{\beta}$ when fitting with ${{\rm d}\Gamma^{}_{\rm eff}}/{{\rm d}K^{}_e}$, and $\chi^{2}_{\beta}(m^{}_{\beta} = m^{\rm bf}_{\beta})|^{}_{\rm min}$ is the minimum value of the $\chi^2$-curve obtained by freely adjusting $m^{}_{\beta}$ with $m^{\rm bf}_{\beta}$ being the best-fit value. The value of
$m^{\rm true}_{\beta}$ can be directly obtained with Eq.~(\ref{eq:mbeta}) once the input value of $m^{}_{1}$ in ${{\rm d}\Gamma^{\prime}_{\rm cl}}/{{\rm d}K^{}_e}$ for simulating the data is given. The difference
$\Delta \chi^2_{\rm true}$ measures how likely one can recover the true value of the model parameter $m^{}_{\beta}$ by fitting with ${{\rm d}\Gamma^{}_{\rm eff}}/{{\rm d}K^{}_e}$. We present $\chi^2_{\beta}$ and $\Delta\chi^2_{\rm true}$ in Fig.~\ref{fig:chi2VSExposureVSDelta} as a function of the exposure $\mathcal{E}$ and the energy resolution $\Delta$.
We fix the lightest neutrino mass as $0~{\rm eV}$ for these plots, as $\chi^{2}_{\beta}$ is maximized in this case according to Fig.~\ref{fig:chi2VSDeltaVSm}.

The experimental configurations of KATRIN, Project 8 and PTOLEMY have been indicated in Fig.~\ref{fig:chi2VSExposureVSDelta}, and their corresponding $\chi^2$-values have been explicitly summarized in Table.~\ref{table:betaDecaysChi2}.
For PTOLEMY the effective beta spectrum can no longer be adopted. The use of the effective spectrum with $m^{}_{\beta}$ would result in a huge error in fitting the neutrino mass compared to the precision that is supposed to be achieved in such an experiment, e.g., $\Delta \chi^2_{\rm true} = 141$ for NO and $\Delta \chi^2_{\rm true} = 81$ for IO for one year of data taking. For KATRIN and Project 8 with one year of exposure, the effective mass $m^{}_{\beta}$ is fortunately applicable with $\chi^2_{\beta},\Delta \chi^2_{\rm true} \lesssim 0.1$. Note that there is a little risk for Project 8 loaded with the atomic tritium. To be more specific, in the extreme case that the data taking time is set to $10$ years and an improvement on the energy resolution is made to $\Delta = 0.03~{\rm eV}$, $\Delta \chi^2_{\rm true}$ for NO can be as large as $1$, indicating that the true value of $m^{}_{\beta}$ is out of the $1\sigma$ CL region by fitting with the effective spectrum ${{\rm d}\Gamma^{}_{\rm eff}}/{{\rm d}K^{}_e}$, hence the description by using the effective spectrum would not be appropriate anymore.

\section{Posterior Distributions}\label{sec:Bayes}

As we have already demonstrated in the previous section, the effective spectrum ${{\rm d}\Gamma^{}_{\rm eff}}/{{\rm d}K^{}_e}$ cannot be used for PTOLEMY, but is safe to use in the KATRIN and Project 8 experiments. Following the Bayesian statistical approach \cite{Trotta:2008qt}, we derive in this section the posterior distributions of the effective neutrino mass $m^{}_\beta$, based on current experimental information from neutrino oscillations, beta decay, neutrinoless double-beta decay ($0\nu \beta\beta$) and cosmology. Since the description of the beta spectrum via the effective neutrino mass is still valid for KATRIN and Project 8, posterior distributions of the effective neutrino mass should be very suggestive for future experiments.
Our results in this section can also be used for the electron-capture experiments ECHo~\cite{Gastaldo:2017edk}, HOLMES~\cite{Alpert:2014lfa} and NuMECS~\cite{Engle:2013qka}, if CPT is assumed
to be conserved in the neutrino sector.
For the similar analysis relevant for the effective neutrino masses in $\beta$ and $0\nu\beta\beta$ decays, see Refs.~\cite{Benato:2015via,DiIura:2016zsx,Caldwell:2017mqu,Ge:2017erv,Agostini:2017jim,Huang:2019qvq,DellOro:2019pqi}. Here we perform an updated analysis for the direct neutrino mass experiments, in light of a good number of experimental achievements.

As usual, two important ingredients for the Bayesian analysis should be specified. First, we have to choose the prior distributions for the relevant model parameters
\begin{eqnarray}\label{eq:modelpara}
\{\sin^2\theta^{}_{12}, \sin^2 \theta^{}_{13}, \Delta m^{2}_{\rm sol}, \Delta m^{2}_{\rm atm}, \rho, \sigma, G^{}_{0\nu}, |\mathcal{M}^{}_{0\nu}|, m^{}_{\rm L}\} \; ,
\end{eqnarray}
where $\Delta m^2_{\rm sol} = \Delta m^2_{21}$ and $\Delta m^2_{\rm atm} = \Delta m^2_{31}$ (or $\Delta m^2_{32}$) in the NO (or IO) case.
For all oscillation parameters $\{\sin^2\theta^{}_{12}, \sin^2\theta^{}_{13}, \Delta m^2_{\rm sol}, \Delta m^2_{\rm atm}\}$, we assume that they are uniformly distributed in the ranges that are wide enough to cover their experimentally allowed values.  For the absolute neutrino mass scale, which is represented by the lightest neutrino mass $m^{}_{\rm L}$ (i.e., $m^{}_1$ in the NO case or $m^{}_3$ in the IO case), we consider the following two possible priors:
\begin{itemize}
\item A flat prior on the logarithm of $m^{}_{\rm L}$ in the range of $(10^{-7}\cdots 10)~{\rm eV}$, namely, ${\rm Log}^{}_{10}(m^{}_{\rm L}/{\rm eV}) \in [-7, 1]$, which will be referred to as the log prior in the following discussion. This prior is scale invariant and motivated by the approximately constant ratios~\cite{Xing:2019vks} of charged fermion masses $m^{}_{\rm u}/m^{}_{\rm c} \sim m^{}_{\rm c}/m^{}_{\rm t} \sim \lambda^2$, $m^{}_{\rm d}/m^{}_{\rm s} \sim m^{}_{\rm s}/m^{}_{\rm b} \sim \lambda$, and $m^{}_e/m^{}_\mu \sim m^{}_\mu/m^{}_\tau \sim \lambda^2$ (where $\lambda = \sin\theta^{}_{\rm C} \approx 0.22$ is the Wolfenstein parameter), as well as by the in general exponential fermion mass hierarchies. Note that an {\it ad hoc} lower cutoff $10^{-7}~{\rm eV}$ for $m^{}_{\rm L}$ has been imposed, which is necessary to bound the prior volume from below.
Decreasing this cutoff is equivalent to putting more and more prior volume  to very small and essentially vanishing values of $m^{}_{\rm L}$.
%For example,  if we take the lower cutoff at $-\infty~{\rm eV}$, it is %equivalent to set $m^{}_{\rm L} = 0~{\rm eV}$, which results in an almost %fixed $m^{}_{\beta}$ as in Fig.~\ref{fig:mbetaVSm1}, i.e., $m^{}_{\beta} %\approx 0.00889~{\rm eV}$ for NO and $m^{}_{\beta} \approx 0.0499~{\rm eV}$ for IO.

\item A flat prior on $m^{}_{\rm L}$ in the range of $(0\cdots 10)~{\rm eV}$. Note that the ratio of the heaviest to the second-heaviest neutrino mass is rather small, at most $\sqrt{\Delta m^2_{31}/\Delta m^2_{21}} \approx 5$ for NO and essentially $1$ for IO, motivating a moderate and non-exponential ordering of neutrino masses.
%Such a kind of prior is expected if neutrino masses are all proportional to a common vacuum expectation value (vev) as in some neutrino mass models. However, it should be noticed that the quark and charged-lepton masses in the standard model are proportional to the vev of the Higgs field.
\end{itemize}
Without a complete theory for neutrino mass generation, we cannot judge which prior is favorable and therefore shall treat both of them on equal footing.
%The strong mass hierarchy of the charged fermions of the same electric charge and the approximate constancy of their mass ratios are long-standing puzzles, calling for further constraints on the flavor structures and the underlying symmetries~\cite{Xing:2019vks}.
The prior dependence of the final posterior distributions reflects that current experimental knowledge on the absolute scale of neutrino masses is still very poor.
If one attempts to set limits on model parameters, a prior-independent approach may be found in Ref.~\cite{Gariazzo:2019xhx}.
%As more precision data on absolute neutrino masses will be accumulated in  future experiments, such a prior dependence can be finally removed.

 The likelihood functions for each type of experiments can be found in Appendix~\ref{sec:appB}. Briefly speaking, the global-fit results of all neutrino oscillation data from Ref.~\cite{Esteban:2018azc} will be used to construct the likelihood function ${\cal L}^{}_{\rm osc}(\sin^2\theta^{}_{12}, \sin^2\theta^{}_{13}, \Delta m^2_{\rm sol}, \Delta m^2_{\rm atm})$. The tritium beta-decay experiments Mainz~\cite{Kraus:2004zw}, Troitsk~\cite{Aseev:2011dq} and KATRIN~\cite{Aker:2019uuj} are taken into account and the likelihood function ${\cal L}^{}_\beta(m^2_\beta)$ involves the model parameters $\{m^{}_{\rm L}, \sin^2\theta^{}_{12}, \sin^2\theta^{}_{13}, \Delta m^2_{\rm sol}, \Delta m^2_{\rm atm}\}$. As for $0\nu\beta\beta$ experiments, the likelihood function ${\cal L}^{}_{0\nu\beta\beta}(m^{}_{\beta\beta}, G^{}_{0\nu}, |{\cal M}^{}_{0\nu}|)$ actually contains all the parameters in Eq.~(\ref{eq:modelpara}). For the two Majorana CP phases $\rho$ and $\sigma$ relevant for $0\nu\beta\beta$ experiments, we shall take flat priors in the range of $[0 \cdots 2\pi)$, as there is currently no experimental constraint on them. For the phase space factor $G^{}_{0\nu}$, a Gaussian prior is assumed with the central value and $1\sigma$ error available from Ref.~\cite{Kotila:2012zza}. The nuclear matrix elements $|\mathcal{M}^{}_{0\nu}|$ take a flat prior in the range spanned by the predictions from different NME models~\cite{Caldwell:2017mqu}. Finally, the upper bound on the sum of three neutrino masses $\Sigma = m^{}_1 + m^{}_2 + m^{}_3$ from cosmological observations will be implemented, and the corresponding likelihood function ${\cal L}^{(i)}_{\rm cosmo}$ depends on $\{m^{}_{\rm L}, \Delta m^2_{\rm sol}, \Delta m^2_{\rm atm}\}$, where $i = 1, 2, 3$ refers respectively to the Planck data on the cosmic microwave background, its combination with gravitational lensing data, and their further combination with baryon acoustic oscillation data, as explained in Appendix \ref{sec:appB}.
%%%%%%%%%%%%%%%%%%%%%%%%%%%%%%%% Fig. 5 %%%%%%%%%%%%%%%%%%%%%%%%%%%%%%%%%
\begin{figure}[t!]
	\begin{center}
		\hspace{-0.2cm}
		\includegraphics[width=0.49\textwidth]{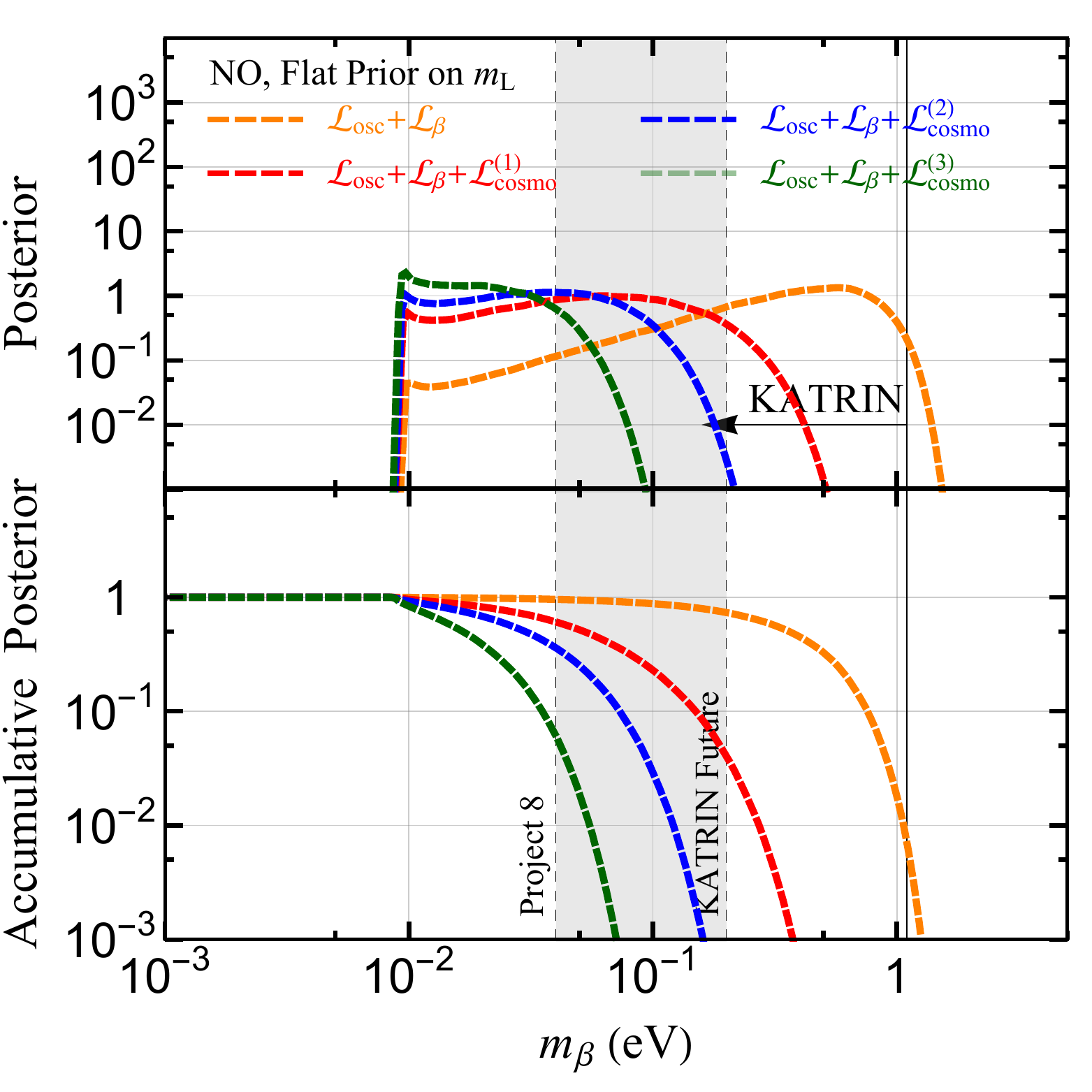}
		\includegraphics[width=0.49\textwidth]{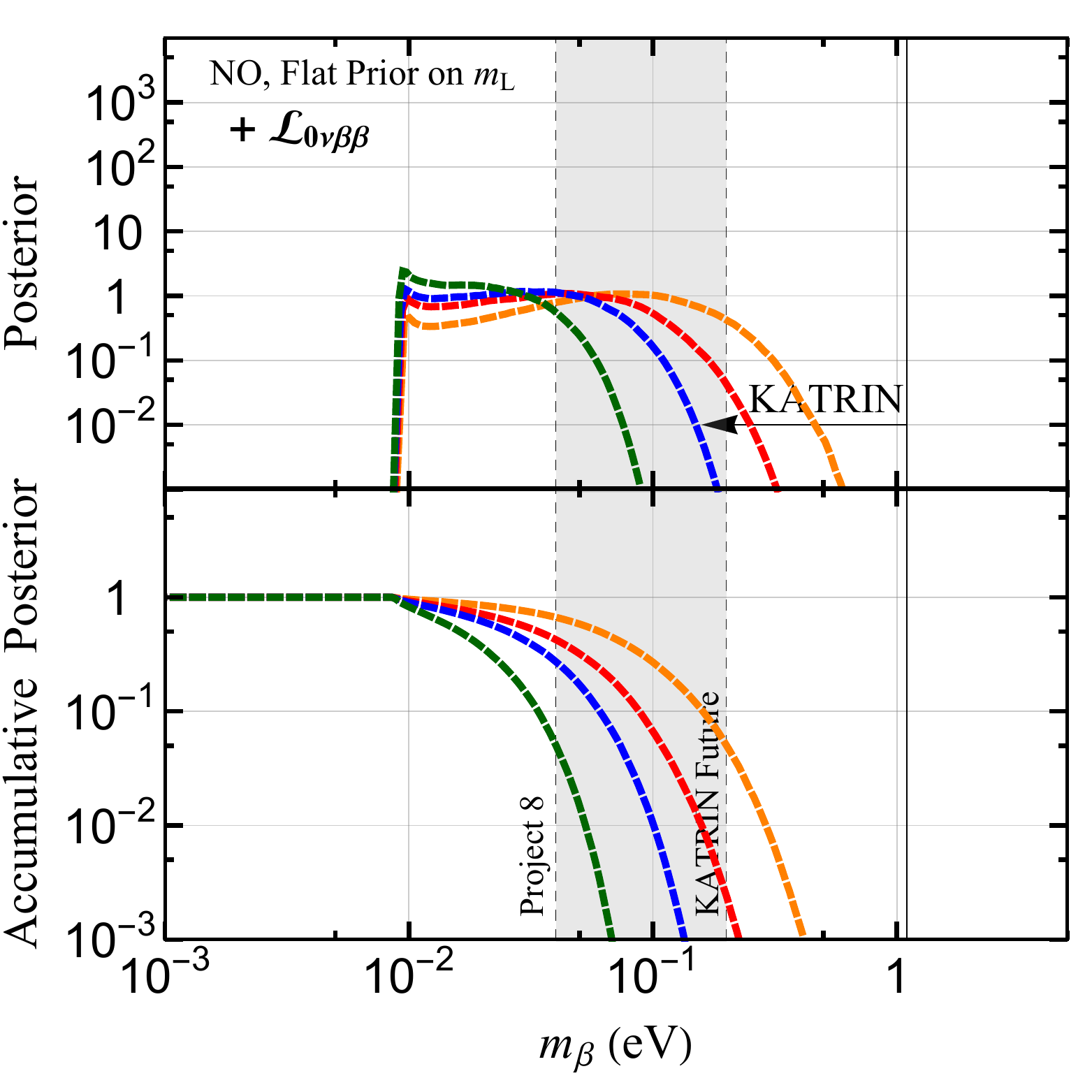}	\\
		\vspace{-0.2cm}
		\includegraphics[width=0.49\textwidth]{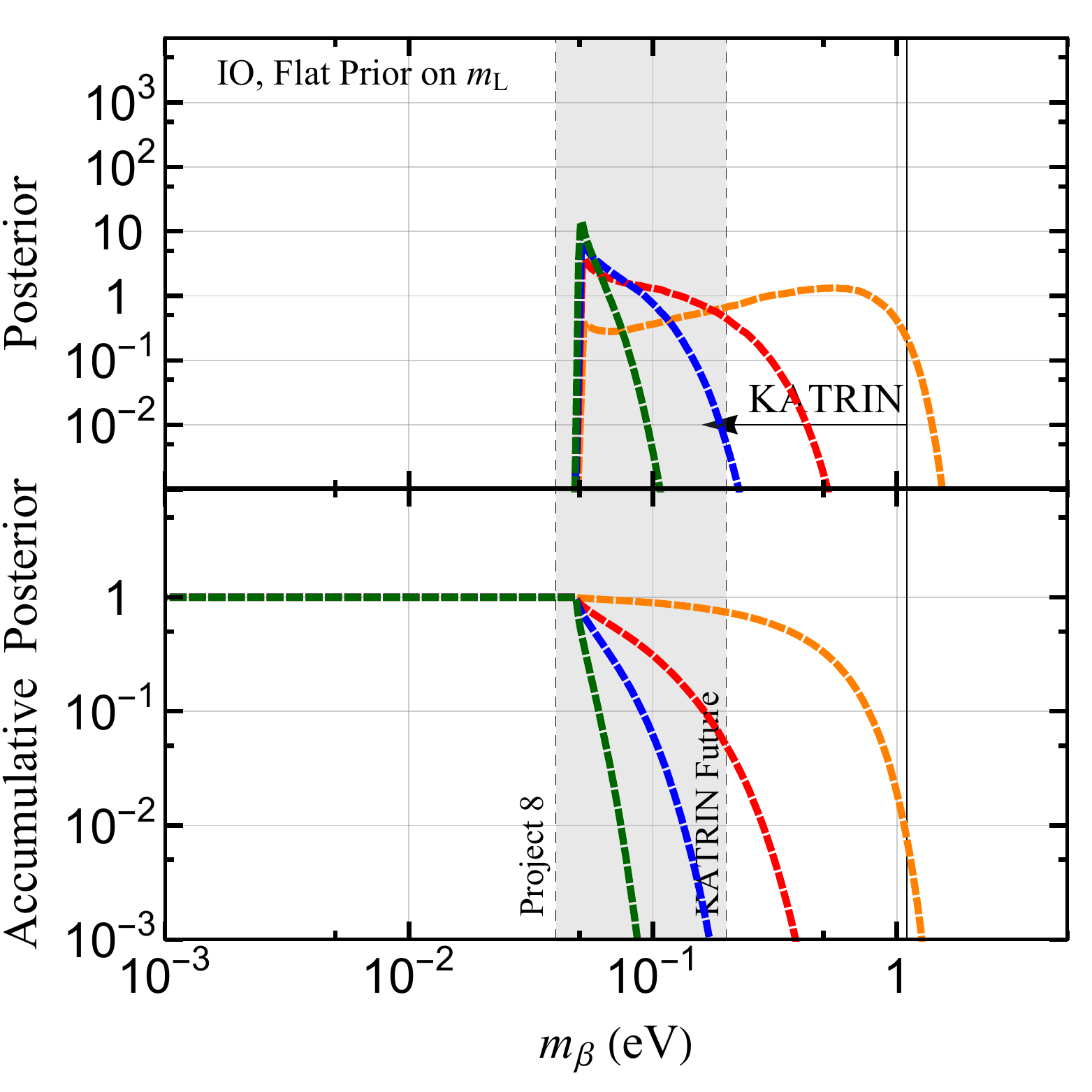}
		\includegraphics[width=0.49\textwidth]{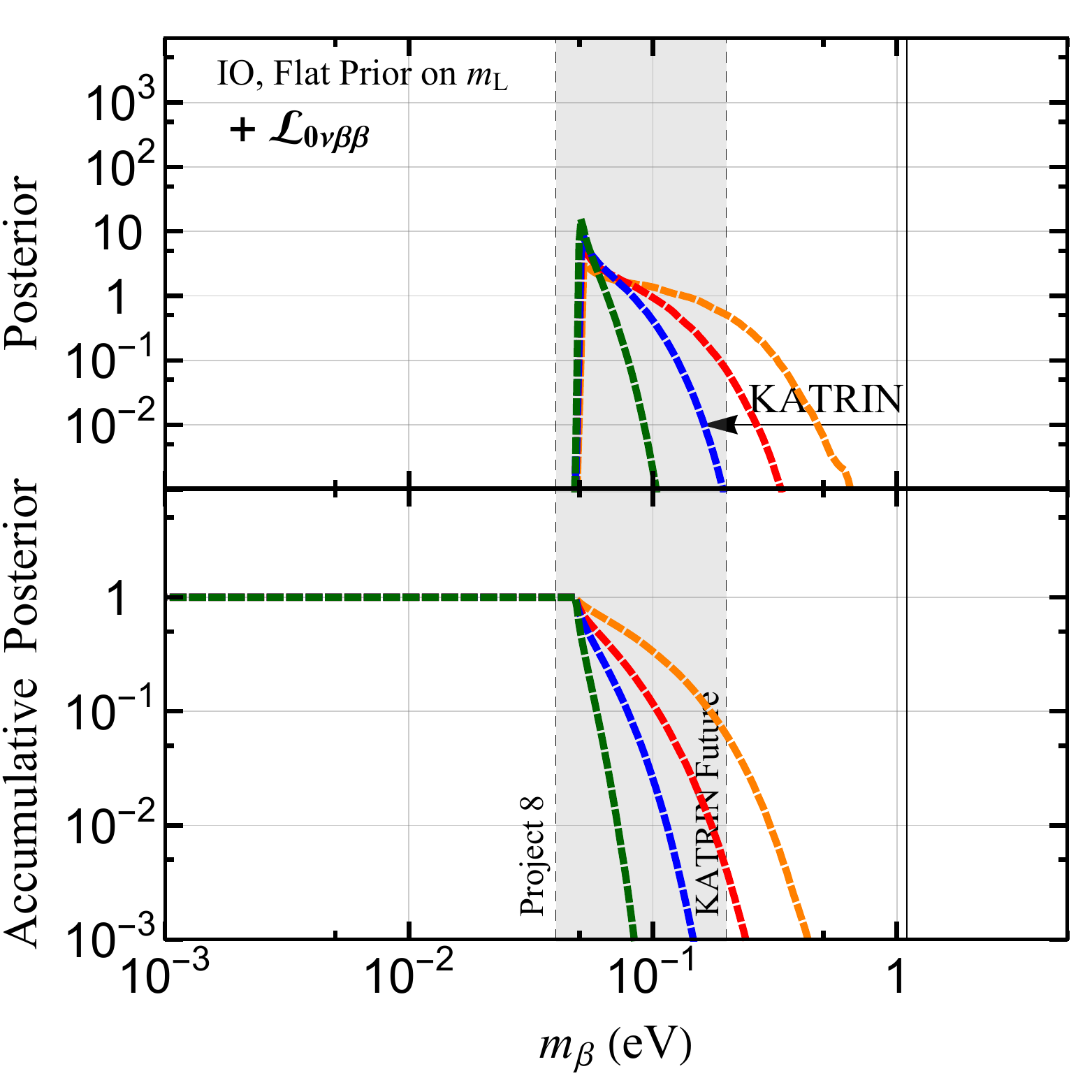}
	\end{center}
	\vspace{-0.5cm}
	\caption{The posterior distributions of the effective neutrino mass $m^{}_{\beta}$ in the NO (upper row) and IO (lower row) cases, given a {\bf flat prior} on the lightest neutrino mass $m^{}_{\rm L}$. In each of the four subfigures, the upper subgraph shows the posterior distributions whereas the lower subgraph gives the accumulative distributions. The results for four different combinations of experimental information have been displayed in the left column: (i) $\mathcal{L}^{}_{\rm osc}+\mathcal{L}^{}_{\beta}$ (orange curves); (ii) $\mathcal{L}^{}_{\rm osc}+\mathcal{L}^{}_{\beta} + \mathcal{L}^{(1)}_{\rm cosmo}$ (red curves); (iii) $\mathcal{L}^{}_{\rm osc}+\mathcal{L}^{}_{\beta} + \mathcal{L}^{(2)}_{\rm cosmo}$ (blue curves); (iv) $\mathcal{L}^{}_{\rm osc}+\mathcal{L}^{}_{\beta} + \mathcal{L}^{(3)}_{\rm cosmo}$ (green curves), while the data from $0\nu\beta\beta$ using $ \mathcal{L}^{}_{0\nu\beta\beta}$ are further included  in the right column. The cosmological bounds on the sum of three neutrino masses corresponding to $\mathcal{L}^{(i)}_{\rm cosmo}$ (for $i=1, 2, 3$) have been summarized in Eq.~(\ref{eq:planckSum}). The latest result $m^{}_\beta < 1.1~{\rm eV}$ from KATRIN is denoted as the vertical solid line, and future sensitivities of KATRIN and Project 8 are represented by two vertical dashed lines.}
	\label{fig:5}
\end{figure}
%%%%%%%%%%%%%%%%%%%%%%%%%%%%%%%%%%%%%%%%%%%%%%%%%%%%%%%%%%%%%%%%%%%%%%%%%%%%%

%%%%%%%%%%%%%%%%%%%%%%%%%%%%%%%% Fig. 6 %%%%%%%%%%%%%%%%%%%%%%%%%%%%%%%%%
\begin{figure}[t!]
	\begin{center}
		\hspace{-0.2cm}
		\includegraphics[width=0.49\textwidth]{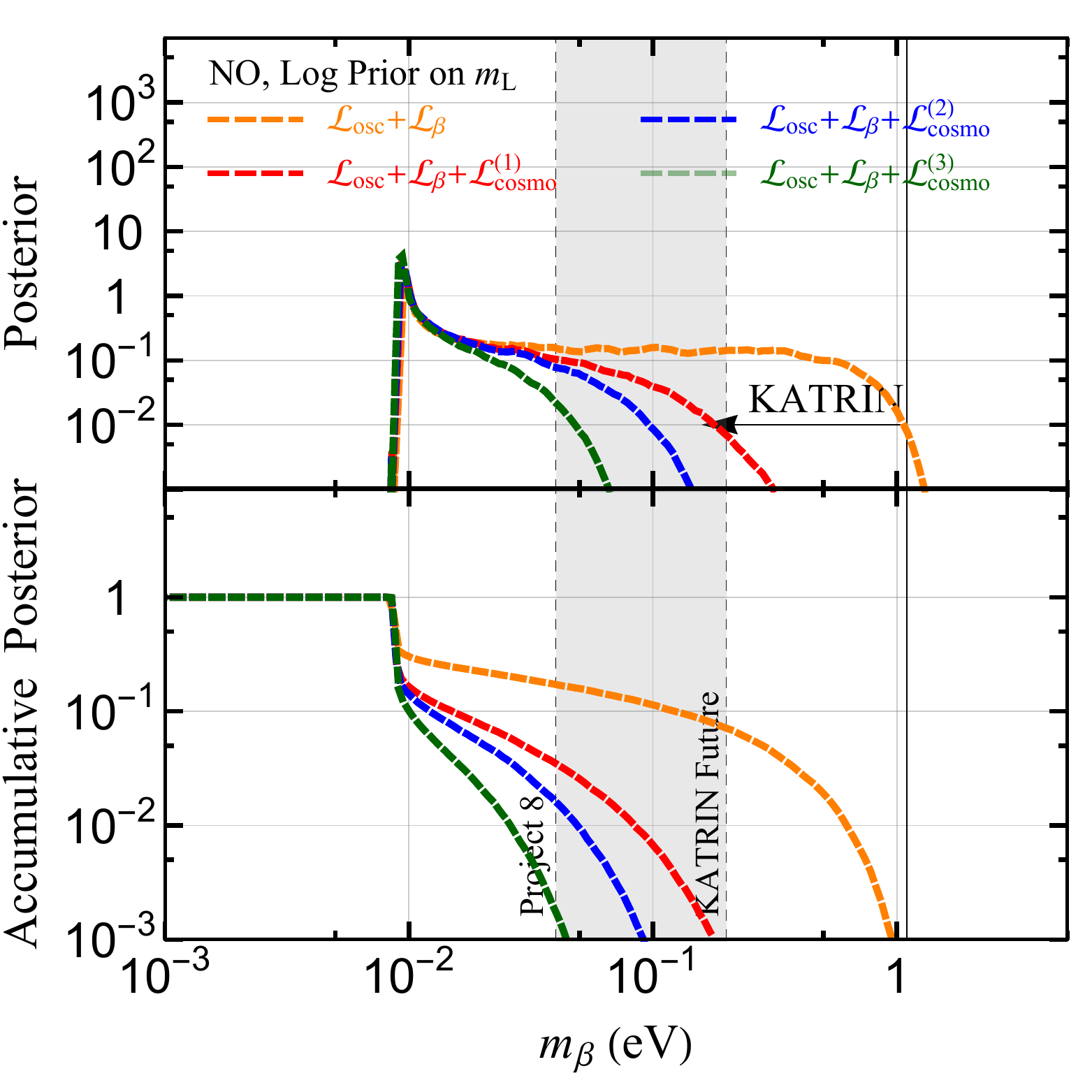}
		\includegraphics[width=0.49\textwidth]{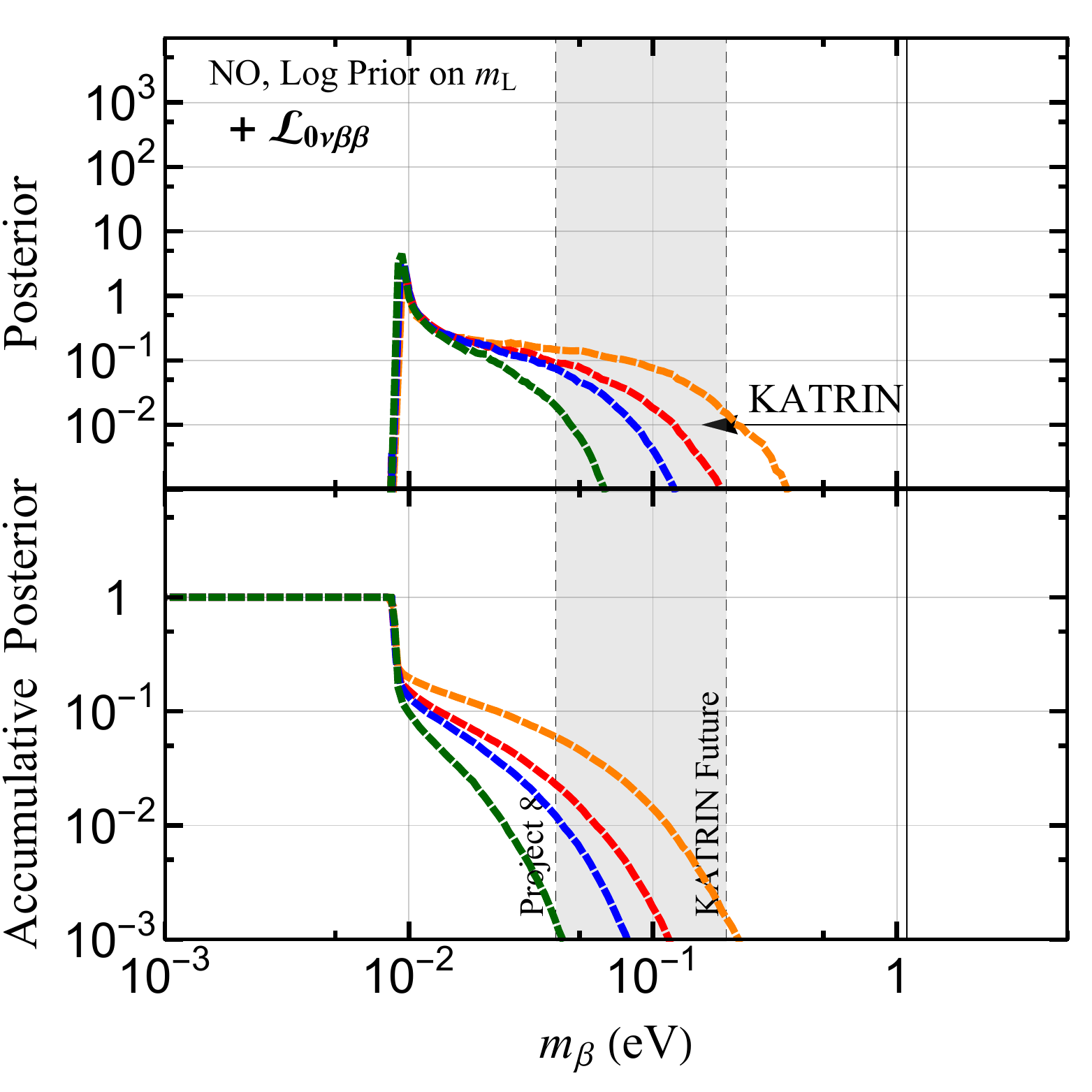}	\\
		\vspace{-0.2cm}
		\includegraphics[width=0.49\textwidth]{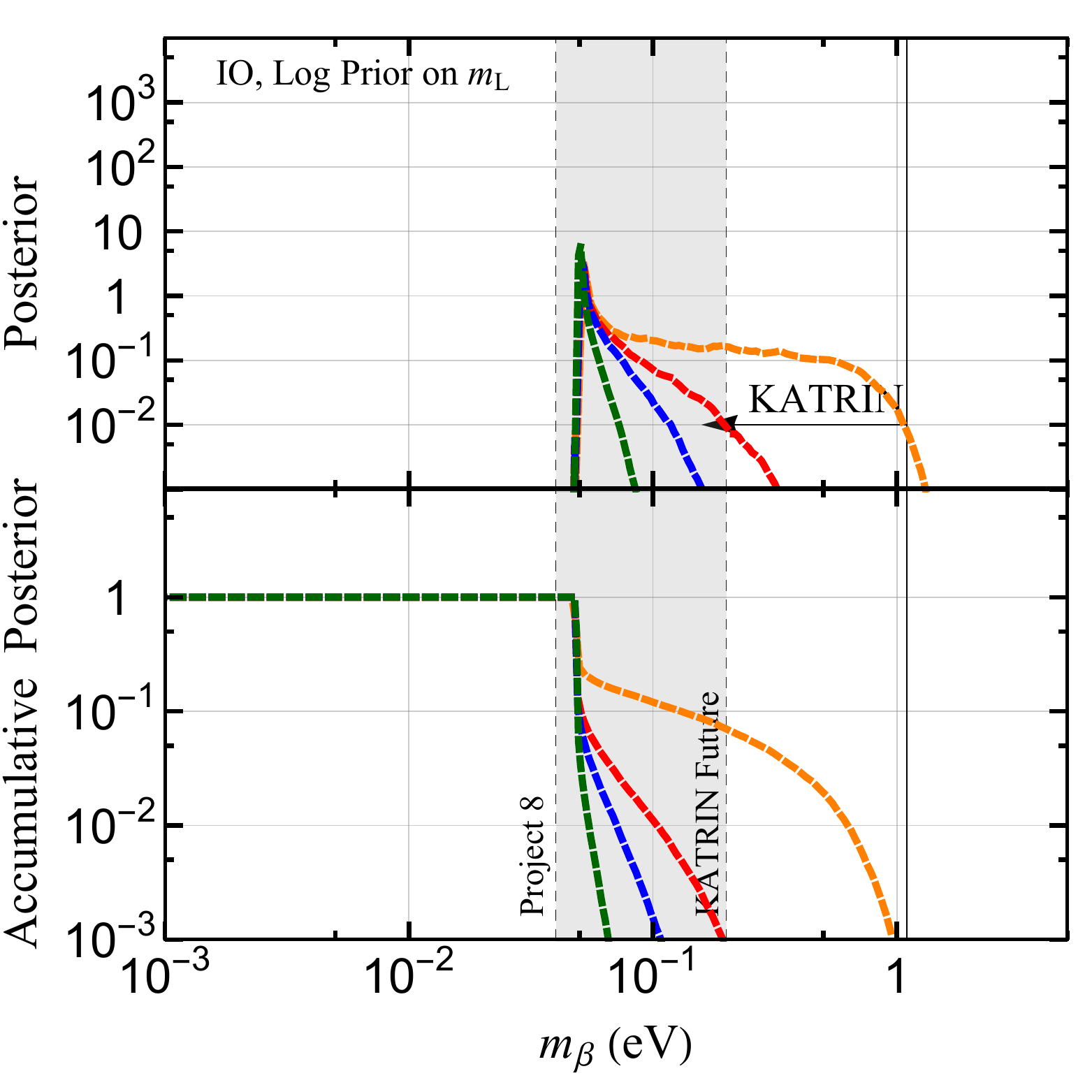}
		\includegraphics[width=0.49\textwidth]{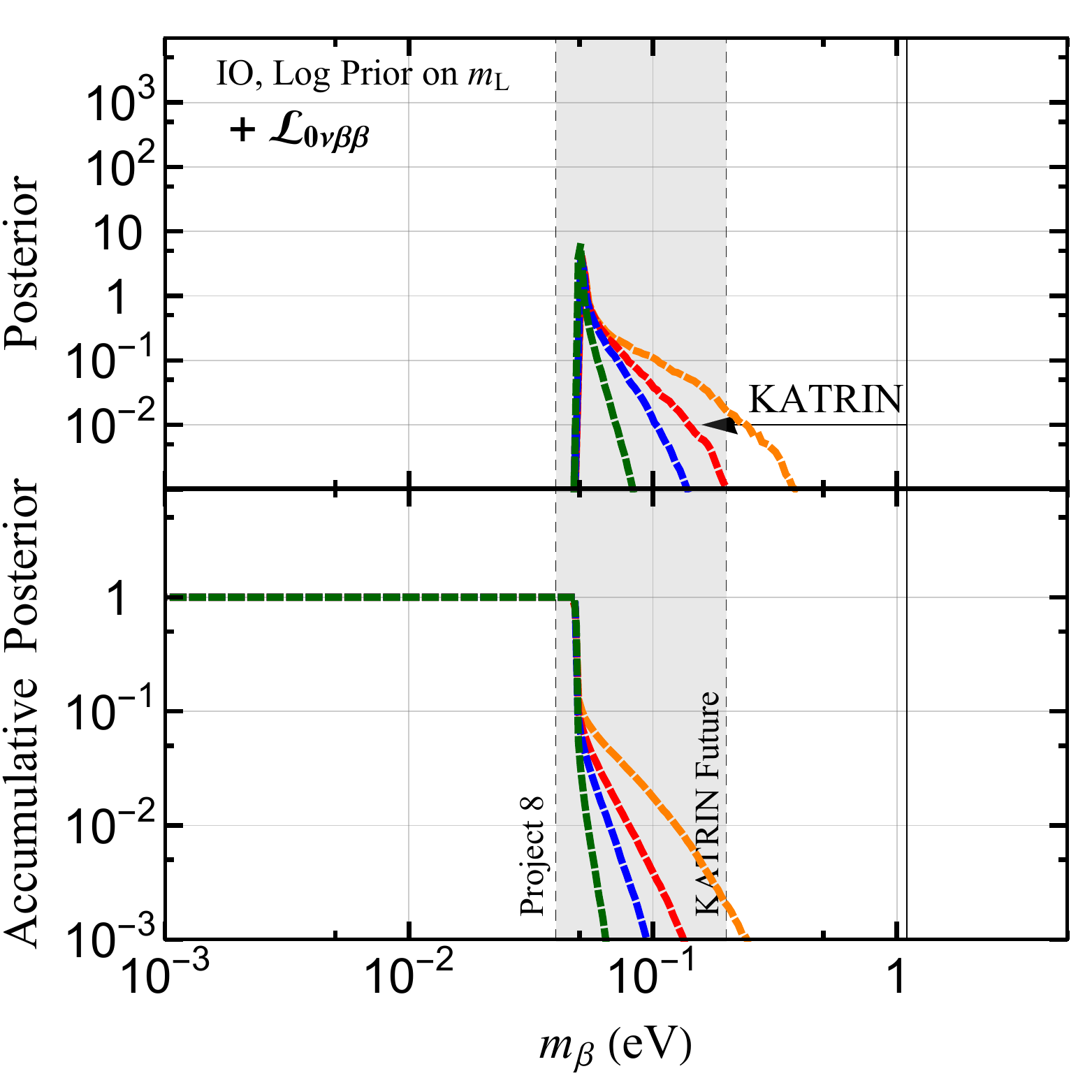}
	\end{center}
	\vspace{-0.5cm}
	\caption{The posterior distributions of the effective neutrino mass $m^{}_{\beta}$ in the NO (upper row) and IO (lower row) cases, given a  {\bf log prior} on the lightest neutrino mass $m^{}_{\rm L}$. In each of the four subfigures, the upper subgraph shows the posterior distributions whereas the lower subgraph gives the accumulative distributions. The results for four different combinations of experimental information have been displayed in the left column: (i) $\mathcal{L}^{}_{\rm osc}+\mathcal{L}^{}_{\beta}$ (orange curves); (ii) $\mathcal{L}^{}_{\rm osc}+\mathcal{L}^{}_{\beta} + \mathcal{L}^{(1)}_{\rm cosmo}$ (red curves); (iii) $\mathcal{L}^{}_{\rm osc}+\mathcal{L}^{}_{\beta} + \mathcal{L}^{(2)}_{\rm cosmo}$ (blue curves); (iv) $\mathcal{L}^{}_{\rm osc}+\mathcal{L}^{}_{\beta} + \mathcal{L}^{(3)}_{\rm cosmo}$ (green curves), while the data from $0\nu\beta\beta$ using $ \mathcal{L}^{}_{0\nu\beta\beta}$ are further included in the right column. The cosmological bounds on the sum of three neutrino masses corresponding to $\mathcal{L}^{(i)}_{\rm cosmo}$ (for $i=1, 2, 3$) have been summarized in Eq.~(\ref{eq:planckSum}). The latest result $m^{}_\beta < 1.1~{\rm eV}$ from KATRIN is denoted as the vertical solid line, and future sensitivities of KATRIN and Project 8 are represented by two vertical dashed lines.}
	\label{fig:6}
\end{figure}
%%%%%%%%%%%%%%%%%%%%%%%%%%%%%%%%%%%%%%%%%%%%%%%%%%%%%%%%%%%%%%%%%%%%%%%%%%%%%
\begin{table}[t]
	\scriptsize
	\centering
	\begin{tabular*}{\textwidth}{l @{\extracolsep{\fill}} c c c c c c c c }
		\hline
		\hline
		Flat Prior&      $\mathcal{L}^{}_{\rm osc} {+} \mathcal{L}^{}_{\beta}$
		& ${+}\mathcal{L}^{(1)}_{\rm cosmo}$ & ${+}\mathcal{L}^{(2)}_{\rm cosmo}$ & ${+}\mathcal{L}^{(3)}_{\rm cosmo}$& ${+}\mathcal{L}^{}_{\rm 0\nu\beta\beta}$& ${+}\mathcal{L}^{}_{\rm 0\nu\beta\beta}{+}\mathcal{L}^{(1)}_{\rm cosmo}$& ${+}\mathcal{L}^{}_{\rm 0\nu\beta\beta}{+}\mathcal{L}^{(2)}_{\rm cosmo}$& ${+}\mathcal{L}^{}_{\rm 0\nu\beta\beta}{+}\mathcal{L}^{(3)}_{\rm cosmo}$
		\\
		\hline
		\hspace{-0.3cm} KATRIN,NO & 73\% & 4.2\% & $6.9{\times} 10^{-5}$ & $<10^{-11}$ & $4.7\%$ & $0.23\%$ & $5.2{\times} 10^{-6}$ & $<10^{-11}$\\
		\hspace{-0.3cm}	KATRIN,IO & 74\% & 4.9\% & $1.2{\times} 10^{-4}$ & $<10^{-11}$ & $6.5\%$ & $0.45\%$ & $1.2{\times} 10^{-5}$ & $<10^{-11}$\\
		\hspace{-0.3cm}	Project 8,NO  & 96\% & 60\% & $35\%$ & $6\%$ & $67\%$ & $43\%$ & $28\%$ & $4.9\%$\\
		\hspace{-0.3cm} Project 8,IO  & $100\%$ & $100\%$ & $100\%$ & $100\%$ & $100\%$& $100\%$& $100\%$ & $100\%$\\
		\hline
		\hline \\
		\hline
		\hline
		Log Prior&  $\mathcal{L}^{}_{\rm osc} {+} \mathcal{L}^{}_{\beta}$
		& ${+}\mathcal{L}^{(1)}_{\rm cosmo}$ & ${+}\mathcal{L}^{(2)}_{\rm cosmo}$ & ${+}\mathcal{L}^{(3)}_{\rm cosmo}$& ${+}\mathcal{L}^{}_{\rm 0\nu\beta\beta}$& ${+}\mathcal{L}^{}_{\rm 0\nu\beta\beta}{+}\mathcal{L}^{(1)}_{\rm cosmo}$& ${+}\mathcal{L}^{}_{\rm 0\nu\beta\beta}{+}\mathcal{L}^{(2)}_{\rm cosmo}$& ${+}\mathcal{L}^{}_{\rm 0\nu\beta\beta}{+}\mathcal{L}^{(3)}_{\rm cosmo}$
		\\
		\hline
		\hspace{-0.3cm}	KATRIN,NO  & 7.2\% & $6.8 \times 10^{-4}$ & $7.9{\times} 10^{-7}$ & $<10^{-11}$ & $0.15\%$ & $3.7{\times} 10^{-5}$ & $6.4{\times} 10^{-8}$ & $<10^{-11}$\\
		\hspace{-0.3cm}	KATRIN,IO  & 7.1\% & $9.3 \times 10^{-4}$ & $1.4{\times} 10^{-6}$ & $<10^{-11}$ & $0.19\%$ & $6.6{\times} 10^{-5}$ & $1.7{\times} 10^{-7}$ & $<10^{-11}$\\
		\hspace{-0.3cm}	Project 8,NO & 17\% & 3.5\% & $1.6\%$ & $0.17\%$ & $6.0\%$ & $2.3\%$ & $1.2\%$ & $0.15\%$\\
		\hspace{-0.3cm}	Project 8,IO  & $100\%$ & $100\%$ & $100\%$ & $100\%$ & $100\%$& $100\%$& $100\%$ & $100\%$\\
		\hline
		\hline
	\end{tabular*}
	\caption{The volume fraction of the $m^{}_{\beta}$ posterior covered by KATRIN with a sensitivity of $m^{}_{\beta} \simeq 0.2~{\rm eV}$ and Project 8 with a sensitivity of $m^{}_{\beta} \simeq 0.04~{\rm eV}$, respectively. A flat and logarithmic prior on the lightest neutrino mass has been assumed for the upper and lower tables, respectively.}
	\label{table:flatPrior}
\end{table}
With the priors of model parameters and the likelihood functions from the relevant experiments, we can compute the posterior distribution of $m^{}_{\beta }$, i.e., $\mathrm{d} \mathcal{P} / \mathrm{d} m^{}_{\beta}$, in the standard way of Bayesian analysis. The sampling is done with the help of the MultiNest routine \cite{Feroz:2007kg,Feroz:2008xx,Feroz:2013hea}. The numerical results for the flat and log priors on $m^{}_{\rm L}$ are shown in Figs.~\ref{fig:5} and \ref{fig:6}, respectively.
A summary of the volume fractions of $m^{}_{\beta}$ posteriors covered by future KATRIN and Project 8 sensitivities has been presented in Table~\ref{table:flatPrior}.
Some comments on the numerical results are in order.
\begin{itemize}
\item In Fig.~\ref{fig:5}, a flat prior on $m^{}_{\rm L}$ is assumed. The plots in the first row are for the NO case, whereas those in the second row are for the IO case. In each row, the upper subgraph in the left column shows the posterior distributions of the effective mass in four different scenarios of adopted experimental information: (1) ${\cal L}^{}_{\rm osc} + {\cal L}^{}_\beta$ for the neutrino oscillation and beta decay data; (2) ${\cal L}^{}_{\rm osc} + {\cal L}^{}_\beta + {\cal L}^{(i)}_{\rm cosmo}$ (for $i = 1, 2, 3$) for a further inclusion of cosmological upper bounds on the sum of three neutrino masses. The lower subgraph gives the accumulative posterior distributions, which are defined as $\mathcal{P}(m^{}_{\beta} > m^{0}_{\beta}) = \int^{\infty}_{m^{0}_{\beta}} \left(\mathrm{d} {\mathcal{P}}/{\mathrm{d} m^{}_{\beta}}\right) \mathrm{d}{m^{}_{\beta}}$. In addition, the plots in the right column differ from those in the left column only by including the experimental information on $0\nu\beta\beta$. Hence, if neutrinos are Dirac particles the results including ${\cal L}^{}_{0\nu\beta\beta}$ do not apply.
  The future sensitivities of KATRIN~\cite{Angrik:2005ep} ($ 0.2~{\rm eV}$) and Project 8~\cite{Esfahani:2017dmu} ($40~{\rm meV}$) are shown as the upper and lower dashed boundaries of the gray bands. This gray region represents the gradual improvement of the sensitivities. In Fig.~\ref{fig:6}, the same computations have been carried out for the log prior on $m^{}_{\rm L}$, where all notations follow those of Fig.~\ref{fig:5}.

\item By comparing among the different scenarios in both Figs.~\ref{fig:5} and \ref{fig:6}, we can make the following important observations. \begin{enumerate}
    \item Let us first focus on the impact of $0\nu\beta\beta$. If the cosmological observations, namely, the upper bounds on neutrino masses, are not considered, then one can make a comparison between the orange curves (corresponding to ${\cal L}^{}_{\rm osc} + {\cal L}^{}_\beta$) in the left column and those (corresponding to ${\cal L}^{}_{\rm osc} + {\cal L}^{}_\beta + {\cal L}^{}_{0\nu\beta\beta}$) in the right column. It is evident that the experimental constraints from $0\nu\beta\beta$ decays lead to a significant shift of the posterior distribution to the region of smaller values of $m^{}_\beta$. Even in this case, it is very likely that the future beta-decay experiments can determine the absolute neutrino mass no matter whether NO or IO is true. For instance, in Fig.~\ref{fig:5} where the flat prior on $m^{}_{\rm L}$ is assumed, Project 8 can cover $67\%$ of the posteriors in the NO case.

     \item One should investigate what role is played by the cosmological observations. For this purpose, we concentrate on the plots in the right columns of Fig.~\ref{fig:5} and Fig.~\ref{fig:6}. When the cosmological observations are considered, one can see that the probability of discovering a nonzero effective neutrino mass in beta-decay experiments drops dramatically. In the worst situation, where the likelihood set $\mathcal{L}^{}_{\rm osc}+\mathcal{L}^{}_{\beta} + \mathcal{L}^{(3)}_{\rm cosmo} + \mathcal{L}^{}_{0\nu\beta\beta}$ is taken in the NO case, even Project 8 can only cover $4.9\%$ of the posterior. Therefore, the detection of a positive signal in this case would imply a tension between the beta-decay experiments and cosmological observations.

     \item For the log prior on $m^{}_{\rm L}$ in Fig.~\ref{fig:6}, it is easy to recognize that the posterior is remarkably reduced in the region of effective neutrino masses covered by the KATRIN and Project 8 experiments. Compared to the flat prior, this can be interpreted as a consequence of the fact that a large fraction of the prior space has been distributed in the neighborhood of a nearly vanishing $m^{}_{\rm L}$. For the NO case, the effective neutrino mass $m^{}_{\beta}$ takes the minimum value $\sim 8.9 \times 10^{-3}~{\rm eV}$ as $m^{}_{\rm L} = m^{}_1 \rightarrow 0~{\rm eV}$. The beta-decay experiments like KATRIN and Project 8 are still far from achieving this sensitivity.
\end{enumerate}

Regardless of the prior and likelihood choices, Project 8 can always cover all the posteriors of the IO case. In this connection, the discrimination between NO and IO seems to be very promising in future beta-decay experiments \cite{Bilenky:2006zd}, e.g. an explicit study of the sensitivity has already been performed for PTOLEMY in Ref.~\cite{Betti:2019ouf}.
\end{itemize}

\section{Summary}\label{sec:conclusion}

The determination of absolute neutrino masses is experimentally challenging, but scientifically very important. As  fundamental parameters in nature, absolute neutrino masses must be precisely measured in order to explore the origin of neutrino masses, which calls for new physics beyond the standard model. Motivated by the latest result from the KATRIN experiment and upcoming tritium beta decay searches, we have performed a detailed study of the exact electron spectrum ${{\rm d}\Gamma^{\prime}_{\rm cl}}/{{\rm d}K^{}_e}$ in Eq.\ (\ref{eq:Gammaprime}), which is a modified relativistic one,
and its difference to the effective electron spectrum ${{\rm d}\Gamma^{}_{\rm eff}}/{{\rm d}K^{}_e}$ in Eq.\ (\ref{eq:Gammaeff}) which includes the usually considered
effective neutrino mass $m^{}_\beta$ or its variants. Moreover, based on current experimental information from neutrino oscillation data, tritium beta decays, neutrinoless double-beta decays and cosmology, we have computed the posterior distributions of the effective neutrino mass $m^{}_\beta$ in Eq.~(\ref{eq:mbeta}). Our main results are summarized as follows.

First, for tritium beta decays, the classical electron spectrum ${\mathrm{d} \Gamma^{}_{\rm cl}}/{\mathrm{d} K^{}_{e}}$ can be modified by replacing $m^{}_i$ with $m^{}_i \cdot \left(m^{}_{{^3}{\rm He}}/m^{}_{{^3}{\rm H}}\right)$ to account for the exact electron spectrum including relativistic corrections. In this case, the difference between the exact relativistic spectrum and the modified classical spectrum ${\mathrm{d} \Gamma^{\prime}_{\rm cl}}/{\mathrm{d} K^{}_{e}}$ can be safely ignored, as the dominant uncertainties in the measurements at KATRIN, Project 8 and PTOLEMY arise from the statistical data fluctuations. Furthermore, it is interesting to compare the exact spectrum with the effective one containing the usually considered observable $m^{}_\beta$. However, as we have demonstrated in a quantitative way, the validity of the effective mass $m^{}_\beta$ actually depends on the energy resolution and the total exposure of a realistic beta-decay experiment. We show that the use of the standard  effective neutrino mass for KATRIN and Project 8 is justified. For the future PTOLEMY experiment with an exposure of $100~{\rm g}\cdot {\rm yr}$, it will be problematic to introduce an effective neutrino mass, and the lightest neutrino mass should be used together with the exact spectrum. While this is known, we have performed here a general analysis with keeping
the exposure and energy resolution as free parameters.

Second, as we have mentioned above, it is justified to describe the exact electron spectrum ${{\rm d}\Gamma^{\prime}_{\rm cl}}/{{\rm d}K^{}_e}$ by the effective one with the effective neutrino mass $m^{}_\beta$ in the KATRIN and Project 8 experiments. Therefore, it does make sense to derive the posterior distributions of the effective neutrino mass, given the latest experimental data on neutrino oscillations, beta decays, neutrinoless double-beta decays and cosmological observations. Although the cosmological upper bound on the sum of three neutrino masses pushes the posterior distribution of $m^{}_\beta$ down to the region almost outside of the sensitivity of Project 8 in the NO case, it does not affect much the situation in the IO case due to the lower bound on $m^{}_\beta \gtrsim 50~{\rm meV}$ even in the limit of $m^{}_3 \to 0$. This also implies that future tritium beta-decay experiments are able to discriminate between neutrino mass orderings.

As KATRIN continues to accumulate more beta-decay events and the development of the techniques to be deployed in Project 8 and PTOLEMY is well in progress, it is timely and necessary to revisit the effective neutrino mass and its validity in future beta-decay experiments. The analysis presented in the present work should be helpful in understanding the approximations made in expressions of the beta spectrum and is suggestive for the improvement on the usage of the effective masses. In light of the precision measurement of the beta spectrum already in the first run of KATRIN, one may go further to extend the analysis to consider the presence of sterile neutrinos and other new physics,
and/or to consider the electron-capture decay of  ${^{163}{\rm Ho}}$.

\section*{Acknowledgments}
The authors would like to thank Yu-Feng Li and Prof.\ Zhi-zhong Xing for useful discussions. This work was supported in part by the National Natural Science Foundation of China under Grant No.~11775232 and No.~11835013, and by the CAS Center for Excellence in Particle Physics.
WR is supported by the DFG with grant RO 2516/7-1 in the Heisenberg program.

\appendix
\numberwithin{equation}{section}

\section{Experimental Setups}\label{sec:appA}
Some experimental details about KATRIN, Project 8 and PTOLEMY experiments are as follows:
%and summarize their main features and projected sensitivities to $m^{}_\beta$ below.
\begin{itemize}
\item The KATRIN experiment~\cite{Angrik:2005ep} implements the so-called MAC-E-Filter (Magnetic Adiabatic Collimation combined with an Electrostatic Filter) to select  electrons from tritium beta decays that can pass through the electrostatic barrier with the potential energy of $E^{}_{\rm V}$. The observable in the MAC-E-Filter is the integrated number of the electrons that have passed through the energy barrier. The sharpness of the filter is characterized by the ratio between the minimum $B^{}_{\rm min} = 3\times 10^{-4}~{\rm T}$ and the maximum $B^{}_{\rm max} = 6~{\rm T}$ of magnetic fields, i.e., $\Delta  = E^{}_e B^{}_{\rm min}/B^{}_{\rm max} \approx 1~{\rm eV}$, where $E^{}_e \approx Q = 18.6~{\rm keV}$ is the electron energy in the range close to the endpoint and $Q \equiv m^{}_{{^3{\rm H}}} - m^{}_{{^3{\rm He}}} - m^{}_e$ is the $Q$-value for tritium beta decay. Since the filter is insensitive to the transverse kinetic energy of electrons, the sharpness denotes roughly the maximum of transverse kinetic energies and thus can be regarded as the energy resolution. Adopting the energy window $E^{}_{\rm V} \in [ Q - 30~{\rm eV}, Q + 5~{\rm eV} ]$ and including all the statistic and systematic uncertainties, the KATRIN experiment~\cite{Angrik:2005ep} with a target tritium mass of $\mathcal{O}(10^{-4}_{})~{\rm g}$ can measure $m^{2}_{\beta}$ with a $1\sigma$ uncertainty of $0.025~{\rm eV^2}$, corresponding to the sensitivity of $m^{}_{\beta} < 0.2~{\rm eV}$ at the $90\%$ CL in the assumption of $m^{}_\beta = 0$ as the true value.
KATRIN can also directly measure the non-integrated beta spectrum by extracting the time of flight information from the source to the detector, operating in the so-called MAC-E-TOF mode. Since the emitting time of the electron at source is not directly measurable, a technique has been devised to infer the emitting time by chopping the source with some high voltage potential frequently. A lower counting rate and a worse energy resolution will be caused by the additional chopping procedure.
The total target mass of $^{3}{\rm H}$ planned to be loaded in the full KATRIN setup can be inferred from the formula of the tritium molecule number $N(T^{}_{2}) = A^{}_{\rm S}\cdot \epsilon^{}_{\rm T}\cdot \rho d \approx 2.518 \times 10^{19}$ with the source cross section $A^{}_{\rm S} = 53~{\rm cm^2}$, the tritium purity $\epsilon^{}_{\rm T} = 0.95$ and the column density $\rho d = 5 \times 10^{17}~{\rm cm^{-2}}$, see Eq.~(25) and Table~7 of Ref.~\cite{Angrik:2005ep} for details. Given the mass per tritium nucleus $\sim 5 \times 10^{-24}~{\rm g}$, we obtain the total target mass of the full KATRIN as $m^{}_{\rm KATRIN} = 2.5 \times 10^{-4}~{\rm g}$. The energy resolution of KATRIN in this work is fixed to $\Delta^{}_{\rm KATRIN} = 1~{\rm eV}$.

\item Unlike the KATRIN experiment, the Project 8 collaboration will utilize the technique of cyclotron radiation emission spectroscopy to measure the electron energies~\cite{Esfahani:2017dmu}. If the magnetic field is uniform in the spectrometer, the cyclotron radiation of accelerating electrons can be observed for a few microseconds and its frequency can be precisely determined, leading to an excellent energy resolution. As has already been shown in Fig.~5 of Ref.~\cite{Esfahani:2017dmu}, with the deployment of $\mathcal{O}(10^{-4})~{\rm g}$ atomic $^3 {\rm H}$ and one year of running time, Project 8 is able to push the upper limit on the effective neutrino mass down to $m^{}_\beta < 40~{\rm meV}$ at the $90\%$ CL, assuming the true value of $m^{}_{\beta} = 0$.
In this work, we will adopt two extreme setups for Project 8: (i) an intermediate phase with the molecular $^{3}{\rm H}$, a target mass of $m^{}_{\rm P8}=5\times 10^{-4}~{\rm g}$ corresponding to $5 \times 10^{19}$ tritium molecules, and an energy resolution $\Delta^{}_{\rm P8m} = 0.36~{\rm eV}$ limited by the irreducible width of the final state molecular excitations \cite{Monreal:2009za};
(ii) an ultimate phase with the atomic $^{3}{\rm H}$, a target mass of $m^{}_{\rm P8}=5\times 10^{-4}~{\rm g}$, and an energy resolution of $\Delta^{}_{\rm P8a} = 0.05~{\rm eV}$ which is limited by the inhomogeneity of the magnetic field $\Delta B/B \sim 10^{-7}$ \cite{Esfahani:2017dmu}.\footnote{The energy resolution of Project 8 with atomic tritium may be roughly obtained by the relation $\Delta E / m^{}_{e} = \Delta f/ f \approx \Delta B/B$, where $f$ is the frequency of the cyclotron radiation and $B$ is the assumed nearly uniform magnetic field.} The target mass $m^{}_{\rm P8}=5\times 10^{-4}~{\rm g}$ can be achieved with a gas volume of $100~{\rm m^2}$ as required by the phase IV of Project 8 and a gaseous tritium number density of $10^{12}~{\rm cm^{-3}}$.

\item The PTOLEMY experiment has been designed to detect the cosmic neutrino background (C$\nu$B)~\cite{Betts:2013uya, Baracchini:2018wwj, Betti:2019ouf} via the electron-neutrino capture on tritium $\nu^{}_e + {^3{\rm H}} \to e^- + {^3{\rm He}}$, as suggested by Steven Weinberg in 1962~\cite{Weinberg:1962zza}. Thanks to the large target mass of $100~{\rm g}$ tritium and the low background rate required for the C$\nu$B detection, PTOLEMY would have an overwhelmingly better sensitivity to the absolute neutrino mass than KATRIN does, namely, the relative uncertainty reaches $\sigma(m^{}_{1})/m^{}_{1} \lesssim 10^{-2}$ for $m^{}_{1}=10~{\rm meV}$ with an energy resolution of $\Delta = 100~{\rm meV}$. In the PTOLEMY experiment, the energy of electrons from tritium beta decays will be measured in three steps. First, the MAC-E-Filter is used to select the electrons close to the endpoint, preventing the calorimeter from being swamped by the huge number of events in the energy range below the endpoint. Second, after passing through the MAC-E-Filter, the electrons are then sent to a long uniform solenoid, undergoing the cyclotron motion in the magnetic field of $2~{\rm T}$. Hence the radio signal can be implemented to track each single electron. Finally, the electrons are decelerated by the electrostatic voltage until their kinetic energies are $100~{\rm eV}$ or so to match the dynamic range of a cryogenic calorimeter. The energy resolution of these electrons can be as low as $50~{\rm meV}$~\cite{Betti:2019ouf}.
\end{itemize}

\section{Experimental Likelihoods}\label{sec:appB}

The likelihood functions from the following different classes of experiments have been used in our analysis: (i) oscillation experiments $\mathcal{L}^{}_{\rm osc}$; (ii) $\beta$-decay experiments $\mathcal{L}^{}_{\beta}$; (iii) $0\nu\beta\beta$ experiments $\mathcal{L}^{}_{\rm 0\nu\beta\beta}$; (iv) cosmological observations $\mathcal{L}^{}_{\rm cosmo}$. To be specific we collect all the details of each likelihood function as follows:
\begin{itemize}
\item $\mathcal{L}^{}_{\rm osc}$---The likelihood information of neutrino oscillation experiments will be taken from the latest global-fit results of the Nu-Fit group \cite{Esteban:2018azc}. The likelihood function can be obtained as
	$\mathcal{L}^{}_{\rm osc} = \exp{(-\Delta \chi^2/2)}$
	with $\Delta \chi^2$ defined as
	\begin{eqnarray} \label{eq:chis}
	\Delta \chi^2 \equiv \sum^{}_{i} \frac{(\Theta^{}_{i}-\Theta^{\rm bf}_{i})^2}{\sigma^{2}_{i}}\;,
	%     (21)
	\end{eqnarray}
	where $\Theta^{}_{i} \in \{ \sin^2\theta^{}_{13}, \sin^2 \theta^{}_{12}, \Delta m^{2}_{\rm sol}, \Delta m^{2}_{\rm atm} \}$, $\Theta^{\rm bf}_{i}$ is the best-fit value of the parameter from the global analysis, and $\sigma^{}_{i}$ is the symmetrized $1\sigma$ error. We take the following central values and symmetrized $1\sigma$ errors of oscillation parameters relevant for the $\beta$ decays:
	\begin{eqnarray} \label{eq:gfno}
	&& \sin^2\theta^{}_{12} = (3.10\pm 0.12) \times 10^{-1} \; , \quad \quad  \Delta m^2_{\rm sol} = (7.39 \pm 0.20) \times 10^{-5} ~{\rm eV}^2  \; ,
	\nonumber \\\nonumber
	&& \sin^2\theta^{}_{13} = (2.241 \pm 0.065) \times 10^{-2}\; , \quad
	\Delta m^2_{\rm atm} = (2.525 \pm 0.032) \times 10^{-3} ~{\rm eV}^2 \; ,\hspace{0.6cm}
	%		(22)
	\end{eqnarray}
	for ${\rm NO}$, and
	\begin{eqnarray}\label{eq:gfio}
	&& \sin^2\theta^{}_{12} = (3.10\pm 0.12) \times 10^{-1} \; , \quad \quad \Delta m^2_{\rm sol} = (7.39 \pm 0.20) \times 10^{-5} ~{\rm eV}^2 \; , 	\nonumber \\\nonumber
	&& \sin^2\theta^{}_{13} = (2.264 \pm 0.066)\times 10^{-2} \; , \quad \Delta m^2_{\rm atm} = (-2.512 \pm 0.033) \times 10^{-3} ~{\rm eV}^2 \; ,\hspace{0.6cm}
	%		(23)
	\end{eqnarray}
	for IO. The preference of NO over IO can be represented by the difference of their $\chi^2$-minima, i.e., $\Delta \chi^2_{\rm MO} = \chi^2_{\rm NO}- \chi^2_{\rm IO} \simeq 9.3$,
	implying a more than $3\sigma$ preference of NO.
\item $\mathcal{L}^{}_{\beta}$---By measuring the endpoint of the $\beta$ decay spectrum, the tritium $\beta$-decay experiments (e.g., Troitsk~\cite{Aseev:2011dq}, Mainz~\cite{Kraus:2004zw} and KATRIN \cite{Aker:2019uuj,Aker:2019qfn}) can already provide us good constraints on the absolute neutrino mass scale via the effective neutrino mass $m^{}_{\beta}\equiv \left(\sum^{}_{i} m^2_{i}|U^{}_{ei}|^2\right)^{1/2}$. The limits of the former two are given as
	\begin{eqnarray}\label{eq:betaconstraints}
	m^2_{\beta} = -0.67 \pm 2.53~{\rm eV^2}~{\rm (Troitsk)}, \quad
    m^2_{\beta} = -0.6 \pm 3.0~{\rm eV^2}~{\rm (Mainz)}.
	\end{eqnarray}
    Similar to Eq.~(\ref{eq:chis}) in the case of neutrino oscillations, the likelihood function can be constructed with the central values and $1\sigma$ errors of $m^2_\beta$ in Eq.~(\ref{eq:betaconstraints}). For KATRIN, we use the likelihood presented in Fig.~4 of Ref.~\cite{Aker:2019uuj}. We find the likelihood can be well approximated by a skewed normal distribution:
	\begin{eqnarray}
	\mathcal{L}^{}_{\rm KATRIN} (m^2_{\beta})\propto \frac{1}{\sqrt{2\pi}\sigma}
	\exp \left[-\frac{(m^2_{\beta}-\mu)^2}{2\sigma^2}\right] {\rm erfc}\left[-\frac{\alpha (m^2_{\beta}-\mu)}{\sqrt{2}\sigma}\right] .
	\end{eqnarray}
where ${\rm erfc}(x)$ is the complementary error function, with $\sigma = 1.506$ eV$^2$, $\mu = 0.0162$ eV$^2$ and $m^{2}_{\beta}$ in units of ${\rm eV}^2$, as well as $\alpha = -2.005$. Since the KATRIN experiment has the highest sensitivity to $m^{}_\beta$, we may have ${\cal L}^{}_\beta \approx {\cal L}^{}_{\rm KATRIN}$.
\item $\mathcal{L}^{}_{0\nu\beta\beta}$---The constraints on the half-life of $0\nu\beta\beta$  are given by the existing $0\nu\beta\beta$ searches. The limits on the effective neutrino mass $|m^{}_{\beta\beta}|$ can be derived by using
	\begin{eqnarray}\label{eq:halflife}
	\left(T^{0\nu}_{1/2}\right)^{-1} = G^{}_{0\nu} \left|{\cal M}^{}_{0\nu}\right|^2 \frac{\left|m^{}_{\beta\beta}\right|^2}{m^2_e} \; ,
	%     (1)
	\end{eqnarray}
	where $G^{}_{0\nu}$ denotes the phase-space factor, ${\cal M}^{}_{0\nu}$ is the nuclear matrix element (NME), and $m^{}_e = 0.511~{\rm MeV}$ is the electron mass. In our numerical analysis we use the likelihood functions from Refs.~\cite{Caldwell:2017mqu, Alduino:2017ehq}, which include the experimental information of GERDA \cite{Agostini:2017iyd}, KamLAND-Zen \cite{KamLAND-Zen:2016pfg}, EXO \cite{Albert:2014awa} and CUORE \cite{Alduino:2017ehq}.
	\item $\mathcal{L}^{}_{\rm cosmo}$---The cosmological observations can set very strong constraints on the sum of the three neutrino masses $\Sigma \equiv m^{}_{1} + m^{}_{2} +m^{}_{3}$. The Planck collaboration has recently updated their results in Ref.~\cite{Aghanim:2018eyx}. For illustration, we will adopt the likelihood functions by combining different datasets which yield the following bounds on the sum of the three neutrino masses at the $95\%$ CL:
\begin{eqnarray} \label{eq:planckSum}
	& & \Sigma < 0.54~{\rm eV} \text{ ($\mathcal{L}^{(1)}_{\rm cosmo}$, $Planck ~{\rm TT}+ {\rm lowE}$)} , \nonumber \\
	& & \Sigma < 0.24~{\rm eV} \text{ ($\mathcal{L}^{(2)}_{\rm cosmo}$, $Planck ~{\rm TT}, {\rm TE}, {\rm EE} + {\rm lowE} + {\rm lensing}$)}, \\
	& & \Sigma < 0.12~{\rm eV} \text{ ($\mathcal{L}^{(3)}_{\rm cosmo}$, $Planck ~{\rm TT}, {\rm TE}, {\rm EE} + {\rm lowE} + {\rm lensing} + {\rm BAO}$)} . \nonumber
\end{eqnarray}
The likelihood functions have been obtained by analyzing the Markov chain files available from the Planck Legacy Archive.
%~\footnote{This is based on the observations with {\it Planck} (\url{http://www.esa.int/Planck}), an ESA science mission with instruments and contributions directly funded by ESA Member States, NASA, and Canada.}.
\end{itemize}

With the likelihood functions listed above, the total likelihood relevant for our analysis can be calculated as $\mathcal{L}^{}_{\rm tot} = \mathcal{L}^{}_{\rm osc} \times \mathcal{L}^{}_{\rm \beta} \times \mathcal{L}^{}_{\rm 0\nu\beta\beta} \times \mathcal{L}^{(i)}_{\rm cosmo}$ (for $i=1,2,3$).

\bibliographystyle{utcaps_mod}
\bibliography{references}

%\printbibliography

%\bibliography{references}

\end{document}